\documentclass[onecolumn,12pt]{article}
\pdfoutput=1
\usepackage{jheppub}
\usepackage{ifpdf}
\usepackage[bb=boondox]{mathalfa}
\graphicspath{ {figures/} }
\usepackage{amsfonts}
\usepackage{amsmath}
\usepackage{amssymb}
\usepackage{epsfig}
\usepackage{tensor}
\usepackage[usenames,dvipsnames]{xcolor}
\usepackage{pdfpages}
\usepackage{bbm}
\usepackage{graphicx,epstopdf}
\usepackage[numbers]{natbib} 
\usepackage[makeroom]{cancel}
\usepackage{bookmark}
\usepackage{array}
\usepackage[export]{adjustbox}
\usepackage[normalem]{ulem}
\usepackage{romannum}
\usepackage{ulem}
\usepackage{subcaption}

\newcommand{\m}[1]{\mathcal{#1}}
\newcommand{\mh}[1]{\hat{\mathcal{#1}}}
\newcommand{\RN}[1]{	\textup{\uppercase\expandafter{\romannumeral#1}}}

\newcommand{\eg}{{\it e.g.,}\ }
\newcommand{\ie}{{\it i.e.,}\ }

\newcommand{\mt}[1]{\textrm{\tiny #1}}

\newcommand{\beq}{\begin{equation}}
	\newcommand{\eeq}{\end{equation}}
\newcommand{\beqa}{\begin{eqnarray}}
	\newcommand{\eeqa}{\end{eqnarray}}

\newcommand{\bea}{\begin{eqnarray}}
	\newcommand{\eea}{\end{eqnarray}}

\newtheorem{thr}{Theorem}

\subheader{\today}
\title{\boldmath Holographic scattering and non-minimal RT surfaces}

\author[a,b]{Jacqueline Caminiti,}
\author[a, b]{Batia Friedman-Shaw,}
\author[a]{Alex May,}
\author[a]{Robert C. Myers}
\author[a,d]{and Olga Papadoulaki}

\affiliation[a]{Perimeter Institute for Theoretical Physics, \\
	Waterloo, ON N2L 2Y5, Canada}
\affiliation[b]{Department of Physics \& Astronomy, University of Waterloo, \\
	Waterloo, ON N2L 3G1, Canada}
 \affiliation[c]{CPHT, CNRS, \'Ecole polytechnique, Institut Polytechnique de Paris,\\
 91120 Palaiseau, France
 }

\emailAdd{jcaminiti@perimeterinstitute.ca}
\emailAdd{bfriedmanshaw@perimeterinstitute.ca}
\emailAdd{amay@perimeterinstitute.ca}
\emailAdd{rmyers@perimeterinstitute.ca}
\emailAdd{olga.papadoulaki@polytechnique.edu}

\abstract{In the AdS/CFT correspondence, the causal structure of the bulk AdS spacetime is tied to entanglement in the dual CFT. 
This relationship is captured by the connected wedge theorem \cite{May:2019odp}, which states that a bulk scattering process implies the existence of $O(1/G_N)$ entanglement between associated boundary subregions. 
In this paper, we study the connected wedge theorem in two asymptotically AdS$_{2+1}$ spacetimes: the conical defect and BTZ black hole geometries. 
In these settings, we find that bulk scattering processes require not just large entanglement, but also additional restrictions related to candidate RT surfaces which are non-minimal. 
We argue these extra relationships imply a certain CFT entanglement structure involving internal degrees of freedom. 
Because bulk scattering relies on sub-AdS scale physics, this supports the idea that sub-AdS scale locality emerges from internal degrees of freedom. 
While the new restriction that we identify on non-minimal surfaces is stronger than the initial statement of the connected wedge theorem, we find that it is necessary but still not sufficient to imply bulk scattering in mixed states. 
}

\begin{document}
\setlength{\parskip}{0pt}

\maketitle

\flushbottom

\section{Introduction}
\label{sec:intro}

A new guiding principle in the search for quantum gravity has become that entanglement builds the geometry of spacetime. 
This idea takes its most explicit form in the context of the AdS/CFT correspondence \cite{Mark_BuildingUpSpacetime, CoolHorizons, Faulkner_2014, swingle2014universality}, but can also be argued for from the perspective of semiclassical gravity \cite{Jacobson_1995, Jacobson_2016}, loop quantum gravity \cite{Donnelly_2008, Donnelly_2012,Bianchi_2023}, and entanglement entropy in QFT \cite{Rob_OnArch, Cooperman_2014}.
In AdS/CFT, semiclassical gravity in an asymptotically AdS$_{d+1}$ spacetime $\m{M}$ is argued to be equivalent to a strongly coupled CFT in $d$ dimensions (or CFT$_d$) defined on its conformal boundary $\partial\m{M}$. 
The link between entanglement and geometry is conventionally captured by the Ryu-Takayanagi (RT) formula \cite{Ryu06_1,Ryu06_2} and its generalizations \cite{Hubeny_2007,Faulkner_2013,Dong_2014}, which relate the entanglement entropy of boundary subregions to areas of minimal surfaces in the bulk spacetime. 
Further, when $d=2$ and the bulk spacetime satisfies the null energy condition, boundary entanglement is also related to bulk causal geometry, as captured by the connected wedge theorem \cite{May:2019yxi,May:2019odp,May:2021nrl,May:2022clu}.

The central concept in the connected wedge theorem is that of a causal discrepancy, first studied in the holographic context by \cite{Gary_2009, Heemskerk_2009, Penedones_2011, maldacena2015looking} to understand the singularity structure of boundary correlators.
A causal discrepancy occurs when a classical scattering process is causally permitted in the bulk spacetime $\m{M}$ but causally forbidden in the boundary $\partial\m{M}$. 
Intuitively, the extra spatial dimension in $\m{M}$ makes shortcuts available, enabling certain processes that may be impossible in $\partial \m{M}$ alone.
By the AdS/CFT correspondence, this means, for example, that two quantum systems living in the boundary theory that could never have collided might display input and output states related by a scattering-like unitary. 
This leads to the surprising conclusion that the holographic boundary theory must be able to implement ``scattering without scattering.''

\begin{figure}
    \centering
    \subfloat[\label{fig:connected-scatteringintro}]{
    \includegraphics[scale=0.5]{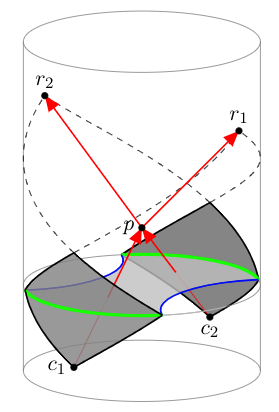}
    }
    \qquad\qquad\qquad\qquad
    \subfloat[\label{fig:disconnected-scatteringintro}]{
    \includegraphics[scale=0.5]{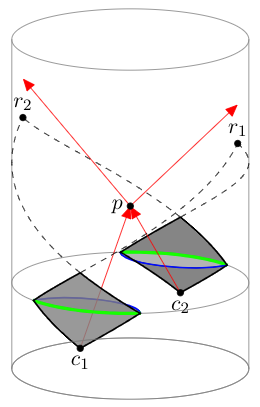}
    }
    \caption{(a) When there exist future-directed causal curves from $c_1, c_2$ and past-directed causal curves from $r_1, r_2$ all meeting at a common point in the bulk, the connected wedge theorem states that associated boundary regions (green) must share large correlation. This also implies the minimal-area extremal surface (blue) homologous to the boundary regions must connect the two boundary regions through the bulk. With some modifications, the points $c_i$ and $r_i$ can be taken away from the asymptotic boundary \cite{May:2021nrl}. (b) When there is no scattering region in the bulk, the boundary regions need not share large correlations. Thus, they can have  disconnected minimal-area extremal surfaces. This figure is reproduced from \cite{May:2019yxi}.}
    \label{fig:phasetransition3dintro}
\end{figure}

The connected wedge theorem states that the ability of the boundary theory to implement scattering-without-scattering tasks relies on its entanglement properties. 
Concretely, the theorem shows that the existence of a holographic causal discrepancy implies an $O(1/G_N)$ correlation between associated boundary subregions, denoted $\mh{V}_1$ and $\mh{V}_2$, as quantified by the mutual information $I(\mh{V}_1:\mh{V}_2)$.  
Via the RT formula, this condition may be expressed in terms of the properties of bulk extremal surfaces. 
To do so, first label the entanglement wedge of a boundary region $\mh{A}$ by $\m{E}(\mh{A})$.
Then, use that $I(\mh{V}_1:\mh{V}_2)=O(1/G_N)$ is equivalent to $\m{E}(\mh{V}_1\cup\mh{V}_2)$ being connected, while $I(\mh{V}_1:\mh{V}_2)=O(1)$ when 
$\m{E}(\mh{V}_1\cup\mh{V}_2)$ consists of two disconnected components. 
See figure \ref{fig:phasetransition3dintro}. 
Translated to bulk language, then, the connected wedge theorem states that bulk scattering implies $\mh{V}_1$ and $\mh{V}_2$ have a connected entanglement wedge. 

While the connected wedge theorem says that bulk scattering implies boundary entanglement, the converse direction is not true: two boundary regions can share $O(1/G_N)$ entanglement without there being any associated bulk scattering. 
In this case,  $\m{E}(\mh{V}_1\cup\mh{V}_2)$ is connected but fails to have a corresponding holographic scattering process. 
While this can be viewed as an unfortunate weakness of the theorem, it raises an interesting question: what distinguishes correlations in holographic states that support scattering from those that do not? 
Towards better understanding this issue, we study which connected entanglement wedges in the conical defect and BTZ black hole geometries correspond to a holographic scattering process.
We do so systematically, and point out various settings where such a correspondence fails to exist.
Note that one can also study this question in higher dimensions; even though it is not known whether the connected wedge theorem holds for $d>2$, it is still true that the connectedness of the entanglement wedge does not always imply bulk scattering.
We comment on the higher-dimensional case in the discussion section.

In the BTZ and heavy conical defect geometries, there are three candidate entanglement wedges of interest: one disconnected configuration and two connected configurations.
In one of the connected configurations, the massive object in the spacetime (the defect or black hole) sits inside the entanglement wedge; in the other, the massive object is outside. 
We call these the $d$ (disconnected), $u$ (excluding the massive object) and $o$ (containing the massive object) configurations --- see figure \ref{fig:dcircs}. 
The connected wedge theorem implies that when holographic scattering occurs, the extremal-area surfaces for either the $u$ wedge, the $o$ wedge, or both are smaller than those for the $d$ wedge, which we summarize by $\min(u,o)<d$. 
We verify this statement in the geometries under consideration, but we also find that a stronger statement holds: $u<d$ whenever bulk scattering exists.
In words, the wedge that excludes the massive object has less area than the disconnected one, whether or not it corresponds to the minimal-area configuration.
Indeed, there exist choices of intervals for which $o<u<d$.
Thus, the existence of bulk scattering is constraining a \emph{non-minimal} extremal surface.

We argue that this condition on a non-minimal RT surface gives a more refined diagnostic of the entanglement structure of the CFT.
To do this, we focus on orbifold geometries, such as defect geometries with $2\pi/n$ opening angle with integer $n$, for which smooth AdS$_3$ provides an $n$-fold covering space. The non-minimal extremal surfaces can then be interpreted as minimal surfaces in the covering geometry. 
Following \cite{Entwinement_2015}, we argue the lengths of minimal surfaces in the covering geometry describe entanglement among between internally, rather than spatially, organized CFT degrees of freedom in the original CFT, and we interpret the $u<d$ condition in this context.  
Our construction suggests bulk scattering relies on a particular pattern of entanglement among internal degrees of freedom in the CFT. 
This provides a step towards understanding what types of correlations distinguish holographic states that support scattering from those that do not.
Further, because bulk scattering relies on sub-AdS locality, this conclusion supports earlier arguments \cite{susskind1999holography, Heemskerk_2009, banks1999m,anous2020areas,Balasubramanian_2005,Entwinement_2015} that have pointed to structure in internal degrees of freedom as responsible for sub-AdS scale physics in the bulk. 

In the most general setting, we consider scattering processes with inputs and outputs at arbitrary points in the bulk spacetime.
To further explore the correlation structure of holographic states, we also consider restricting to scattering processes with inputs and outputs at the conformal boundary, referred to as ``points-based'' scattering processes.\footnote{We refer to scattering processes which begin and end at the conformal boundary as ``points-based,'' as opposed to ``regions-based,'' scattering processes, since the natural language for discussing quantum information stored deep in the bulk is entanglement wedge reconstruction, which identifies certain codimension-1 subregions of the bulk with codimension-1 subregions of the boundary. 
Similarly, we refer to the special case of the connected wedge theorem involving points-based scattering as the ``points-based'' theorem, as opposed to the (more general, ``regions-based'') connected wedge theorem.}
In this setting, we find that many holographic states do not support bulk scattering. 
For example, in the BTZ black hole, whenever there is a bulk scattering process with inputs and outputs at the conformal boundary, there is also a scattering process within the boundary geometry. 
The same occurs for the conical defect geometry with defects above a threshold mass of $M=-4/9$. 
From the CFT perspective, bulk scattering processes which begin and end at the conformal boundary correspond to the application of local operators. 
Thus, our observations suggest that correlations in these states cannot be harnessed to perform bulk-only scattering by the application of local operators alone. 
Instead, their correlation is such that it can only be exploited for this purpose by applying delocalized operators.
Further motivations to study the points-based theorem include that it can be easily argued for at the level of quantum information without assuming entanglement wedge reconstruction \cite{May:2019yxi} and that, as hinted above, it seems closely related to the behaviour of boundary four-point functions, which are the objects usually studied in the context of understanding the bulk causal structure from a boundary perspective.

The remainder of the paper is organized as follows: in section \ref{sec:geometryandwedges}, we introduce the conical defect and BTZ black hole spacetimes, and we discuss the structure of their entanglement wedges. 
In section \ref{sec:holscat}, we introduce the connected wedge theorem and apply it to these spacetimes, uncovering a connection between holographic scattering and non-minimal RT surfaces.
We close with a discussion of our results and possible future directions in section \ref{sec:discuss}.
In particular, we interpret the role of non-minimal RT surfaces, consider higher-dimensional cases, and provide further discussion of CFT correlations that support bulk scattering versus those that do not.
In appendix \ref{sec:pointsappendix}, we apply the points-based version of the connected wedge theorem to the defect and BTZ geometries, highlighting similarities and differences to section \ref{sec:holscat}.
In appendix \ref{sec:lightconeappendix}, we justify various geometric claims about the AdS$_3$ spacetimes under consideration. 
In appendix \ref{sec:regionsappendix}, we derive the holographic scattering inequalities presented in section \ref{sec:geometryandwedges}.
In appendix \ref{sec:higherdappendix}, we present the holographic scattering problem for strip regions in Poincar\'e AdS$_4$.

\subsection{Summary of notation}

\begin{itemize}
\itemsep0.25em 
\item We use curly capital letters $\m{A}$, $\m{B}$, $\m{C}$, \ldots for codimension-0 bulk spacetime regions.
\item We use curly capital letters with hats $\mh{A}$, $\mh{B}$, $\mh{C}$, \ldots for codimension-0 boundary regions.
\item We use plain capital letters $A$, $B$, $C$, \ldots for codimension-1 boundary regions.
\item We use $\m{J}^{\pm}(\m{S})$ to denote the causal future or past of the bulk region $\m{S}$ and $\mh{J}^{\pm}(\mh{S})$ to denote the causal future or past of $\mh{S}$ taken in the boundary. Bulk sets like $\m{J}^+(\m{S})$ are taken within the conformally completed spacetime that includes the asymptotic boundary, so they can include boundary points.
\item The RT surface associated to region $A$ is denoted $\gamma_A$.
\item The entanglement wedge of a boundary region $\mh{S}$ is denoted by $\m{E}(\mh{S})$.
\item The causal wedge $\m{J}^+(\mh{S})\cap \m{J}^-(\mh{S})$ of a boundary region $\mh{S}$ is denoted by $\m{C}(\mh{S})$.
\item The domain of dependence of a bulk or boundary region is denoted by $\m{D}(\cdot)$.
\end{itemize}
\section{\texorpdfstring{Entanglement wedges in defect and BTZ geometries}{Entanglement wedges in defect and BTZ geometries}}
\label{sec:geometryandwedges}

In this section, we introduce the conical defect and BTZ black hole geometries. We present the three relevant entanglement wedge configurations for boundary subregions consisting of two intervals $A$ and $B$, as well as the competition among the areas of their associated extremal surfaces.

\subsection{Defect and BTZ geometries}
\label{sec:sec21}

We will consider static, axisymmetric, asymptotically AdS$_{2+1}$ spacetimes described by the following metric:
\begin{equation}
\begin{aligned}
ds^2 &= -\left(r^2-M\right)dt^2 + \frac{dr^2}{r^2-M}+r^2d\phi^2 \,.
\end{aligned}
\label{eq:mumetric}
\end{equation}
Here $r\in[0,\infty)$ and $t\in(-\infty,\infty)$, while $\phi$ is a periodic angular coordinate with $\phi\sim\phi+2\pi$.\footnote{We work in units where the AdS curvature scale is set to one, \ie $\ell_{AdS}=1$.}
The parameter $M$ determines the ADM energy according to $\mu_\mt{ADM}=M/(8G_N)$. Setting $M=-1$ yields pure AdS$_3$ in global coordinates, while setting $M\geq0$ yields a (nonrotating) BTZ black hole with the horizon at $r=\sqrt{M}$ \cite{BTZ_1992,BTZ_1993}. 

\begin{figure}[htbp]
\centering
\includegraphics[width=.5\textwidth]{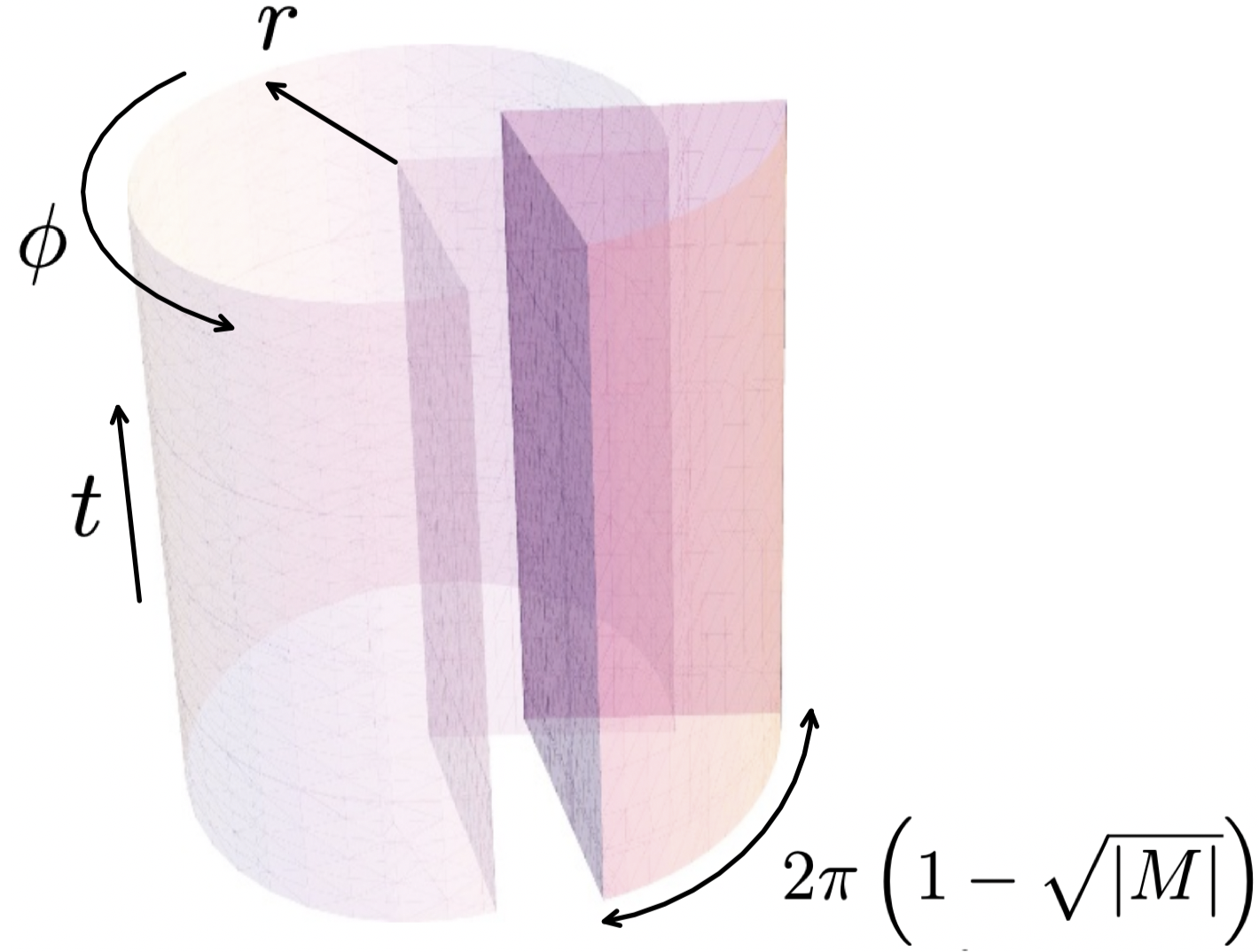}
\caption{Penrose diagram of AdS$_3$ with a wedge of angular extent $\delta\tilde\phi=2\pi\left(1-\sqrt{|M|}\right)$ removed. The conical defect spacetime corresponding to eq.~\eqref{eq:mumetric} with $-1<M<0$ is obtained by identifying the faces left exposed by this cutting procedure.\label{fig:wedge}}
\end{figure}

Also of interest is the regime $-1<M<0$, which corresponds to the geometry produced by a point particle of mass $\mu:=\frac{1}{4G_N}\left(1-\sqrt{|M|}\right)$ sitting at $r=0$ \cite{Deser1984, Deser19842}. This spacetime has a conical defect at $r=0$, a feature best understood by performing the following change of coordinates \cite{Entwinement_2015}:
\begin{equation}
\begin{aligned}
\tilde{r} &= \frac{r}{\sqrt{|M|}} \,,\qquad \tilde{t} &= \sqrt{|M|}\,t \,,\qquad \tilde{\phi} &= \sqrt{|M|}\,\phi \,.
\end{aligned}
\label{eq:defecttopure}
\end{equation}
In the new coordinates, the metric reads
\begin{equation}
\begin{aligned}
ds^2 &= -\left(\tilde{r}^2+1\right)d\tilde{t}^2 + \frac{d\tilde{r}^2}{\tilde{r}^2+1}+\tilde{r}^2d\tilde{\phi}^2 \,,
\end{aligned}
\label{eq:primemetric}
\end{equation}
which is pure AdS$_3$ except for the identification $\tilde{\phi}\sim\tilde{\phi}+2\pi\sqrt{|M|}$. Therefore, the conical defect spacetime is what results from removing a wedge of angle $2\pi \left(1-\sqrt{|M|}\right)$ from pure AdS$_3$ and gluing together the exposed faces --- see figure \ref{fig:wedge}.

Let us mention two special cases of the conical defect geometry.
First, setting $M=-1/n^2$ with integer $n$, one obtains the orbifold geometry AdS$_3$/$\mathbb{Z}_n$. 
This will be useful in section \ref{sec:discuss} when discussing that a certain entanglement structure exists among internal degrees of freedom of the dual CFT. 
Second, in the limit of increasing the particle mass towards $\mu=1/(4G_N)$, we obtain the BTZ black hole with $M=0$. 
In this sense, the conical defect geometries can be thought of as interpolating between the pure AdS$_3$ ($M=-1$) and BTZ ($M\geq0$) geometries.

For all choices of $M$, the bulk spacetime $\m{M}$ may be assigned the same conformal boundary $\partial\m{M}$, namely $S^1\times\mathbb{R}$ equipped with metric
\begin{equation}
   ds^2=-dt^2+d\phi^2\,.
   \label{boundmetric}
\end{equation} By the AdS/CFT correspondence, the different bulk spacetimes correspond to different states in the CFT$_2$ living on $\partial\m{M}$.
In what follows, we probe the entanglement structure of these states via the RT formula and connected wedge theorem.

\subsection{Entanglement wedges in defect geometry}
\label{sec:sec22}

In this section, we specialize to the conical defect geometry. We explain the relationship between three candidates for the RT surface and hence for the entanglement wedge for a boundary subregion comprised of two intervals of equal size.

To fix notation, recall that the RT formula relates the entanglement entropy $S(A)$ of a subregion $A$ in $\partial\m{M}$ to the area of a minimal surface $\gamma_{A}$ in the bulk via
\begin{equation}
S(A)=\frac{1}{4 G_N}\mathrm{Area}(\gamma_{A}) + O(G_N^0) \,,
\label{eq:RTformula}
\end{equation}
where $\gamma_{A}$ is the minimal-area surface satisfying the homology condition $\gamma_{A}\cup A=\partial m_A$, and $m_A$ can be any spacelike codimension-1 bulk subregion. 
In particular, in the present context of a three-dimensional bulk, $\gamma_A$ is a geodesic, and $\mathrm{Area}(\gamma_A)$ is its length.
Hence, we denote the latter as $\mathrm{Length}(\gamma_A)$ in what follows.
For simplicity, we will limit our considerations to configurations where $A$ and $\gamma_A$ lie on a constant-$t$ slice.
The entanglement wedge of $A$ is defined as $\m{E}(A)=\m{D}(m_A)$, where $\m{D}(\cdot)$ denotes the domain of dependence. 

\begin{figure}[htbp]
\centering
\begin{subfigure}[b]{0.3\textwidth}
    \centering
    \includegraphics[width=\textwidth]{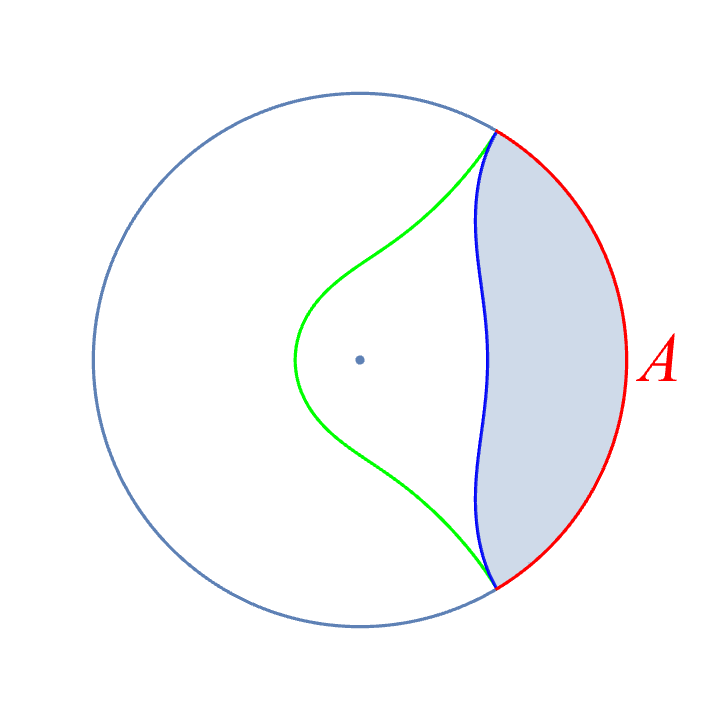}
    \caption{}
\end{subfigure}
\hspace{0.1\textwidth}
\begin{subfigure}[b]{0.3\textwidth}
    \centering
    \includegraphics[width=\textwidth]{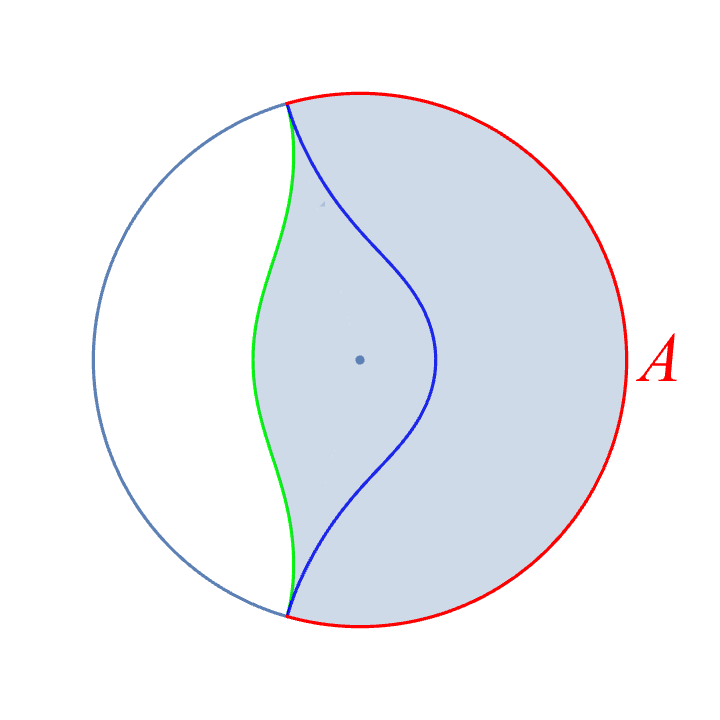}
    \caption{}
\end{subfigure}
\caption{Candidates $\gamma_{A,1}$ (blue) and $\gamma_{A,2}$ (light green) for the RT surface $\gamma_A$ compete according to the inequality \eqref{eq:gammaconicalAdS}: (a) $\gamma_A=\gamma_{A,1}$ for $\Delta\phi\le\pi$, and (b) $\gamma_A=\gamma_{A,2}$ for $\Delta\phi\ge\pi$. 
In both panels, the shaded region corresponds to the entanglement wedge $\m{E}(A)$.
\label{fig:conicgeods}}
\end{figure}

With the aim of finding the RT surfaces for a boundary subregion comprised of two intervals $A$ and $B$, let us solve for extremal surfaces in the conical defect spacetime, \ie $-1<M<0$. 
We begin by considering a single interval $A$ with endpoints $\phi=\pm\frac{\Delta\phi}{2}$ in a constant-$t$ slice. 
Hence, $\Delta\phi$ denotes the angular size of the interval, with $0<\Delta\phi\leq2\pi$. 
There are two candidates for the extremal surface $\gamma_A$, shown in figure \ref{fig:conicgeods}, which we denote $\gamma_{A,1}$ and $\gamma_{A,2}$. 
We must determine which of these yields the minimal length. 

The surface $\gamma_{A,1}$ is described by the following curve \cite{Entwinement_2015}:
\begin{equation}
r(\phi) = 
\frac{\sqrt{|M|}\ \sec\!\left(\sqrt{|M|} \, \phi\right)}{\sqrt{\tan^2\left(\sqrt{|M|}\,\Delta\phi \,/\,2 \right)-\tan^2\left(\sqrt{|M|}\, \phi\right)}} \,, \quad \phi\in\left(-\frac{\Delta\phi}{2},\frac{\Delta\phi}{2}\right) \,,
\label{eq:sol1conical}
\end{equation}
and the corresponding candidate for $\m{E}(A)$ does not include the defect. 
Note that in the above expression, $\tan\left(\sqrt{|M|}\, \Delta\phi/2\right)$ diverges for $\Delta \phi = \pi/\sqrt{|M|}$ yielding a senseless solution. To avoid this problem, we typically focus on the case of heavy defects with $|M|<1/4$.\footnote{To elaborate, the $\gamma_{A,1}$ candidate does not exist for $\Delta \phi > \pi/\sqrt{|M|}$, and the $\gamma_{A,2}$ candidate does not exist for $\Delta \phi < 2\pi- \pi/\sqrt{|M|}$ --- see also figure \ref{fig:circCW} below.
Restricting to $|M|<1/4$ ensures that both candidates exist in the full range $\Delta \phi \in (0, 2\pi)$.
\label{foot:nonexistence}}
The length of $\gamma_{A,1}$ is given by
\begin{equation}
\mathrm{Length}(\gamma_{A,1}) = 2\ln\left[\frac{2}{\epsilon\sqrt{|M|}}\,\sin\!\left(\frac{\sqrt{|M|}\Delta\phi}{2}\right)\right] \,,
\label{eq:lengthgammaconical}
\end{equation}
where we have introduced a UV cutoff surface at $r=\frac{1}{\epsilon}$.

The analogous results for $\gamma_{A,2}$ are obtained from eqs.~\eqref{eq:sol1conical} and \eqref{eq:lengthgammaconical} for $\gamma_{A,1}$ by replacing $\Delta\phi\mapsto2\pi-\Delta\phi$. In eq.~\eqref{eq:sol1conical}, we must also perform a $\pi$-rotation about the defect, \ie $\phi\mapsto\phi+\pi$. It follows that the corresponding candidate for $\m{E}(A)$ includes the defect at $r=0$.

Comparing the lengths above, we find the RT surface for $A$ is given by:
\begin{equation}
\begin{aligned}
\gamma_A &= \begin{cases}
\gamma_{A,1},&  \Delta\phi\leq\pi \,,\\
\gamma_{A,2},&  \Delta\phi\geq\pi \,.
\end{cases}
\end{aligned}
\label{eq:gammaconicalAdS}
\end{equation}
As shown in in figure \ref{fig:conicgeods}, the entanglement wedge (shaded) extends between the RT surface in the bulk and the interval $A$ on the boundary. Note that in the limiting case of pure AdS$_3$ ($M=-1$), we have $\gamma_{A,1}=\gamma_{A,2}$.

\begin{figure}[htbp]
\centering
\includegraphics[width=.27\textwidth]{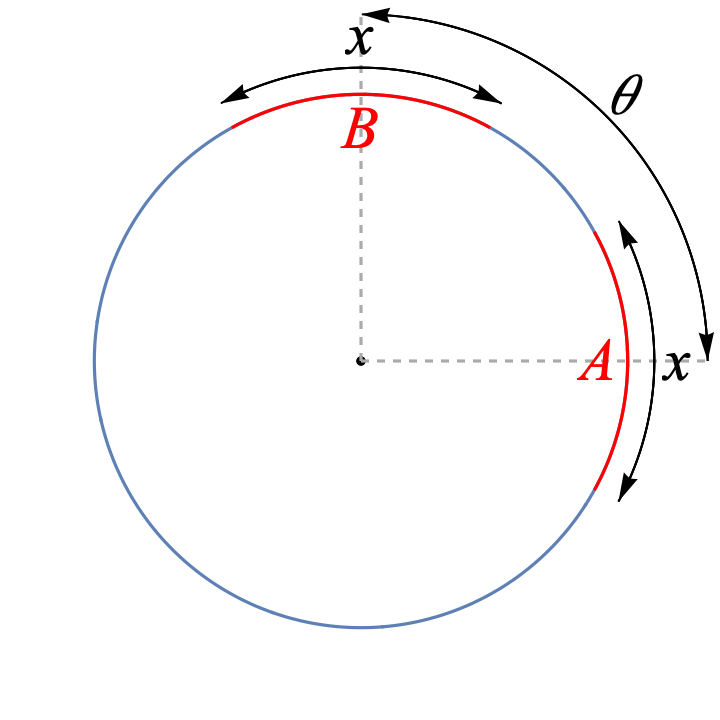}
\caption{Diagram of the region $A\cup B$ under consideration in the CFT.
\label{fig:alphax}}
\end{figure}

We may now study the candidates for the entanglement wedge of a region comprised of two intervals $A$ and $B$. 
For simplicity, we restrict our attention to the case where $A$ and $B$ have equal width $(\Delta \phi)_A=(\Delta \phi)_B=x$. 
We denote the angular separation of the midpoints of the two regions as $\theta$ --- see figure \ref{fig:alphax}.
Note that for $A$ and $B$ not to overlap, we must have $x\in(0,\pi)$ and $\theta\in(x,\pi]$.\footnote{Without loss of generality, we restrict to $\theta\le\pi$. For $\pi<\theta\le2\pi$, we may replace $\theta\mapsto 2\pi-\theta$ in the following.} 
It follows from eq.~\eqref{eq:gammaconicalAdS} that $\gamma_A = \gamma_{A,1}$ and $\gamma_B = \gamma_{B,1}$, so that the defect lies outside of the entanglement wedges for the individual regions. 

\begin{figure}[htbp]
\centering
\begin{subfigure}[b]{0.3\textwidth}
    \centering
    \includegraphics[width=\textwidth]{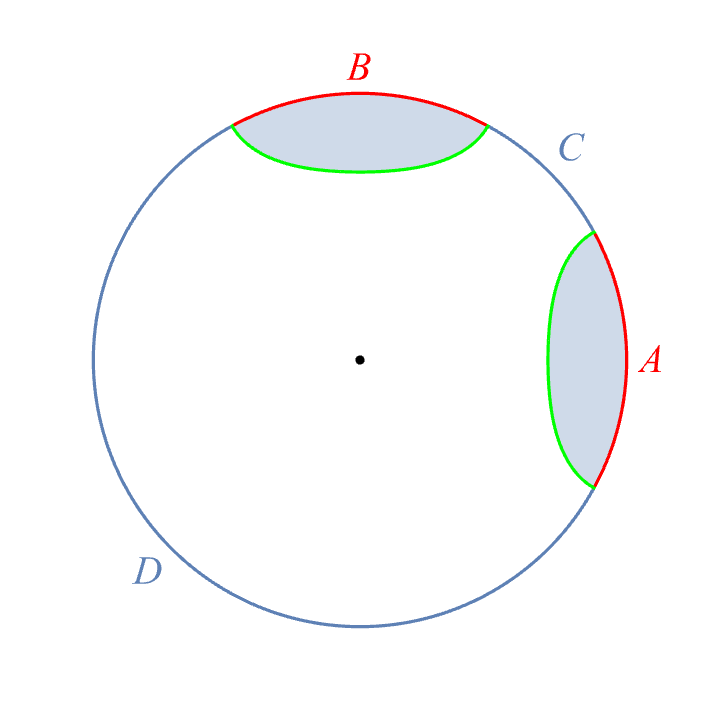}
    \caption{}
    \label{fig:dcircs_d}
\end{subfigure}
\hfill
\begin{subfigure}[b]{0.3\textwidth}
    \centering
    \includegraphics[width=\textwidth]{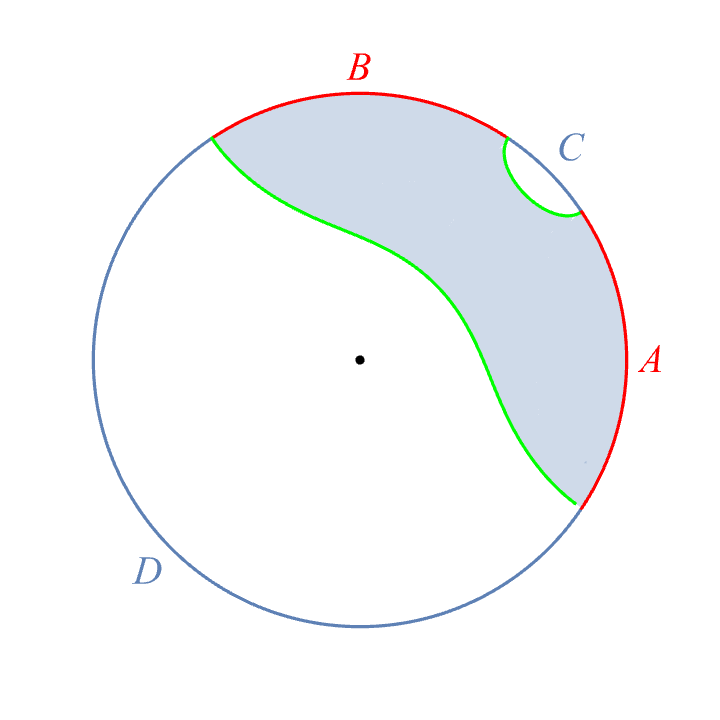}
    \caption{}
    \label{fig:dcircs_u}
\end{subfigure}
\hfill
\begin{subfigure}[b]{0.3\textwidth}
    \centering
    \includegraphics[width=\textwidth]{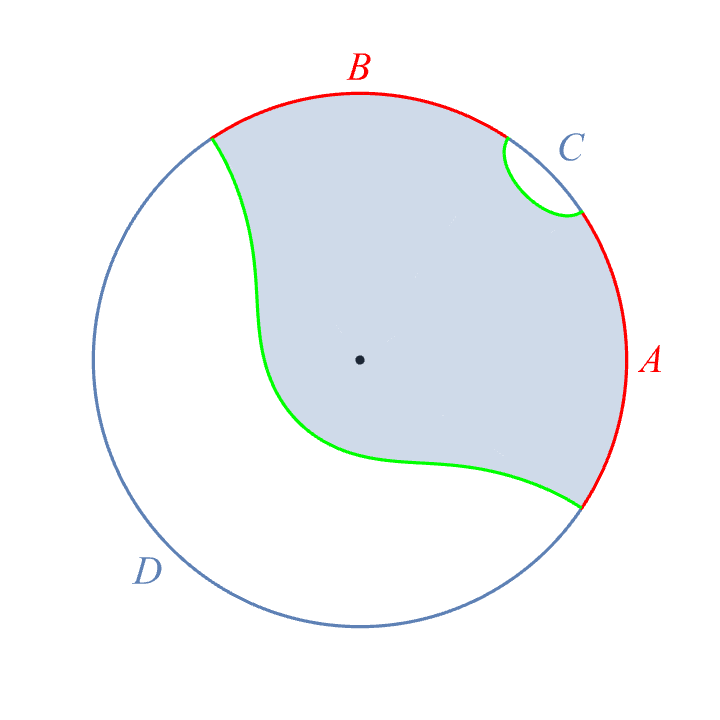}
    \caption{}
    \label{fig:dcircs_o}
\end{subfigure}
\caption{A constant-$t$ slice of the conical defect geometry showing, from left to right, the $d$, $u$, and $o$ candidates for $\m{E}(A\cup B)$.
\label{fig:dcircs}}
\end{figure}

There are three candidates of interest for $\m{E}(A\cup B)$, which we label $d$, $u$ and $o$ following the notation from the introduction. 
The first candidate $d$ is disconnected, with extremal surface $\gamma_A\cup\gamma_B$ --- see figure \ref{fig:dcircs_d}. 
The remaining two candidates for the entanglement wedge are both connected. 
The first, $u$, does not contain the defect, while the second, $o$, does.
The names $u$ and $o$ are chosen because the $o$-wedge is punctured by the defect singularity, so it is topologically like the letter $o$ --- see figures \ref{fig:dcircs_u} and \ref{fig:dcircs_o}.

\begin{figure}[htbp]
\centering
\begin{subfigure}[b]{0.3\textwidth}
    \centering
    \includegraphics[width=\textwidth]{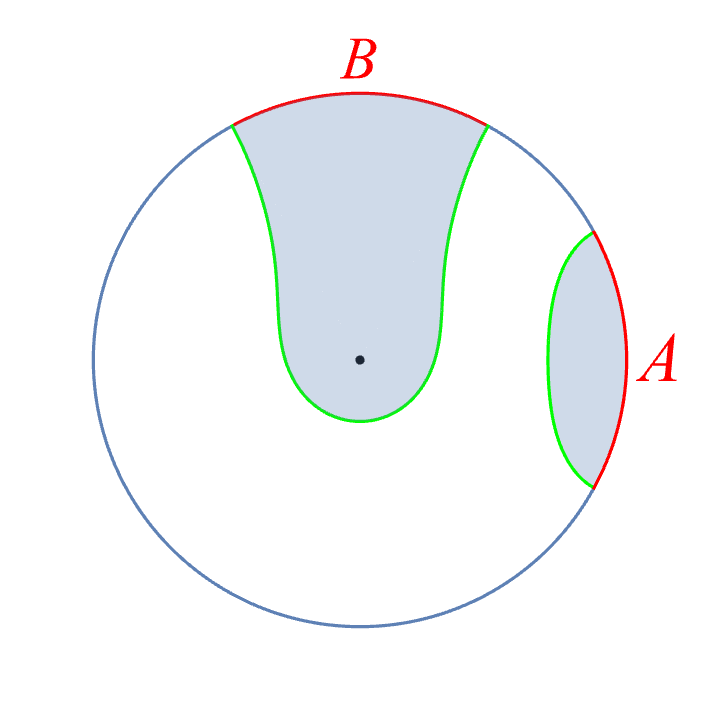}
    \caption{}
    \label{fig:dcircs_t}
\end{subfigure}
\qquad \qquad \qquad
\begin{subfigure}[b]{0.3\textwidth}
    \centering    \includegraphics[width=\textwidth]{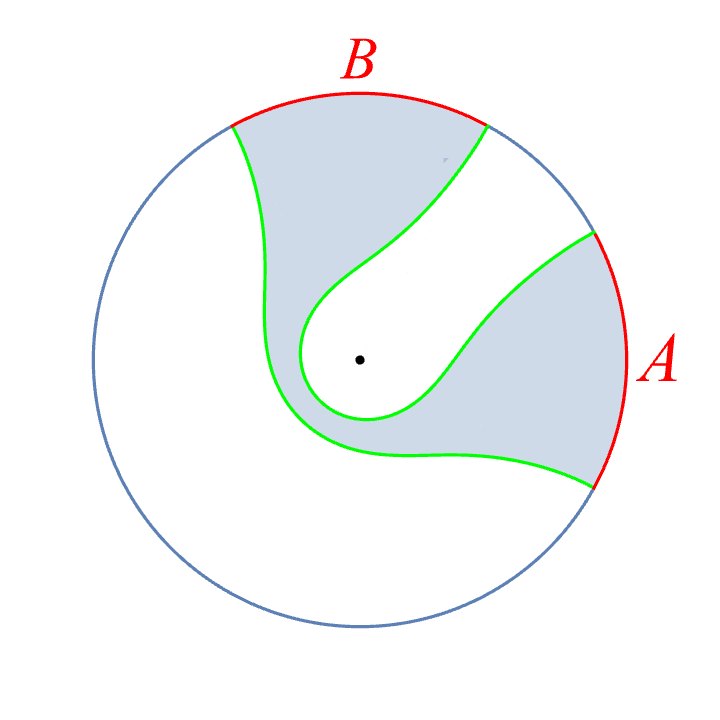}
    \caption{}
    \label{fig:dcircs_h}
\end{subfigure}
\caption{In our analysis, we do not consider the (a) $d'$ and (b) $u'$ configurations shown here as candidates for the entanglement wedge, because it is clear that they never provide the RT surface for $A\cup B$.
\label{fig:dcircsunusual}}
\end{figure}

As an aside, note that we are overlooking a variety of other candidates for the entanglement wedge.
Two examples are shown in figure \ref{fig:dcircsunusual}.
These configurations may be overlooked because they never provide the RT surface for $A\cup B$.
For example, $d$ generally has less area than the configuration $d'$ defined in figure \ref{fig:dcircs_t}, and $o$ has less area than the configuration $u'$ defined in figure \ref{fig:dcircs_h}.
These conclusions may be deduced by recalling that each of the individual surfaces is a candidate RT surface for a single interval, and then applying eq.~\eqref{eq:gammaconicalAdS}.

Note also that the restriction to heavy defects ($|M|<1/4$) is important here to ensure that all three candidates in figure \ref{fig:dcircs} exist.
As described in footnote \ref{foot:nonexistence}, there are choices of $\theta$ and $x$ for which either the $u$ or $o$ candidate does not exist when $|M|>1/4$; \eg the $u$ configuration would disappear for $\theta+x>\pi/\sqrt{M}$.\footnote{We thank Caroline Lima for pointing out this feature, which was not explained in the initial version of our paper.\label{foot:Carol}}

Let us now study the relationship between the $d$, $u$, and $o$ candidate configurations. 
We define an ordering relation on them by writing, for example, $u<d$ if the total length of the extremal surfaces in $u$ is less than that in $d$. 
Notice that with this notation, the entanglement wedge is in the connected phase if and only if $\min(u,o)<d$.

First, let us examine when $u<o$. 
A useful fact is that both the $u$ and $o$ configurations always have one extremal surface in common: the RT surface $\gamma_C$ for $C$, the smaller of the two intervals separating $A$ and $B$  --- see figures \ref{fig:dcircs_u} and \ref{fig:dcircs_o}.
The interval $C$ has extent $(\Delta\phi)_C=\theta-x < \pi$; hence, $\gamma_C=\gamma_{C,1}$ by eq.~\eqref{eq:gammaconicalAdS}. 
Thus, the ordering of $u$ and $o$ is governed by the choice of RT surface for the boundary interval $D$ in figures \ref{fig:dcircs_u} and \ref{fig:dcircs_o}, and we find $u<o$ if and only if
\begin{equation}
(\Delta\phi)_D=2\pi-(\theta+x) > \pi \,.
\label{eq:ouineq}
\end{equation}

\begin{figure}[htbp]
\centering
\begin{subfigure}[b]{0.43\textwidth}
    \centering
\includegraphics[width=\textwidth]{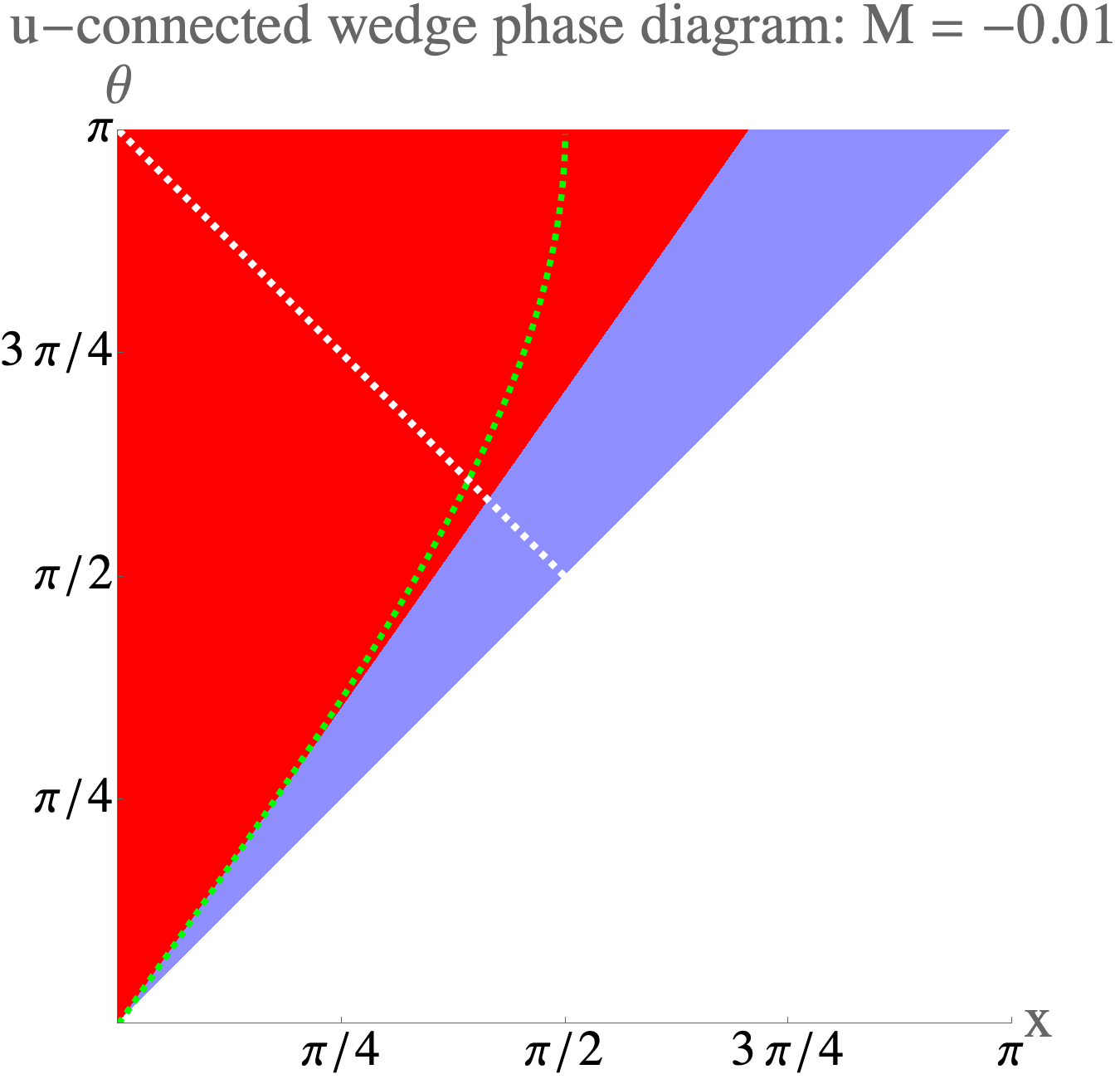}
    \caption{}
    \label{fig:uphasea}
\end{subfigure}
\qquad\qquad
\begin{subfigure}[b]{0.43\textwidth}
    \centering
    \includegraphics[width=\textwidth]{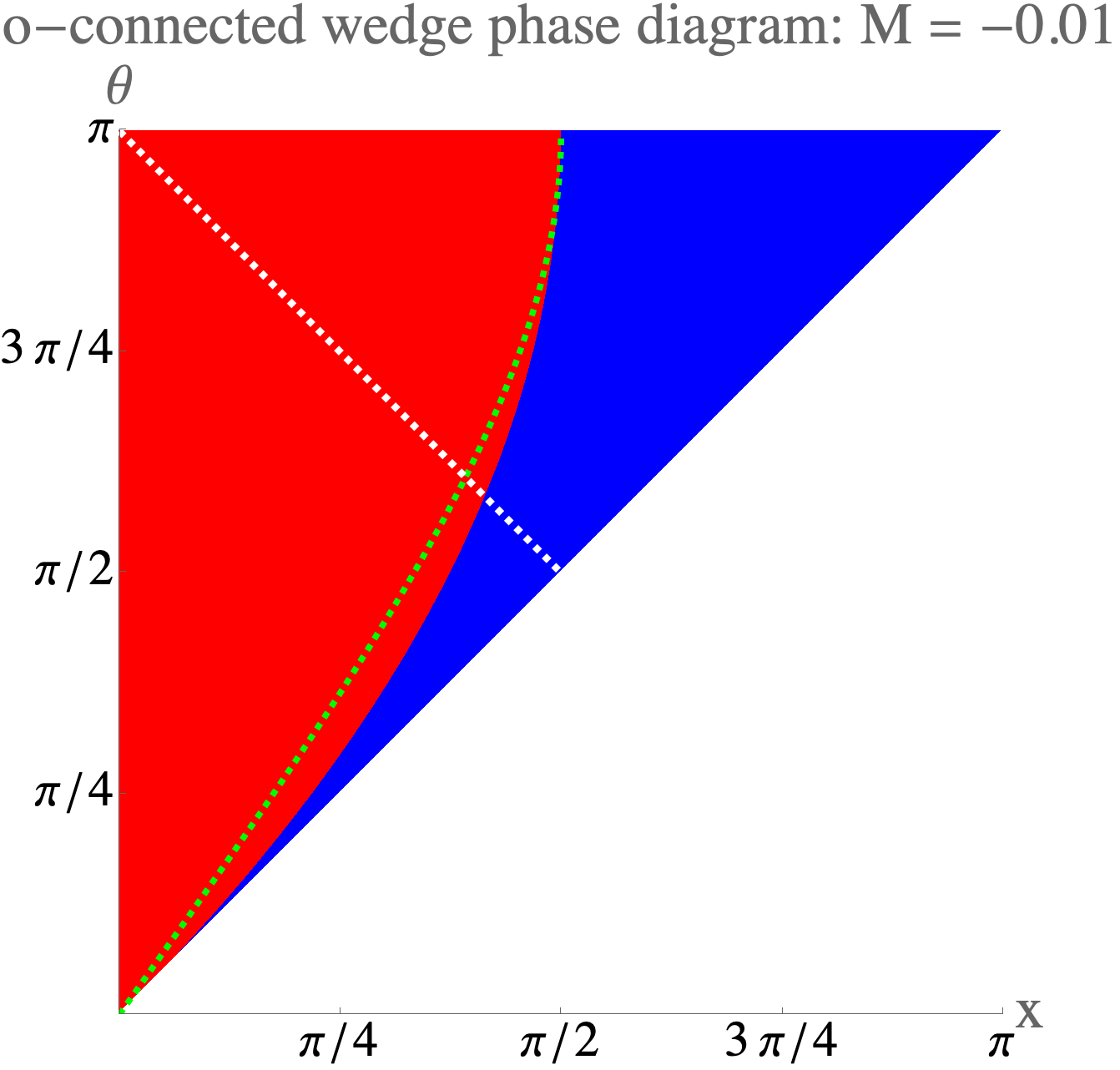}
    \caption{}
     \label{fig:ophaseb}
\end{subfigure}
\caption{
(a) Competition between the $u$ and $d$ configurations, as described by eq.~\eqref{eq:uineq}. The red and light blue regions correspond to $d<u$ and $u<d$, respectively. The saturation curve for pure AdS ($M=-1$) is shown in green for reference, as well as the curve $o=u$ (white).  (b) Competition between the $o$ and $d$ configurations, as described by eq.~\eqref{eq:oineq}. The red and dark blue regions correspond to $d<o$ and $o<d$, respectively.
Note that we are not interested in cases where $A$ and $B$ overlap, so the region $\theta \le x$ is left blank --- see figure \ref{fig:alphax}.
\label{fig:ophase}}
\end{figure}

The remaining inequalities of interest, namely $u<d$ and $o<d$, follow directly from the length formula \eqref{eq:lengthgammaconical}. 
We have $u<d$ if and only if
\begin{equation}
\sin^2\frac{\sqrt{|M|}\,x}{2}>\sin\frac{\sqrt{|M|}\,(\theta-x)}{2}\sin\frac{\sqrt{|M|}\,(\theta+x)}{2} \,.
\label{eq:uineq}
\end{equation}
\noindent
In the phase diagram of figure \ref{fig:uphasea}, the range of $(x,\theta)$ values satisfying eq.~\eqref{eq:uineq} is shown in blue, while the range of $(x,\theta)$ values violating eq.~\eqref{eq:uineq} is shown in red. 
Similarly, we have $o<d$ if and only if
\begin{equation}
\begin{aligned}
&\sin^2\frac{\sqrt{|M|}\,x}{2}>\sin\frac{\sqrt{|M|}\,(\theta-x)}{2}\sin\frac{\sqrt{|M|}\,\left(2\pi-(\theta+x)\right)}{2} \,.\\
\end{aligned}
\label{eq:oineq}
\end{equation}
The corresponding phase diagram is shown in figure \ref{fig:ophaseb}.
In both plots, we take the same mass for the defect ($M=-0.01$).

\begin{figure}[htbp]
\centering
\begin{subfigure}[b]{0.43\textwidth}
    \centering
    \includegraphics[width=\textwidth]{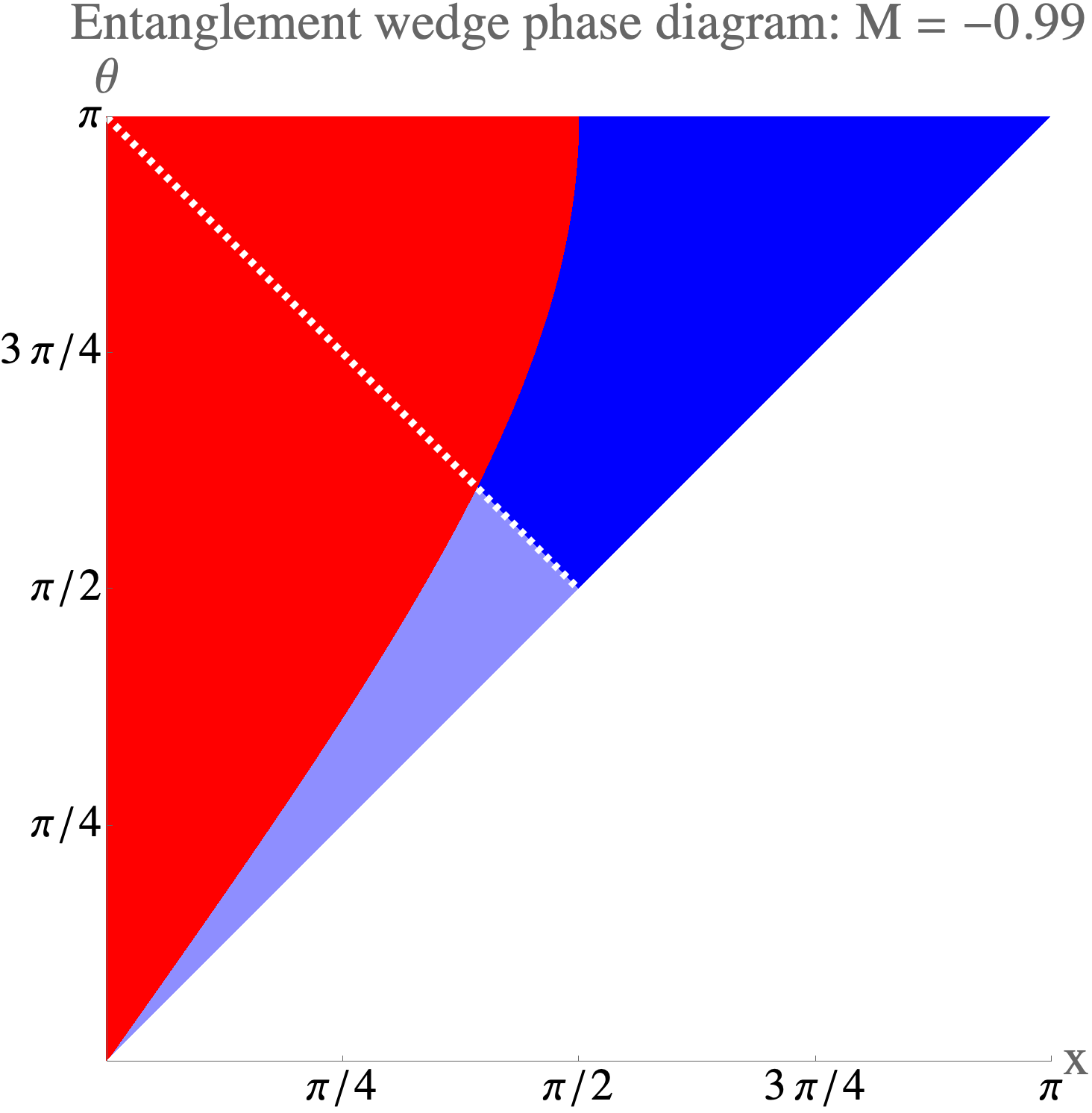}
    \caption{}
    \label{fig:ewplot0_99}
\end{subfigure}
\qquad \qquad
\begin{subfigure}[b]{0.43\textwidth}
    \centering
    \includegraphics[width=\textwidth]{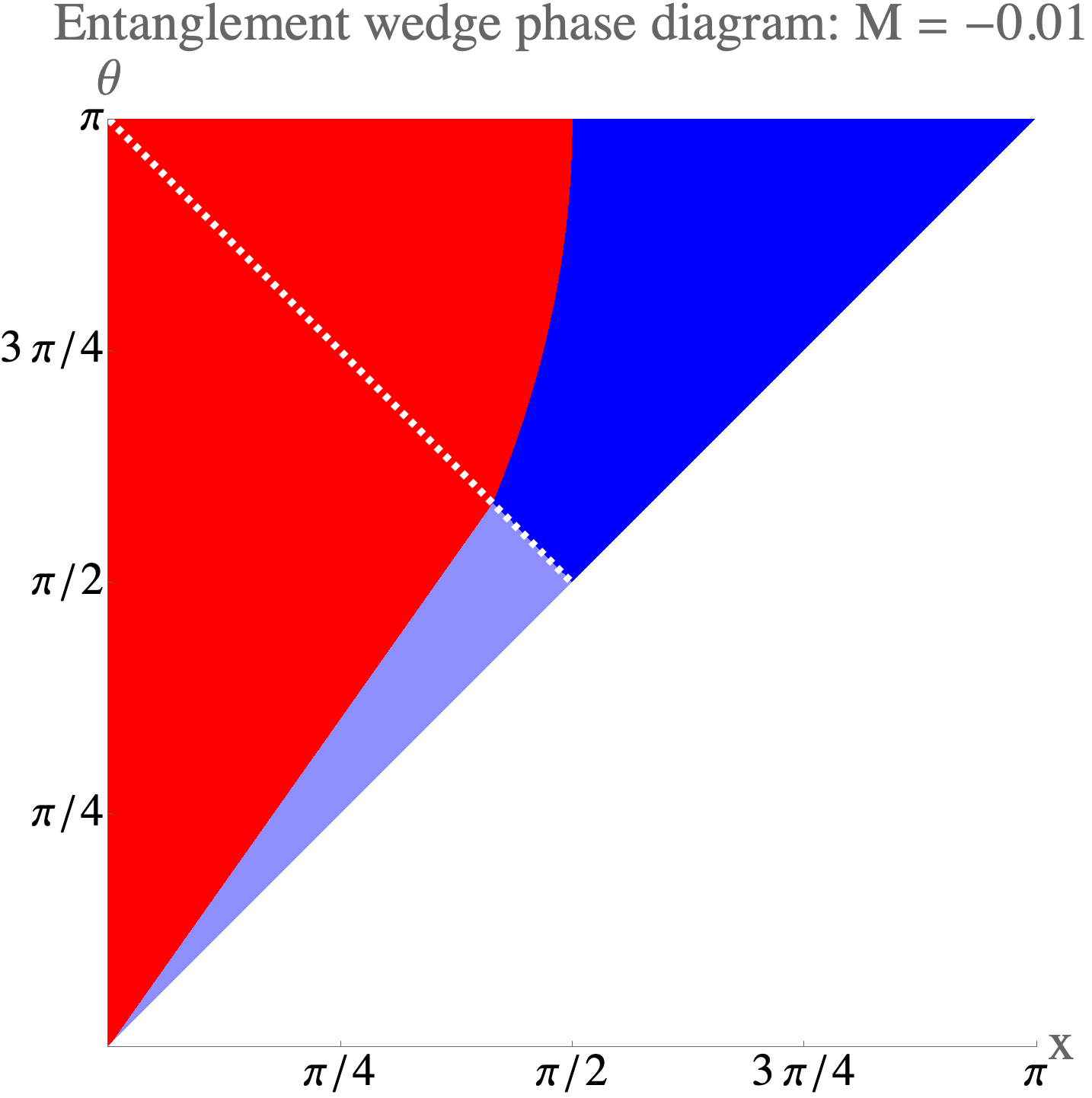}
    \caption{}
    \label{fig:ewplot0_01}
\end{subfigure}
\caption{Phase diagrams for the entanglement wedge in the defect geometries with (a) $M=-0.99$ and (b) $M=-0.01$. The parameter spaces where the $d$ (disconnected) and $u$ and $o$ (connected) phases are minimal are shown in red, light blue, and dark blue, respectively. The $o=u$ curve is shown in white; the regions below and above it correspond to figures  \ref{fig:uphasea} and \ref{fig:ophaseb}, respectively. The similarity between the $M=-1$ curve in figure \ref{fig:ophase} and the boundary of the disconnected phase in figure \ref{fig:ewplot0_99} ($M=-0.99$) is a consequence of the smooth transition to pure AdS, where $\gamma_1$ of eq.~\eqref{eq:gammaconicalAdS} equals $\gamma_2$.
\label{fig:conicalphasediagram}}
\end{figure}

Generally, we wish to know whether the entanglement wedge is connected, \ie whether $\min(u,o)<d$. 
Recall from eq.~\eqref{eq:ouineq} that $u<o$ when $\theta+x<\pi$, while $u>o$ when $\theta+x>\pi$.
Hence, the desired phase diagram, shown in figure \ref{fig:conicalphasediagram}, results from stitching together figures \ref{fig:uphasea} and \ref{fig:ophaseb} along the $u=o$ line at $\theta+x=\pi$, shown in white. Note that this stitching produces a kink in the boundary of the disconnected phase (shown in red) at $\theta+x=\pi$. Note that for any $x > \frac{\pi}{2}$, the entanglement wedge is always connected. This may be deduced from symmetry arguments.\footnote{To see this, first consider the case $\theta=\pi$. As in figure 9a of \cite{May:2019odp}, the disconnected and connected configurations have the same length when $x=\pi/2$. Since the geodesic length \eqref{eq:lengthgammaconical} is increasing in $\Delta \phi$, we conclude $\m{E}(A\cup B)$ becomes connected under any small enlargement of the regions. Further, decreasing $\theta$ from $\theta=\pi$ will not cause the wedge to become disconnected. To see why, a hint is to notice that the $\Delta \phi$-derivative of the length \eqref{eq:lengthgammaconical} is decreasing in $\Delta \phi$.}

Another interesting feature of the phase diagrams in figure \ref{fig:conicalphasediagram} is that increasing the mass of the defect increases the region for the disconnected phase (red) and decreases the region where the entanglement wedge is connected (blue). Indeed, by increasing the mass $M$ into the black hole regime ($M\geq0$), the connected region shrinks even further, as will be discussed in the next section.

\subsection{Entanglement wedges in BTZ geometry}
\label{sec:sec23}

In this section, we repeat the analysis of section \ref{sec:sec22} to study the relationship between entanglement wedge candidates ($d$, $u$, and $o$)  in the BTZ black hole spacetime.

\begin{figure}[htbp]
\centering
\begin{subfigure}[b]{0.3\textwidth}
    \centering
    \includegraphics[width=\textwidth]{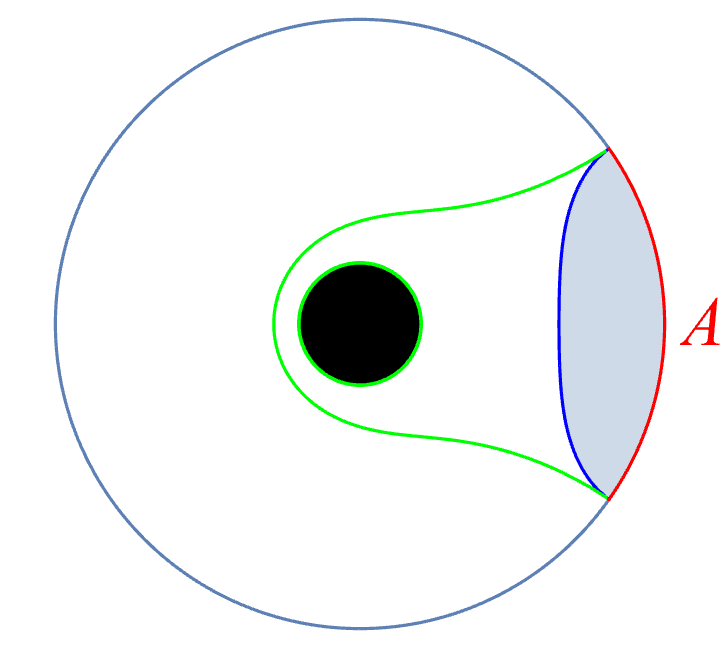}
    \caption{}
\end{subfigure}
\hfill
\begin{subfigure}[b]{0.3\textwidth}
    \centering
    \includegraphics[width=\textwidth]{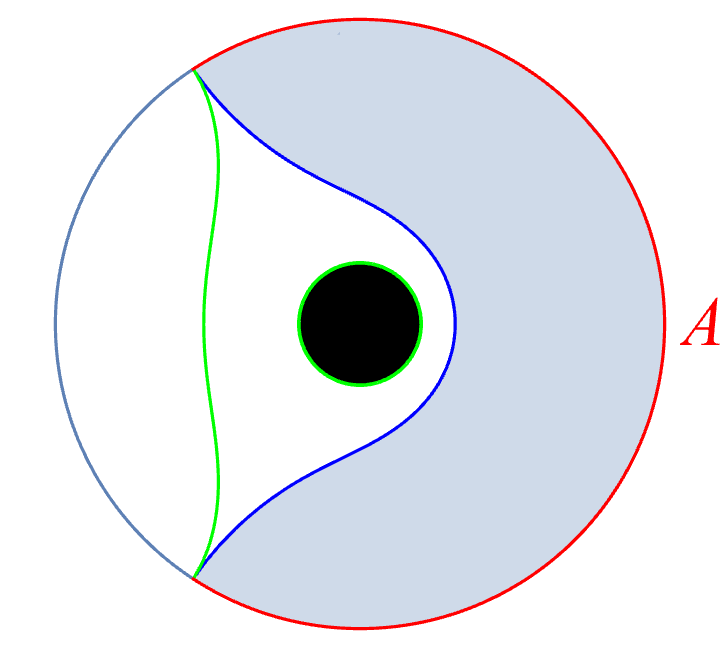}
    \caption{}
\end{subfigure}
\hfill
\begin{subfigure}[b]{0.3\textwidth}
    \centering
    \includegraphics[width=\textwidth]{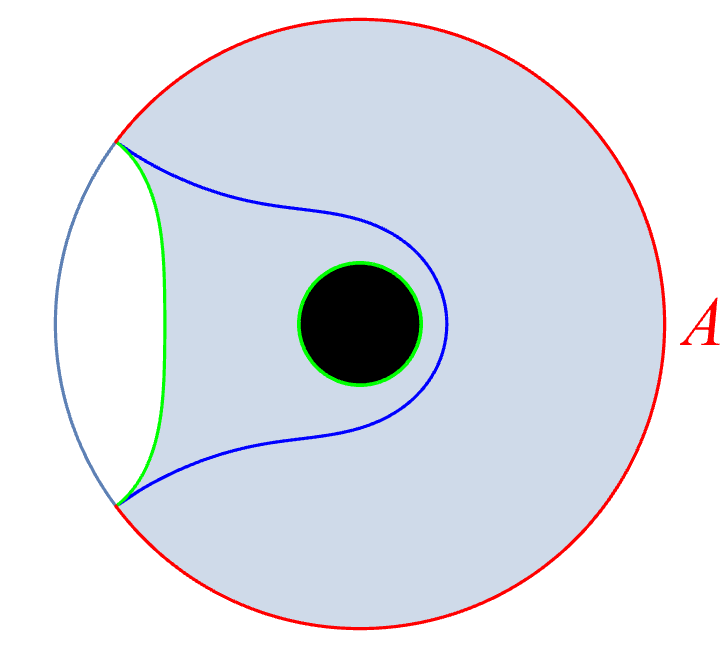}
    \caption{}
\end{subfigure}
\caption{Candidates $\gamma_{A,1}$ (blue) and $\gamma_{A,2}$ (light green) for the RT surface $\gamma_A$ compete according to the inequality \eqref{eq:gammaBTZ}. The shaded region is $\m{E}(A)$. In (a) and (b), $\Delta\phi<\Delta\phi^*(M)$, so $\gamma_A=\gamma_{A,1}$. In (c), $\Delta\phi>\Delta\phi^*(M)$, so $\gamma_A=\gamma_{A,2}$.
\label{fig:BTZgeodesics}}
\end{figure}

Recall that given a boundary subregion $A$ in the defect spacetime, there exist two candidates for $\gamma_A$, which we called $\gamma_{A,1}$ and $\gamma_{A,2}$. An analogous statement holds in the BTZ spacetime --- see figure \ref{fig:BTZgeodesics}. In the BTZ case, $\gamma_{A,1}$ is described by the following curve \cite{Tsujimura_2020}:
\begin{equation}
\begin{aligned}
r(\phi) &= \frac{\sqrt{M}\,\mathrm{sech}\left(\sqrt{M}\,\phi\right)}{\sqrt{\tanh ^2\left(\sqrt{M}\,\Delta\phi\,/\,2\right)-\tanh ^2\left(\sqrt{M}\,\phi\right)}} \,, \quad 
\phi\in\left(-\frac{\Delta\phi}{2},\frac{\Delta\phi}{2}\right) \,,
\end{aligned}
\label{eq:gammaBTZ1}
\end{equation}
and the corresponding candidate for $\m{E}(A)$ does not include the black hole (horizon). Integrating yields
\begin{equation}
\text{Length}(\gamma_{A,1})=2\ln\left(\frac{2}{\epsilon\sqrt{M}}\sinh\frac{\sqrt{M}\,\Delta\phi}{2}\right)\,.
\label{eq:lengthgammaBTZAdS1}
\end{equation}

The surface $\gamma_{A,2}$ has two components. The first component is obtained from $\gamma_{A,1}$ via $\Delta\phi\mapsto2\pi-\Delta\phi$ and $\phi\mapsto\phi+\pi$, analogously to the defect case. The second component is the black hole horizon $r=\sqrt{M}$, which is included in order to satisfy the homology condition. It follows that
\begin{equation}
\text{Length}(\gamma_{A,2})=2\pi \sqrt{M} + 2\ln\left(\frac{2}{\epsilon\sqrt{M}}\sinh\frac{\sqrt{M}(2\pi-\Delta\phi)}{2}\right) \,.
\label{eq:lengthgammaBTZAdS2}
\end{equation}

The RT surface for $A$ is given by the minimal-area candidate as follows:
\begin{equation}
\begin{aligned}
\gamma_A &= \begin{cases}
\gamma_{A,1},&  \Delta\phi \leq \Delta\phi^*(M) \,,\\
\gamma_{A,2},&  \Delta\phi \geq \Delta\phi^*(M) \,,
\end{cases}
\end{aligned}
\label{eq:gammaBTZ}
\end{equation}
\noindent
where the threshold $\Delta\phi=\Delta\phi^*(M)$ between an entanglement wedge that includes and one that  excludes the black hole is given by \cite{Umemoto_2019}
\begin{equation}
\Delta\phi^*(M) = \pi+
\frac1{\sqrt{M}}\,\log\cosh\! \left(\sqrt{M}\pi\right)\,.
\label{eq:btzthresh}
\end{equation}
The function $\Delta\phi^*(M)$ is monotonically increasing. Notice that for $M\to0$, one retrieves the conical defect threshold, $\Delta\phi^*(0)=\pi$, but for $M\to\infty$, the threshold approaches $2\pi$.

The entanglement wedge for $A\cup B$ has $d$, $u$, and $o$ candidates as in the defect case, but now with the additional feature that the extremal surface for the $o$ candidate includes the black hole horizon.

By reasoning identical to that above eq.~\eqref{eq:ouineq}, we have $u<o$ if and only if
\begin{equation}
    \theta+x<\Delta\phi^*(M) \,.
\label{eq:ouineqbtz}
\end{equation}
This inequality is always satisfied in the strict $M\to\infty$ limit, where $\Delta\phi^*(M)\to2\pi$.

From the length formulae \eqref{eq:lengthgammaBTZAdS1} and \eqref{eq:lengthgammaBTZAdS2}, we have $u<d$ if and only if
\begin{equation}
    \sinh^2\frac{\sqrt{M}\,x}{2} > \sinh\frac{\sqrt{M}\,(\theta-x)}{2}\sinh\frac{\sqrt{M}\,(\theta+x)}{2} \,.
    \label{eq:uineqbtz}
\end{equation}
For large $M$, this inequality reads $\theta \lesssim x$. By definition, $\theta > x$, so in the strict $M\to\infty$ limit, we have $d<u$ for all choices of $x$ and $\theta$.

Similarly, we have $o<d$ if and only if
\begin{equation}
    \sinh^2\frac{\sqrt{M}\,x}{2} > e^{\pi\sqrt{M}}\sinh\frac{\sqrt{M}\,(\theta-x)}{2}\sinh\frac{\sqrt{M}\,\left(2\pi-(\theta+x)\right)}{2} \,.
    \label{eq:oineqbtz}
\end{equation}
For large $M$, this inequality reads $x \gtrsim \pi$. By definition, $x < \pi$, so in the strict $M\to\infty$ limit, we always have $d<o$.

\begin{figure}[htbp]
\centering
\includegraphics[width=.3\textwidth]{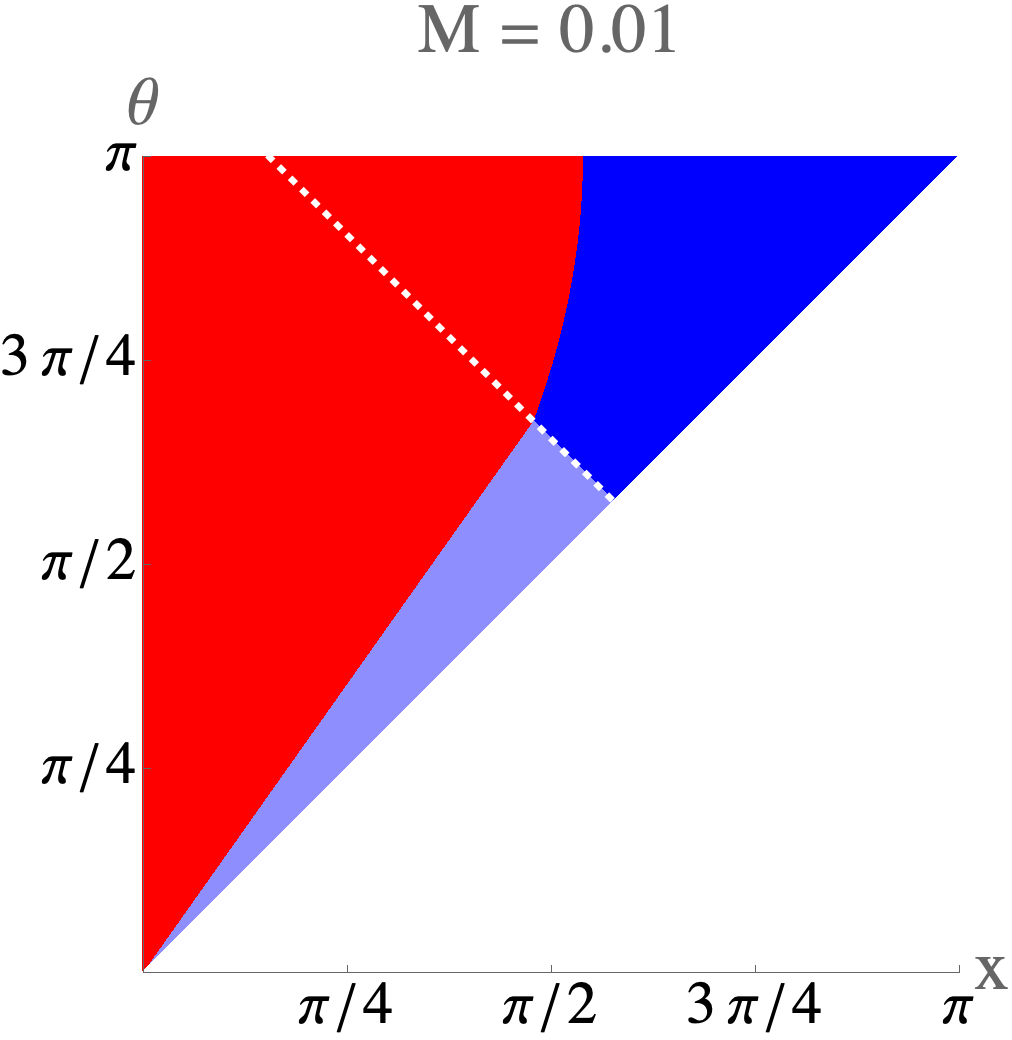}
\hfill
\includegraphics[width=.3\textwidth]{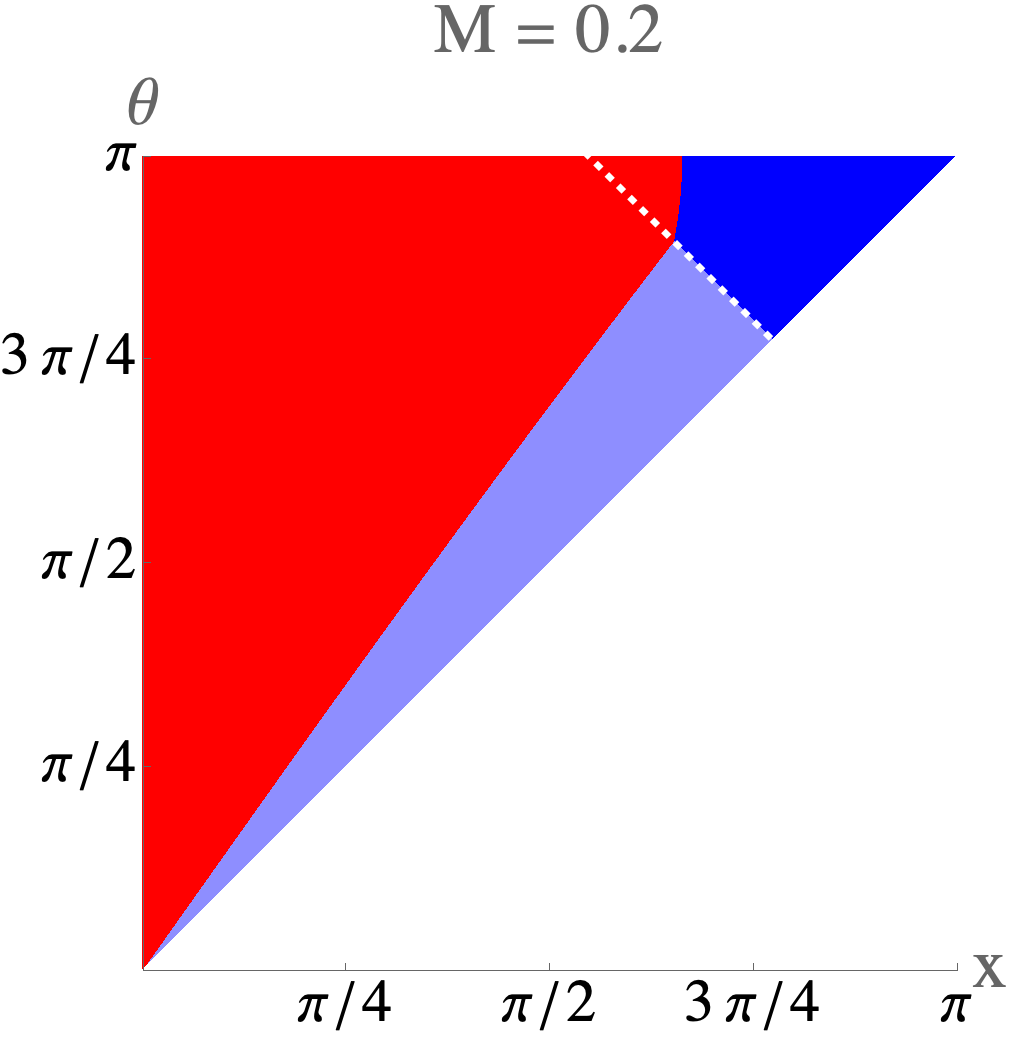}
\hfill
\includegraphics[width=.3\textwidth]{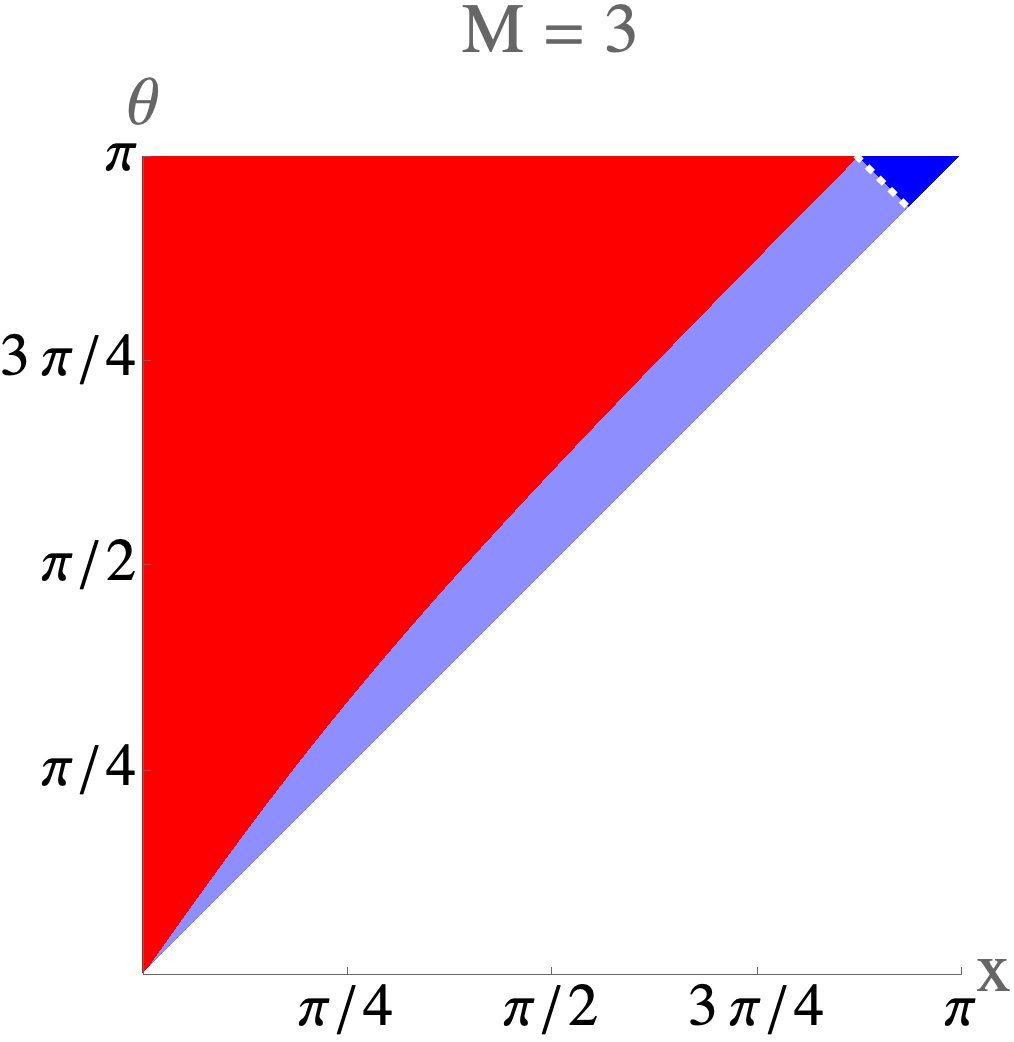}
\caption{Entanglement wedge phase diagram for BTZ geometries with $M=0.01, 0.2,$ and $3$. As in figure \ref{fig:conicalphasediagram}, red, dark blue, and light blue correspond to $d$, $o$, and $u$ phases, respectively. The $o=u$ line (\ie $\theta+x=\Delta\phi^*(M)$) is shown in white; above and below it, the relevant inequalities are  eq.~\eqref{eq:oineqbtz} ($o<d$) and  eq.~\eqref{eq:uineqbtz} ($u<d$), respectively.
\label{fig:BTZphasediagram}}
\end{figure}

Entanglement wedge connectivity phase diagrams are provided in figure \ref{fig:BTZphasediagram} for various values of $M$.
Note that in the conical defect geometry, the entanglement wedge is always connected when $x > \pi/2$ --- see figure \ref{fig:conicalphasediagram}.
However, in the BTZ case, disconnected entanglement wedges can have $x > \pi/2$, as the connected configurations increase in length due to the black hole horizon.
In particular, for $M\to\infty$, we approach $d<u<o$ for all $x$ and $\theta$ values by the discussions below eqs.~\eqref{eq:ouineqbtz} and \eqref{eq:uineqbtz}.
In the phase diagrams \ref{fig:BTZphasediagram}, this corresponds to the sizes of the dark blue ($o$-phase) and light blue ($u$-phase) regions shrinking as $M$ increases, with the $o$ region shrinking faster than the $u$ region. 
Note that the diagram for $M=0.01$ in figure \ref{fig:BTZphasediagram} nearly coincides with that for $M=-0.01$ in figure \ref{fig:ewplot0_01}, which is a consequence of the smooth transition between the defect and BTZ geometries across $M=0$. 

A further consistency check of our analysis comes from examining the regime where both $\theta$ and $x$ are small.
From eqs.~\eqref{eq:ouineq} and \eqref{eq:ouineqbtz}, we have $u<o$ in this regime.
Hence, from eqs.~\eqref{eq:uineq} and \eqref{eq:uineqbtz}, we find that the connected wedge inequality $u<d$ reduces to 
\begin{equation}
\theta<\sqrt{2}\,x
\label{eq:vacineq}
\end{equation}
for small $\theta$ and $x$ for both the defect and BTZ geometries. We note that this result \eqref{eq:vacineq} is independent of $M$. 
In fact, the boundary between $u$ and $d$ must reduce to $\theta=\sqrt{2}\,x$ for small $x$ and $\theta$ in all asymptotically AdS$_3$ spacetimes. This occurs because in this regime, the regions $A$ and $B$ become smaller and closer together, and hence the relevant extremal surfaces lie increasingly close to the asymptotic boundary, where the geometry approaches that of pure AdS$_3$. From the boundary QFT perspective, this is the familiar statement that every state looks like the vacuum at short distances.
Hence, our analysis captures the entanglement structure of the vacuum.

To summarize, increasing $M$ from $-1$ towards infinity decreases the region of phase space where the entanglement wedge is connected.
In the following section, we aim to discover which of these connected wedges correspond to a holographic scattering process via the connected wedge theorem.

\section{Applying the connected wedge theorem}
\label{sec:holscat}

In this section, we review the connected wedge theorem of \cite{May:2019odp, May:2021nrl, May:2022clu} and apply it to the conical defect and BTZ black hole geometries of section \ref{sec:geometryandwedges}. We find that bulk scattering processes require not just large entanglement, \ie $\min(u,o)<d$, but also the stronger condition $u < d$.

\subsection{The connected wedge theorem}
\label{sec:sec31}

It is most natural to introduce the connected wedge theorem in the context of relativistic quantum tasks.
Relativistic quantum tasks are quantum computations with inputs and outputs occurring at specified spacetime locations \cite{Kent_2012}.
The inputs and outputs can be encoded in point-like systems such as single particles or into extended spacetime regions in a delocalized way.
We focus on extended regions in this section and discuss points in appendix \ref{sec:pointsappendix}.

In the holographic setting, such quantum tasks can be considered from a bulk as well as a boundary perspective. 
A key expectation from holography is that any task that can be completed in the bulk should have a dual boundary description. 
Further, the corresponding boundary task should be completed with the same success probability. 

\begin{figure}[htbp]
\centering
\begin{subfigure}[b]{0.3\textwidth}
    \centering
    \includegraphics[width=\textwidth]{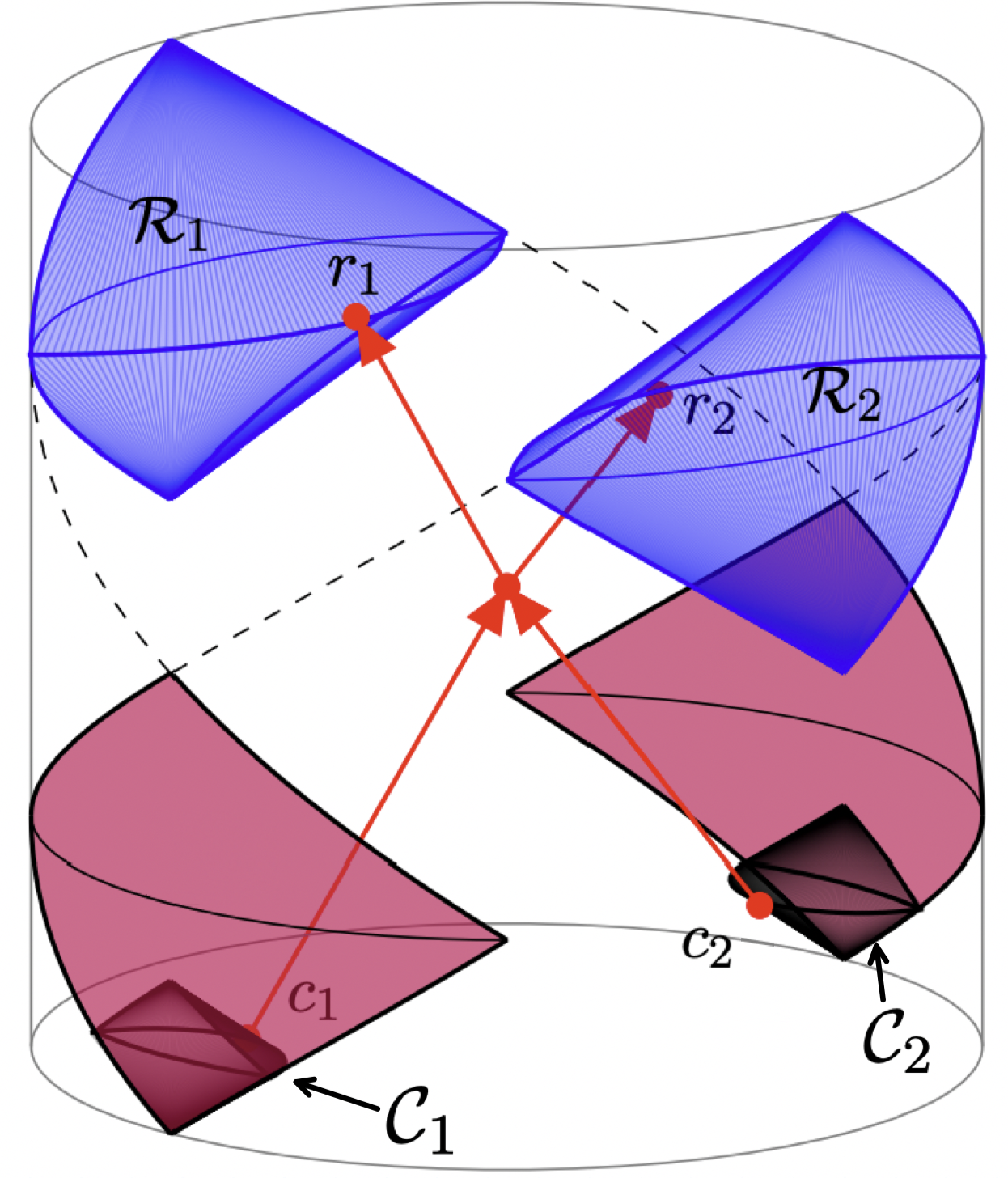}
    \caption{}
    \label{fig:C1C2plot_bulk}
\end{subfigure}
\hfill
\begin{subfigure}[b]{0.6\textwidth}
    \centering
    \includegraphics[width=\textwidth]{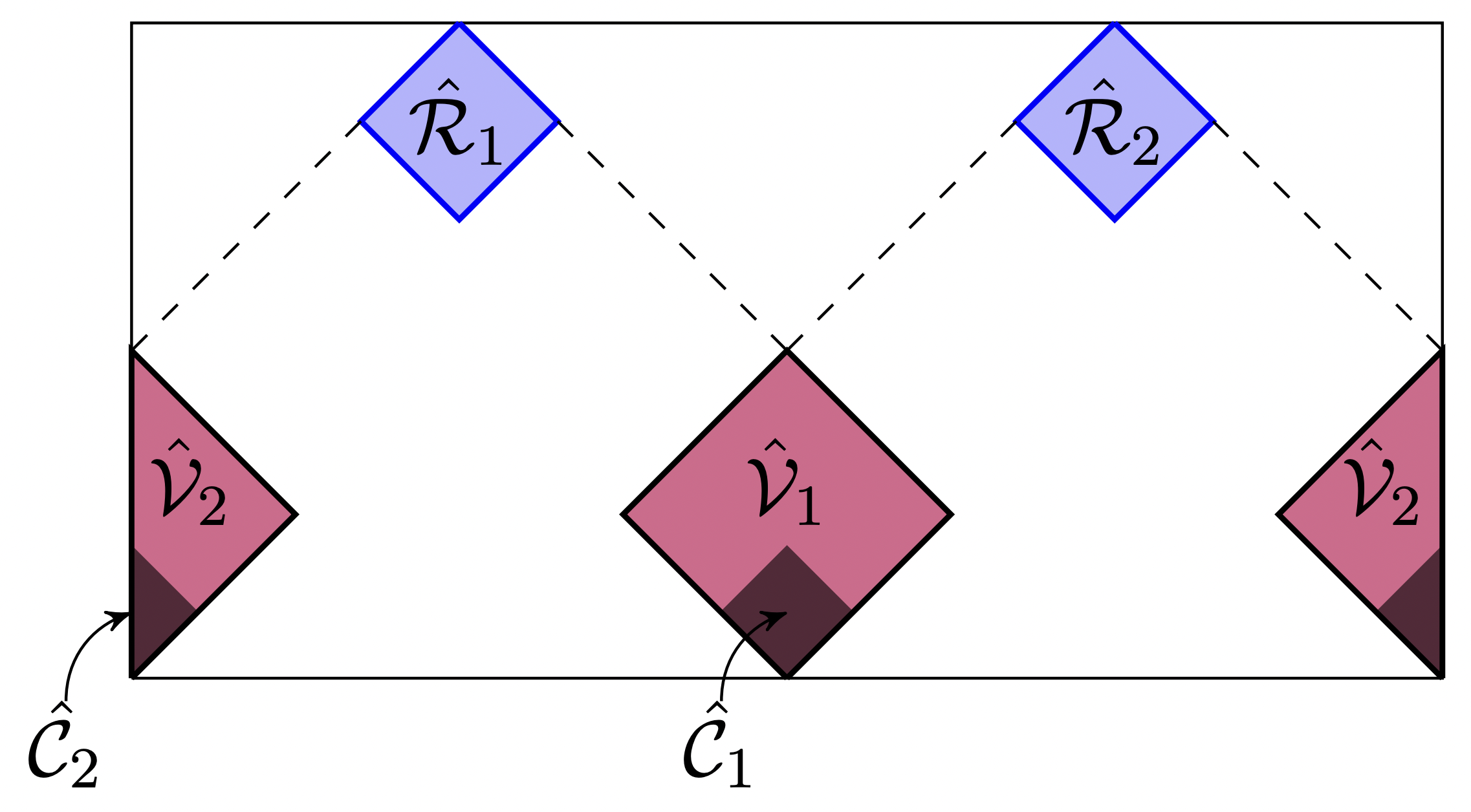}
    \caption{}
    \label{fig:C1C2plot_bdry}
\end{subfigure}

\caption{(a) An illustration of  choices for $\m{C}_1$, $\m{C}_2$, $\m{R}_1$, and $\m{R}_2$ for a quantum task in $\m{M}$. As discussed above eq.~\eqref{eq:scatteringregionbdry}, we take these regions to be entanglement wedges. In this example, two subsystems (red) at $c_1 \in \m{C}_1$ and $c_2 \in \m{C}_2$ can scatter to $r_1 \in \m{R}_1$ and $r_2 \in \m{R}_2$, implying $J_{12\to 12}$ of eq.~\eqref{eq:scatteringregion} is nonempty. (b) Input and output regions for the corresponding boundary quantum task. These figures are reproduced from \cite{May:2021nrl}.
\label{fig:C1C2plot}}
\end{figure}

Let us begin with the bulk perspective, defining two input regions $\m{C}_1$ and $\m{C}_2$ and two output regions $\m{R}_1$ and $\m{R}_2$ for a quantum task in $\m{M}$. Following \cite{May:2021nrl}, we always take input and output regions to be causally complete, \ie $\m{D}(\m{S}) = \m{S}$ for all $\m{S}\in\{\m{C}_1,\m{C}_2,\m{R}_1,\m{R}_2\}$, where $\m{D}(\cdot)$ denotes the domain of dependence. This choice is natural when discussing resources for a task because all classical and quantum data in $\m{D}(\m{S})$ is fixed by the data in $\m{S}$ by time evolution.
A typical setup is shown in figure \ref{fig:C1C2plot_bulk}.

When describing causal constraints on this task, it is useful to denote the causal future and past of $\m{S}$ taken in the bulk by $\m{J}^+(\m{S})$ and $\m{J}^-(\m{S})$, respectively. Causal futures and pasts restricted to the conformal boundary will be denoted by $\hat{\m{J}}^{\pm}(\cdot)$.

With this notation established, let us assume 
\begin{equation}
    \m{C}_1\subset\m{J}^-(\m{R}_1)\cap\m{J}^-(\m{R}_2) \,,
    \qquad
    \m{C}_2\subset\m{J}^-(\m{R}_1)\cap\m{J}^-(\m{R}_2) \,,
    \label{eq:signal}
\end{equation}
\noindent
which says any subsystem in either of the input regions can signal to both output regions. However, once such a subsystem leaves the decision regions, defined as 
\begin{equation}
\begin{aligned}
    &\m{V}_1:=\m{J}^+(\m{C}_1)\cap\m{J}^-(\m{R}_1)\cap\m{J}^-(\m{R}_2) \,, \\
    &\m{V}_2:=\m{J}^+(\m{C}_2)\cap\m{J}^-(\m{R}_1)\cap\m{J}^-(\m{R}_2) \,,
\end{aligned}
\label{eq:decisionregions}
\end{equation}
it can signal to $\m{R}_1$ or $\m{R}_2$ (or neither of the two), but not to both. Notice that the $\m{V}_i$ contain the $\m{C}_i$.

We define the scattering region $J_{12\to12}$ as follows:
\begin{equation}
    J_{12\to12}:=\m{J}^+(\m{C}_1)\cap\m{J}^+(\m{C}_2)\cap\m{J}^-(\m{R}_1)\cap\m{J}^-(\m{R}_2) =\m{V}_1 \cap \m{V}_2 \,.
    \label{eq:scatteringregion}
\end{equation}
Notice that two subsystems, one starting from the input region $\m{C}_1$ and the other starting from the input region $\m{C}_2$, can scatter and reach output regions $\m{R}_1$ and $\m{R}_2$, respectively, if and only if $J_{12\to12}$ is nonempty. 

By AdS/CFT, the bulk task has a boundary description. 
The input and output regions $\mh{C}_i$ and $\mh{R}_i$ for the boundary task should be chosen such that $\m{C}_i \subset \m{E}(\mh{C}_i)$ and $\m{R}_i \subset \m{E}(\mh{R}_i)$, since entanglement wedge reconstruction implies that all data in $\m{E}(\mh{S})$ is encoded in $\mh{S}$, and vice versa. 
In particular, one can define $\m{C}_i := \m{E}(\mh{C}_i)$ and $\m{R}_i := \m{E}(\mh{R}_i)$, where $\mh{C}_i$ and $\mh{R}_i$ are arbitrary diamonds satisfying boundary analogs of eqs.~\eqref{eq:signal} and \eqref{eq:decisionregions}. 
Refer to figure \ref{fig:C1C2plot_bdry} for a diagram of the $\mh{R}_i$ and $\mh{C}_i\subset \mh{V}_i$. The boundary scattering region is defined as
\begin{equation}
\hat{J}_{12\to12}:=\mh{J}^+(\mh{C}_1)\cap\mh{J}^+(\mh{C}_2)\cap\mh{J}^-(\mh{R}_1)\cap\mh{J}^-(\mh{R}_2)= \mh{V}_1 \cap \mh{V}_2 \,.
\label{eq:scatteringregionbdry}
\end{equation}

The connected wedge theorem concerns the case where $\hat{J}_{12\to12}=\emptyset$ and yet $J_{12\to12}\neq\emptyset$. In this case, even though scattering is forbidden in $\partial\m{M}$, holography demands the boundary theory must be able to reproduce the outputs of bulk tasks in which the two systems make causal contact.
This is called holographic scattering. The connected wedge theorem, stated below, says holographic scattering places a lower bound on correlations between $\mh{V}_1$ and $\mh{V}_2$. The intuition from quantum information theory is that some or all of these correlations serve as a resource for reproducing the bulk interaction.

\begin{thr}[The connected wedge theorem]
Adapted from \cite{May:2021nrl} and \cite{May:2022clu}.
Pick four regions $\mh{C}_1$, $\mh{C}_2$, $\mh{R}_1$, and $\mh{R}_2$ on the boundary of an asymptotically AdS$_{2+1}$ spacetime with a holographic dual.
Assume the spacetime satisfies the null energy condition and is AdS-hyperbolic.
Define the decision regions
\begin{equation}
\begin{aligned}
    &\mh{V}_1:=\mh{J}^+(\mh{C}_1)\cap\mh{J}^-(\mh{R}_1)\cap\mh{J}^-(\mh{R}_2) \,, \\
    &\mh{V}_2:=\mh{J}^+(\mh{C}_2)\cap\mh{J}^-(\mh{R}_1)\cap\mh{J}^-(\mh{R}_2) \,.
\end{aligned}
\label{eq:decisionregionsthm}
\end{equation}
Assume that $\mh{C}_i\subset \mh{V}_i$. Define the (bulk) scattering region 
\begin{equation}
J_{12\to12}:=\m{J}^+(\m{C}_1)\cap\m{J}^+(\m{C}_2)\cap\m{J}^-(\m{R}_1)\cap\m{J}^-(\m{R}_2) \,,
\label{eq:scatteringregionbulkthm}
\end{equation}
where $\m{C}_i=\m{E}(\mh{C}_i)$ and $\m{R}_i=\m{E}(\mh{R}_i)$.
Assume that the RT surface for $\m{E}(\mh{V}_1\cup\mh{V}_2)$ can be found by a maximin procedure \cite{Wall_2014}.
Then, $J_{12\to12}\neq \emptyset$ implies $\m{E}(\mh{V}_1\cup\mh{V}_2)$ is connected.
\end{thr}

Note that the theorem holds trivially when $\hat{J}_{12\to12}\neq\emptyset$. In this case, ${J}_{12\to12}\neq\emptyset$, so the theorem says the entanglement wedge is connected, but this is already implied by the fact that $\mh{V}_1$ and $\mh{V}_2$ overlap. In what follows, we will always choose the $\mh{V}_i$ to be non-overlapping. 

Also note that the theorem only goes one way. That is, it is possible that $\m{E}(\mh{V}_1\cup\mh{V}_2)$ is connected and yet $J_{12\to12}$ is empty for all compatible choices of the $\mh{C}_i$ and $\mh{R}_i$. 
In this case, we say that $\m{E}(\mh{V}_1\cup\mh{V}_2)$ does not have a corresponding holographic scattering process. Physically, this means the boundary correlations do not support a bulk causal scattering process despite being large. 

\begin{figure}[htbp]
\centering

\begin{subfigure}[b]{0.57\textwidth}
    \centering
    \includegraphics[width=\textwidth]{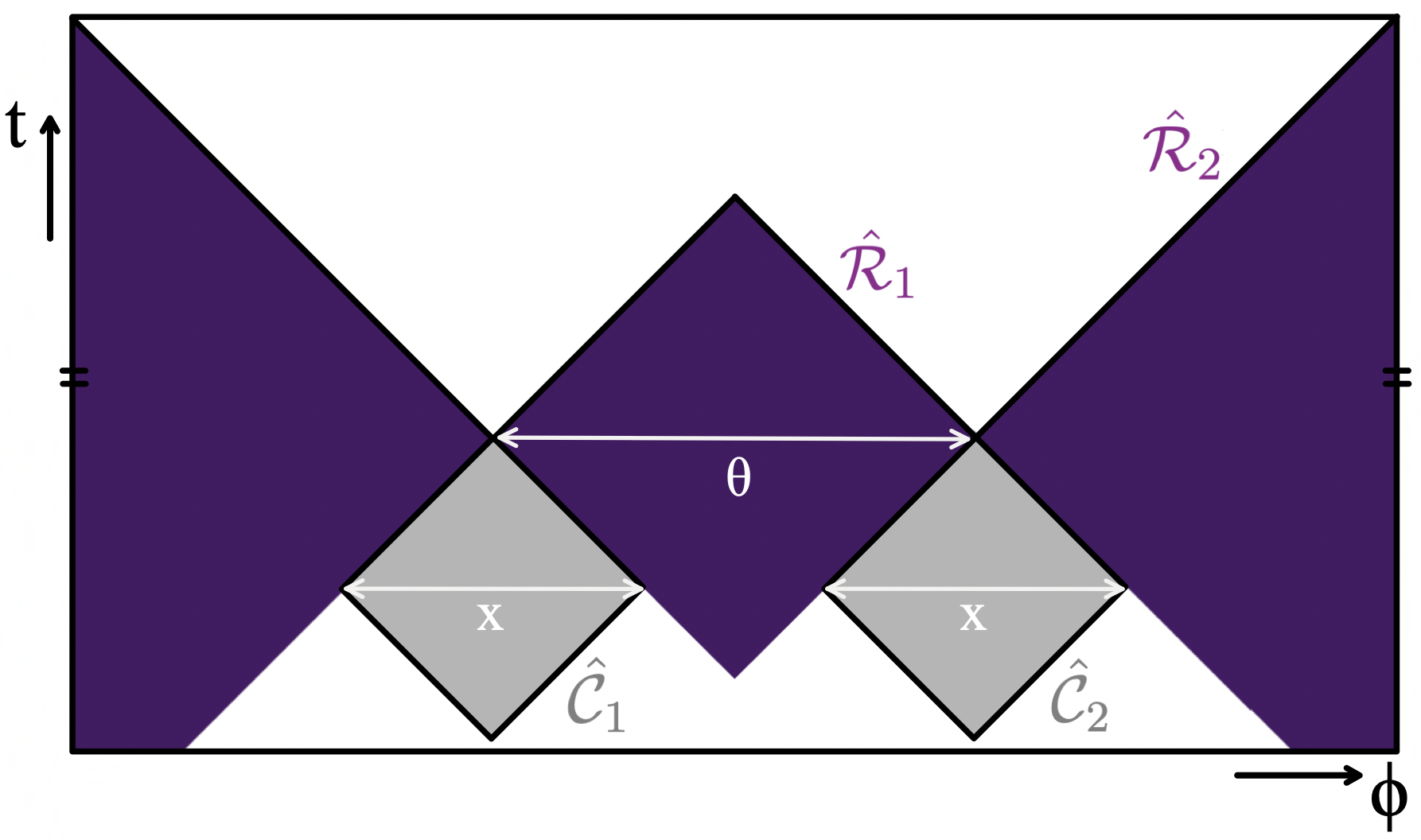}
    \caption{}
\end{subfigure}
\hfill
\begin{subfigure}[b]{0.36\textwidth}
    \centering
    \includegraphics[width=\textwidth]{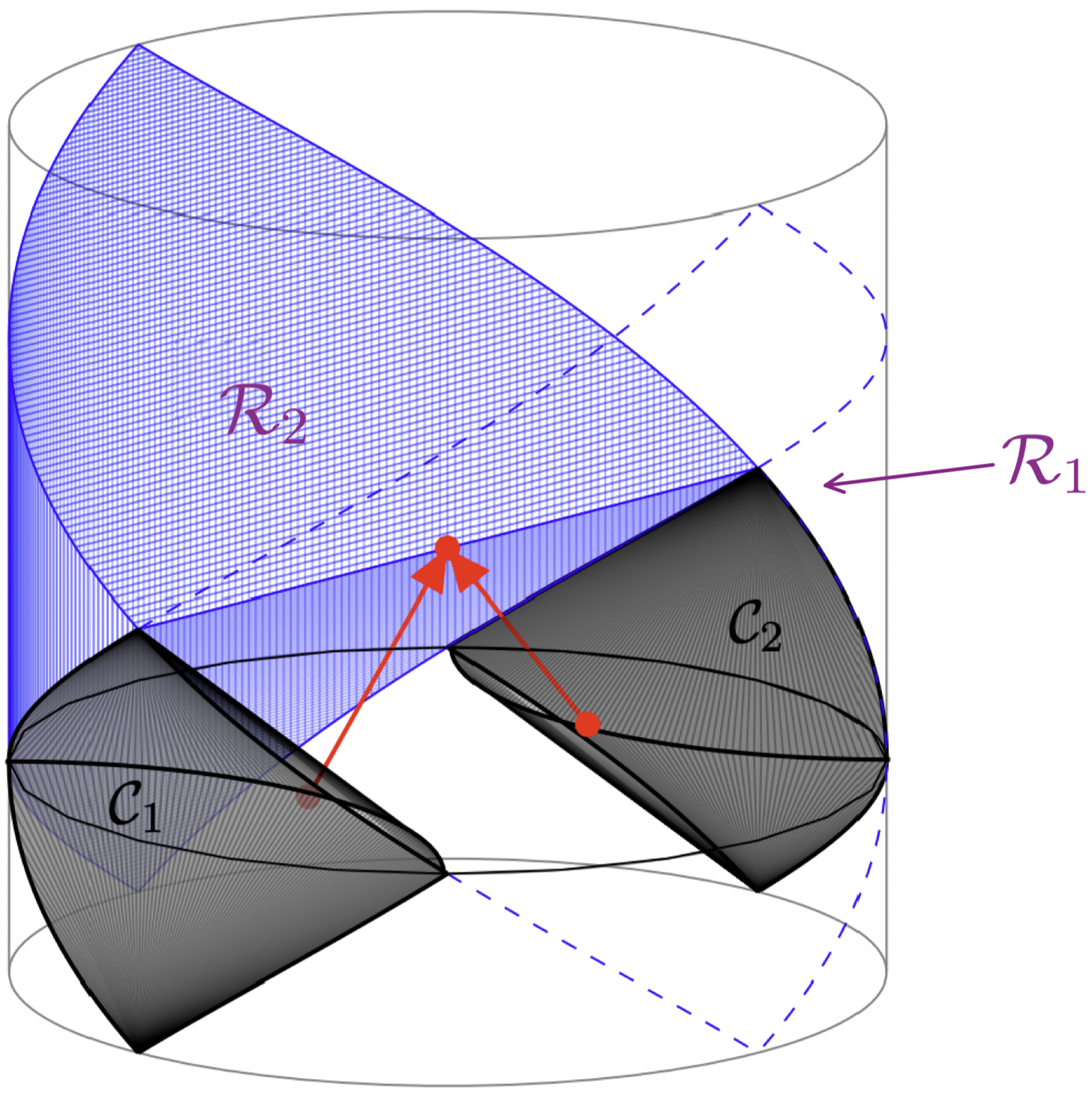}
    \caption{}
\end{subfigure}

\caption{Maximal choices of $\mh{C}_1$, $\mh{C}_2$, $\mh{R}_1$, and $\mh{R}_2$ corresponding to $\m{E}(\mh{V}_1\cup\mh{V}_2)$ from the (a) boundary and (b) bulk perspectives. In particular, $\mh{C}_i=\mh{V}_i$. A non-maximal example is provided by figure \ref{fig:C1C2plot}. Figure (b) is reproduced from \cite{May:2021nrl}.}
\label{fig:R1R2plot}
\end{figure}

To check whether $\m{E}(\mh{V}_1\cup\mh{V}_2)$ has a corresponding holographic scattering process, it suffices to check whether $J_{12\to12}$ is nonempty for a special choice of the $\mh{C}_i$ and $\mh{R}_i$.
Setting $\mh{C}_i = \mh{V}_i$, we see that this contains all other compatible choices of $\mh{C}_i$, which implies the entanglement wedge of $\mh{C}_i = \mh{V}_i$ contains the entanglement wedge of all other choices of $\mh{C}_i$. 
This means that if scattering is possible, it will be possible for $\mh{C}_i = \mh{V}_i$. 
Similarly, we take the $\mh{R}_i$ as large as possible, as shown in figure \ref{fig:R1R2plot}, so that they contain all other possible $\mh{R}_i$ as subregions. 
With this choice, $\mh{R}_1\cup\mh{R}_2$ contains a Cauchy slice in the boundary.

In the next sections, we apply these ideas to the conical defect and BTZ black hole geometries. We determine which connected wedges have a corresponding holographic scattering process, revealing a connection between holographic scattering and nonminimal extremal surfaces whose physical meaning will be discussed in section \ref{sec:discuss}.

\subsection{Holographic scattering in defect geometry}
\label{sec:sec32}

In this section, we consider two boundary regions $A$ and $B$ determined by the parameters $x$ and $\theta$ as in section \ref{sec:geometryandwedges}, and examine which connected wedges $\m{E}(A\cup B)$ in the conical defect spacetime have a corresponding holographic scattering process. 
We find that holographic scattering comes with not just $\min(u,o)<d$, as implied by the connected wedge theorem, but also the stronger condition $u < d$.

To determine whether a connected wedge $\m{E}(A\cup B)$ has a corresponding scattering process, we identify $\mh{C}_1 = \mh{V}_1 = \m{D}(A)$ and $\mh{C}_2 = \mh{V}_2 = \m{D}(B)$. In particular, the $\mh{V}_i$ have width $x$. We also take the $\mh{R}_i$ as in figure \ref{fig:R1R2plot}. Then, $\mh{R}_1$ and $\mh{R}_2$ have widths $\theta$ and $2\pi-\theta$, respectively. By the constraints on $x$ and $\theta$ discussed below eq.~\eqref{eq:gammaconicalAdS}, boundary scattering is forbidden, and we are checking if there is scattering in the bulk.

Let us break down the statement $J_{12\to12}\neq \emptyset$ into something computable, starting from the definition \eqref{eq:scatteringregionbulkthm} of the scattering region.
In appendix \ref{sec:defectwedgesappendix}, we demonstrate that in a defect geometry, the entanglement wedge $\m{E}(\mh{S})$ for a causal diamond $\mh{S}$ equals its causal wedge $\m{C}(\mh{S}) := \m{J}^+(\mh{S})\cap \m{J}^-(\mh{S})$ if and only if the width of $\mh{S}$ is less than or equal to $\pi$.\footnote{We restrict to the case $\mh{S}=\m{D}(A)$, where $A$ lies in a constant-time slice. Thus, the width of $\mh{S}$ is defined as the angular extent of $A$.} Since $x,\,\theta \leq \pi$, this allows us to rewrite eq.~\eqref{eq:scatteringregionbulkthm} as follows:
\begin{equation}
J_{12\to12}=\m{J}^+(\m{C}(\mh{V}_1))\cap\m{J}^+(\m{C}(\mh{V}_2))\cap\m{J}^-(\m{C}(\mh{R}_1))\cap\m{J}^-(\m{E}(\mh{R}_2)) \,.
\label{eq:threeCs}
\end{equation}
Further, because the $\mh{V}_i$ are causal diamonds, we have $\m{J}^+(\m{C}(\mh{V}_1)) = \m{J}^+(c_i)$, where $c_i$ are the past-most points of $\mh{V}_i$. In words, $J_{12\to12}\neq\emptyset$ if and only if particles leaving from $c_1$ and $c_2$ are able to meet at some point $P$ in $\m{J}^-(\m{C}(\mh{R}_1))\cap\m{J}^-(\m{E}(\mh{R}_2))$.

A final simplification comes from writing 
\begin{equation}
\m{J}^-(\m{C}(\mh{R}_1))\cap\m{J}^-(\m{E}(\mh{R}_2))=\m{J}^-(\gamma_{\m{R}}) \,,
\label{eq:rtpast}
\end{equation}
where $\gamma_{\m{R}}$ is the RT surface shared by the boundary subregions $\mh{R}_1$ and $\mh{R}_2$.
To see why this equality is true, let us restrict our attention to the past of the constant-$t$ slice in which $\gamma_R$ lives. In this region, $\partial\m{J}^-(\m{C}(\mh{R}_1))$ consists of past-directed null rays launched orthogonally from $\gamma_{\m{R}}$ in the direction away from $\mh{R}_1$.\footnote{For an explanation of why the boundary of the past of a codimension-two spacelike surface is bounded by such null rays, refer to \cite{Witten_2020}, end of Section 5.3.} $\m{J}^-(\m{E}(\mh{R}_2))$ also consists of past-directed null rays launched orthogonally from $\gamma_{\m{R}}$, but in the direction toward $\mh{R}_1$. Thus, the intersection $\m{J}^-(\m{C}(\mh{R}_1))\cap\m{J}^-(\m{E}(\mh{R}_2))$ has as its boundary the two past-directed orthogonal null congruences emanating from $\gamma_{\m{R}}$, which precisely define $\partial\m{J}^-(\gamma_{\m{R}})$. Hence, we have eq.~\eqref{eq:rtpast}.

\begin{figure}[htbp]
\centering
\begin{subfigure}[b]{0.36\textwidth}
    \centering
    \includegraphics[width=\textwidth]{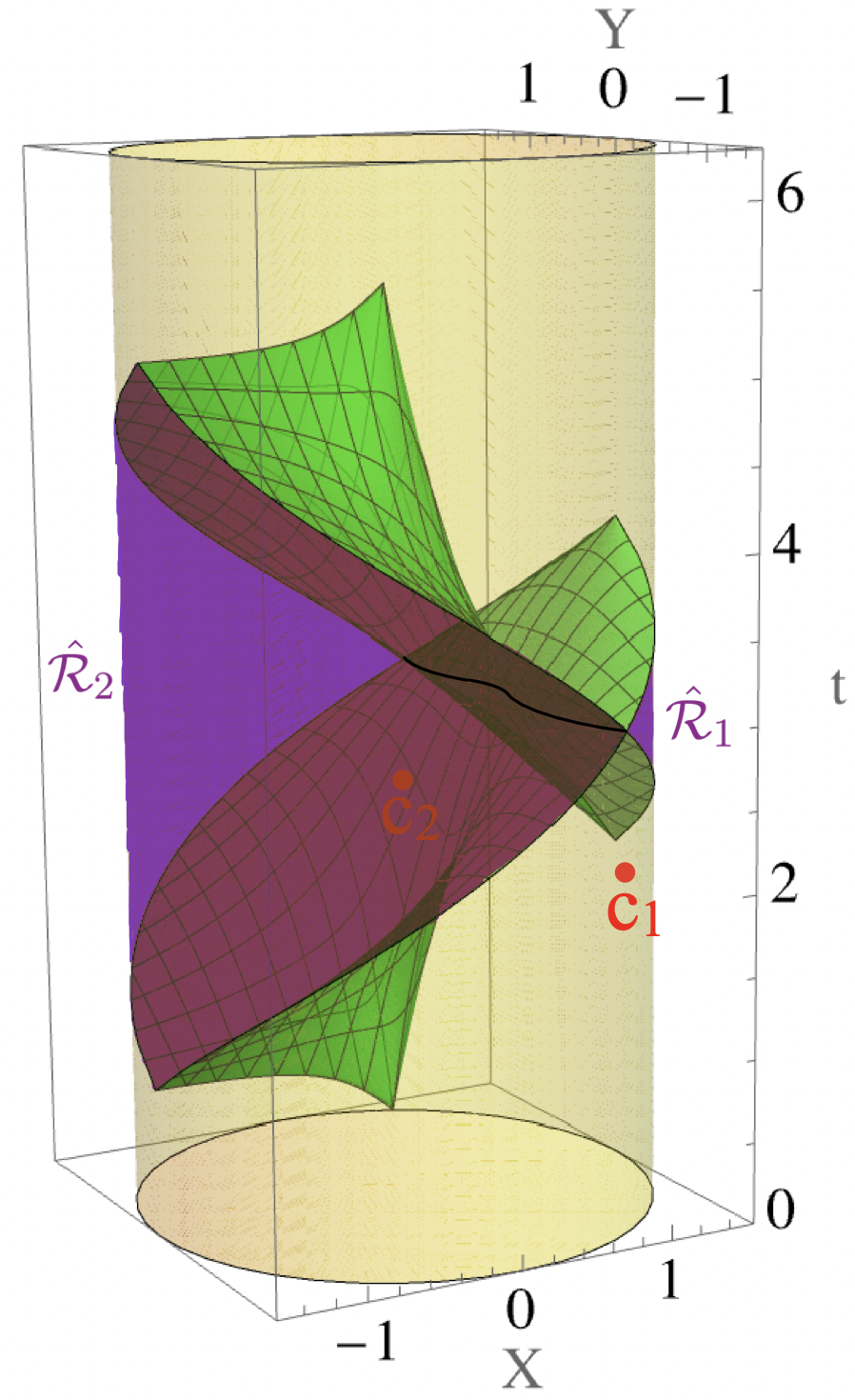}
    \caption{}
    \label{fig:prettywedges_conical}
\end{subfigure}
\qquad \qquad \qquad
\begin{subfigure}[b]{0.32\textwidth}
    \centering
    \includegraphics[width=\textwidth]{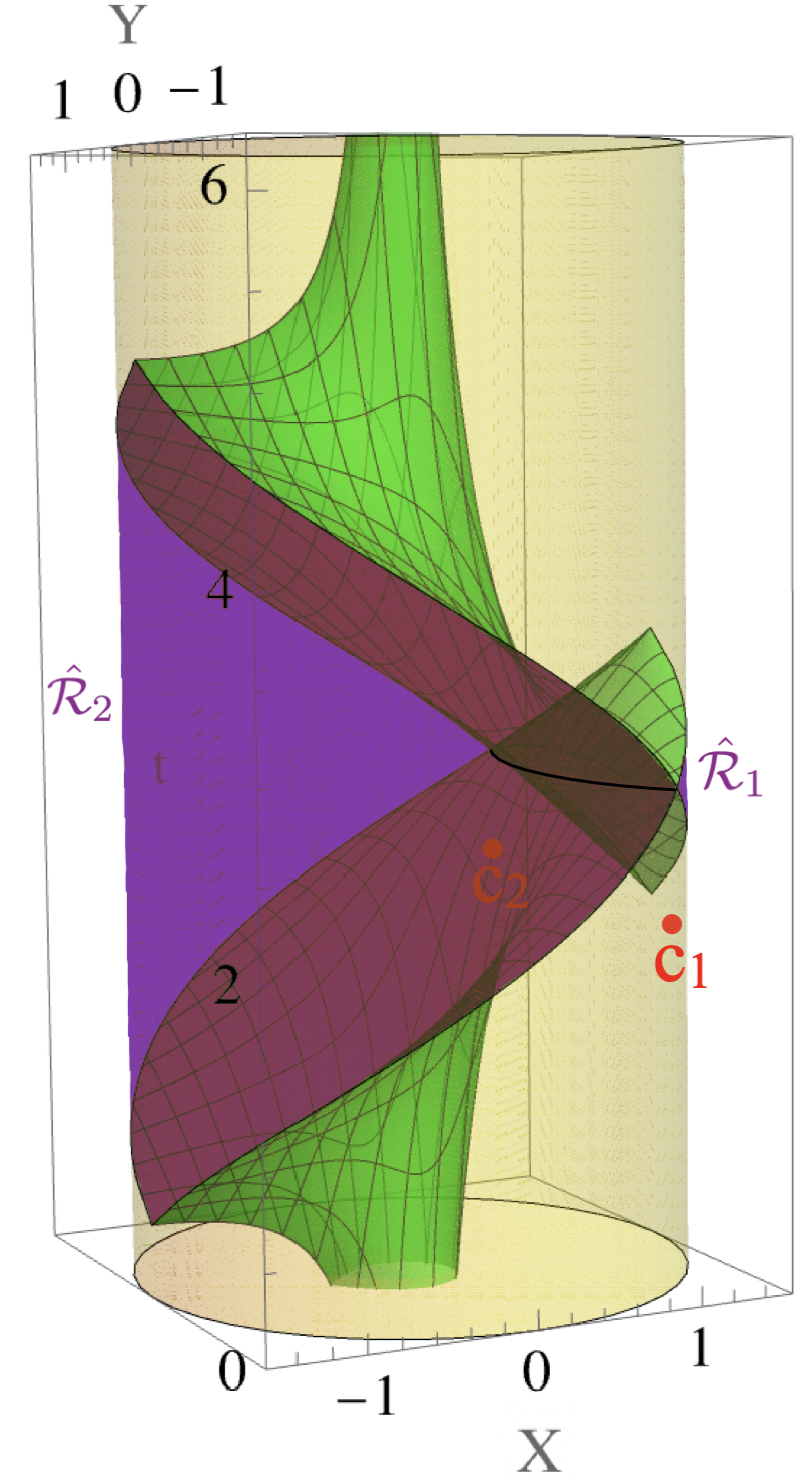}
    \caption{}
    \label{fig:prettywedges_BTZ}
\end{subfigure}
\caption{Setup for computing the holographic scattering inequalities \eqref{eq:regionsineqconical} and \eqref{eq:regionsineqbtz} in the (a) defect and (b) BTZ geometries, respectively. The boundary regions $\mh{R}_1$ and $\mh{R}_2$ are shown in purple. Their entanglement wedges are shown in green, and the two entanglement wedges meet at $\gamma_{\mh{R}_1}=\gamma_{\mh{R}_2}$, shown in black.}
\label{fig:prettywedges}
\end{figure}

Using eq.~\eqref{eq:rtpast}, we show in appendix \ref{sec:meetingatP} that $J_{12\to12}\neq\emptyset$ if and only if the particles leaving from $c_1$ and $c_2$ are able to meet at $P_*$, where $P_*:=(t_*, r_*, \phi_*)$ is defined as the deepest (minimal-radius) point of $\gamma_{\m{R}}$. 
This enables a simple method for computing the conditions on $x$ and $\theta$ for holographic scattering.
First, let $\phi=0$ coincide with the center of $\mh{R}_1$.
Note that $c_1$ and $c_2$ are placed symmetrically about $\phi=0$ --- see figure \ref{fig:prettywedges_conical}.
Then, let $\Delta t$ be the coordinate time required for a null geodesic to reach $(r=r_*, \phi=0)$, which are coordinates of $P_*$, from the point $(r=\infty, \phi=\theta/2)$, which are coordinates of $c_2$. If $\Delta t \leq t_*-t_{c_2}=x$, then massless particles from $c_1$ and $c_2$ have enough time to reach $P_*$, and therefore the scattering region is nonempty. Conversely, if $t_*>x$, then the scattering region is empty.

\begin{figure}[htbp]
\centering
\includegraphics[width=.5\textwidth]{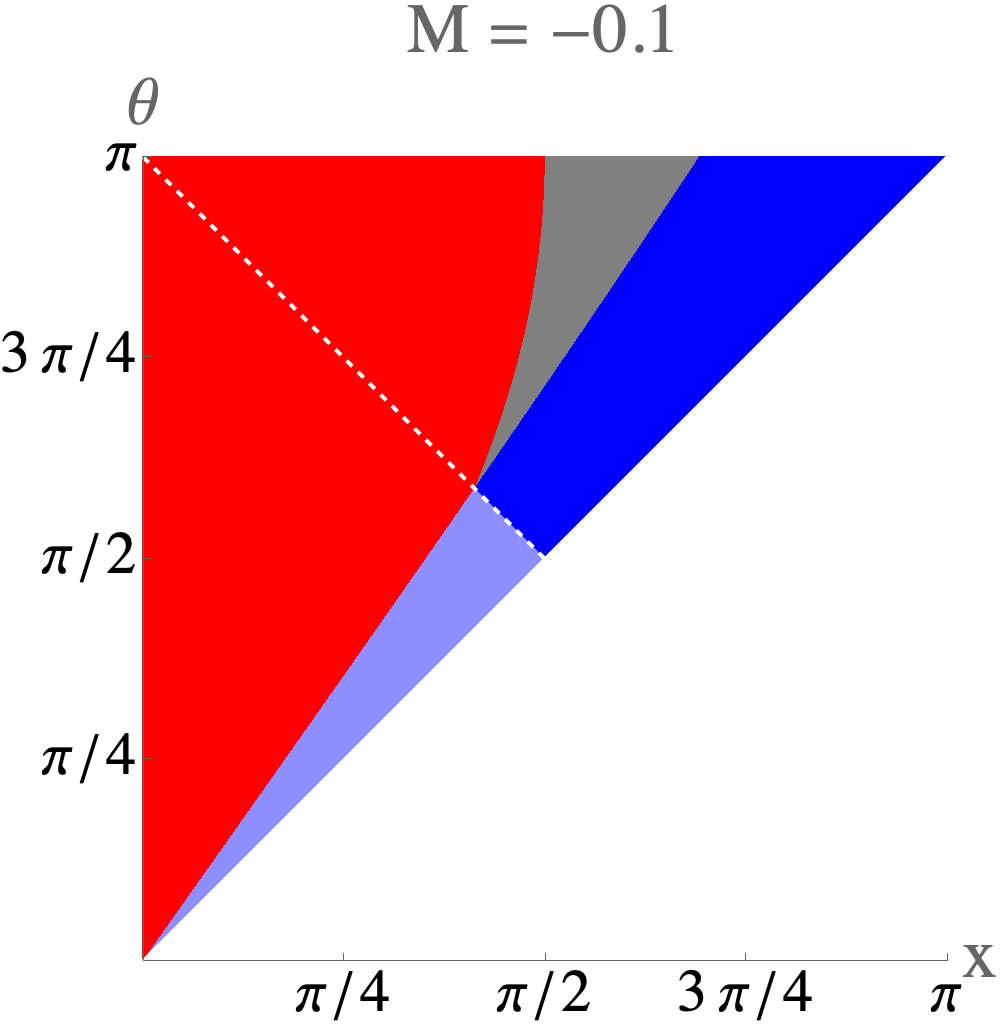}
\caption{Holographic scattering phase diagram for the defect spacetime with $M=-0.1$. Red, dark blue, and light blue regions correspond to $d$, $o$, and $u$ phases of the entanglement wedge, respectively. Part of the $o$-phase region is shaded grey, indicating that in this region, the entanglement wedge is connected but there is no corresponding holographic scattering process. By the discussion below eq.~\eqref{eq:regionsineqconical}, the holographic scattering region (both light and dark blue) matches the $u<d$ region of figure \ref{fig:uphasea}.
\label{fig:tempregionsconical}}
\end{figure}

The inequality $t_*<x$ is computed in appendix \ref{sec:regionsappendix}. The result is that holographic scattering occurs if and only if
\begin{equation}
\cos^2\left(\sqrt{|M|}\, \theta\, /\, 2\right) > \cos \left(\sqrt{|M|}\, x\right) \,,
\label{eq:regionsineqconical}
\end{equation}
which is equivalent to the $u<d$ inequality \eqref{eq:uineq}. 
See figure \ref{fig:tempregionsconical} and caption.
This agrees with the connected wedge theorem insofar as holographic scattering implies the connectivity of the entanglement wedge. However, a connected entanglement wedge is a condition on minimal surfaces only, \ie $\min(o,u)<d$.
We have found a stronger condition, namely $u<d$.
We emphasize that holographic scattering requires this condition even in situations where $o<u$.
Hence, holographic scattering appears to constrain extremal surfaces that are not necessarily minimal.\footnote{Note that the scattering inequality \eqref{eq:regionsineqconical} holds for all choices of the defect parameter $M$. However, the discussion in footnotes \ref{foot:nonexistence} and \ref{foot:Carol} implies that the connection to nonminimal surfaces requires either we consider a heavy defect with $|M| < 1/4$ or, in the case of light defects ($|M| > 1/4$), we must also constrain $\theta+x<\pi\sqrt{|M|}$. Otherwise the $u$ configuration does not exist.}

Note that since our setup is symmetric under $\theta \mapsto 2\pi-\theta$, an extended phase diagram with $\theta \in (0, 2\pi)$ would be symmetric under reflection under the $\theta = \pi$ axis. However, in the regime $\theta > \pi$, the holographic scattering inequality reads not $u<d$ but rather $u'<d$, where the $u'$ configuration is defined in figure \ref{fig:dcircsunusual}.
We will revisit this point in section \ref{sec:sec34}, since there we must break the $\theta \mapsto 2\pi-\theta$ symmetry.

In the following section, we find that holographic scattering with $\theta\le\pi$ also comes with $u<d$ in the black hole case. 
We discuss these results in section \ref{sec:discuss}.

\subsection{\texorpdfstring{Holographic scattering in BTZ geometry}{Holographic scattering in BTZ geometry}}
\label{sec:sec33}

In this section, we detail which connected wedges $\m{E}(A\cup B)$ in the BTZ geometry have a corresponding holographic scattering process. 
We again find that holographic scattering comes with not just $\min(u,o)<d$, but also the stronger condition $u < d$.
Further, we find an additional constraint on holographic scattering besides $u<d$ which involves the size of the output regions $\mh{R}_i$.

As in the defect case, we identify $\mh{C}_1 = \mh{V}_1 = \m{D}(A)$ and $\mh{C}_2 = \mh{V}_2 = \m{D}(B)$ and choose the $\mh{R}_i$ to be as large as possible. 
Then, $\m{E}(A\cup B)$ has a corresponding holographic scattering process for the particular choice of $x$ and $\theta$ if and only if $J_{12\to12}\neq \emptyset$.

Recall that in the defect analysis of section \ref{sec:sec32}, we simplified the expression \eqref{eq:scatteringregionbulkthm} for $J_{12\to12}$ using that the entanglement wedge $\m{E}(\mh{S})$ for a causal diamond $\mh{S}$ equals its causal wedge $\m{C}(\mh{S}) := \m{J}^+(\mh{S})\cap \m{J}^-(\mh{S})$ if and only if the width of $\mh{S}$ is less than or equal to $\pi$. 
As shown in appendix \ref{sec:lightconeappendix}, a similar statement holds in the BTZ case: the entanglement wedge $\m{E}(\mh{S})$ for a causal diamond $\mh{S}$ equals its causal wedge $\m{C}(\mh{S}) := \m{J}^+(\mh{S})\cap \m{J}^-(\mh{S})$ if and only if the width of $\mh{S}$ is less than or equal to $\Delta \phi^*(M)$.

Since $x,\theta\leq \pi\leq \Delta \phi^*(M)$, we can again rewrite the definition \eqref{eq:scatteringregionbulkthm} for the bulk scattering region $J_{12\to 12}$ as in eq.~\eqref{eq:threeCs}. In the defect case, we were unable to write $\m{E}(\mh{R}_2)=\m{C}(\mh{R}_2)$, since the condition $2\pi-\theta < \pi$ would require $\theta > \pi$. 
In the BTZ case, however, as long as $2\pi-\theta \leq \Delta \phi^*(M)$, we also have $\m{E}(\mh{R}_2)=\m{C}(\mh{R}_2)$ and so all entanglement wedges in the definition for $J_{12\to 12}$ can be replaced by causal wedges. 

\begin{figure}
\centering
\includegraphics[width=.5\textwidth]{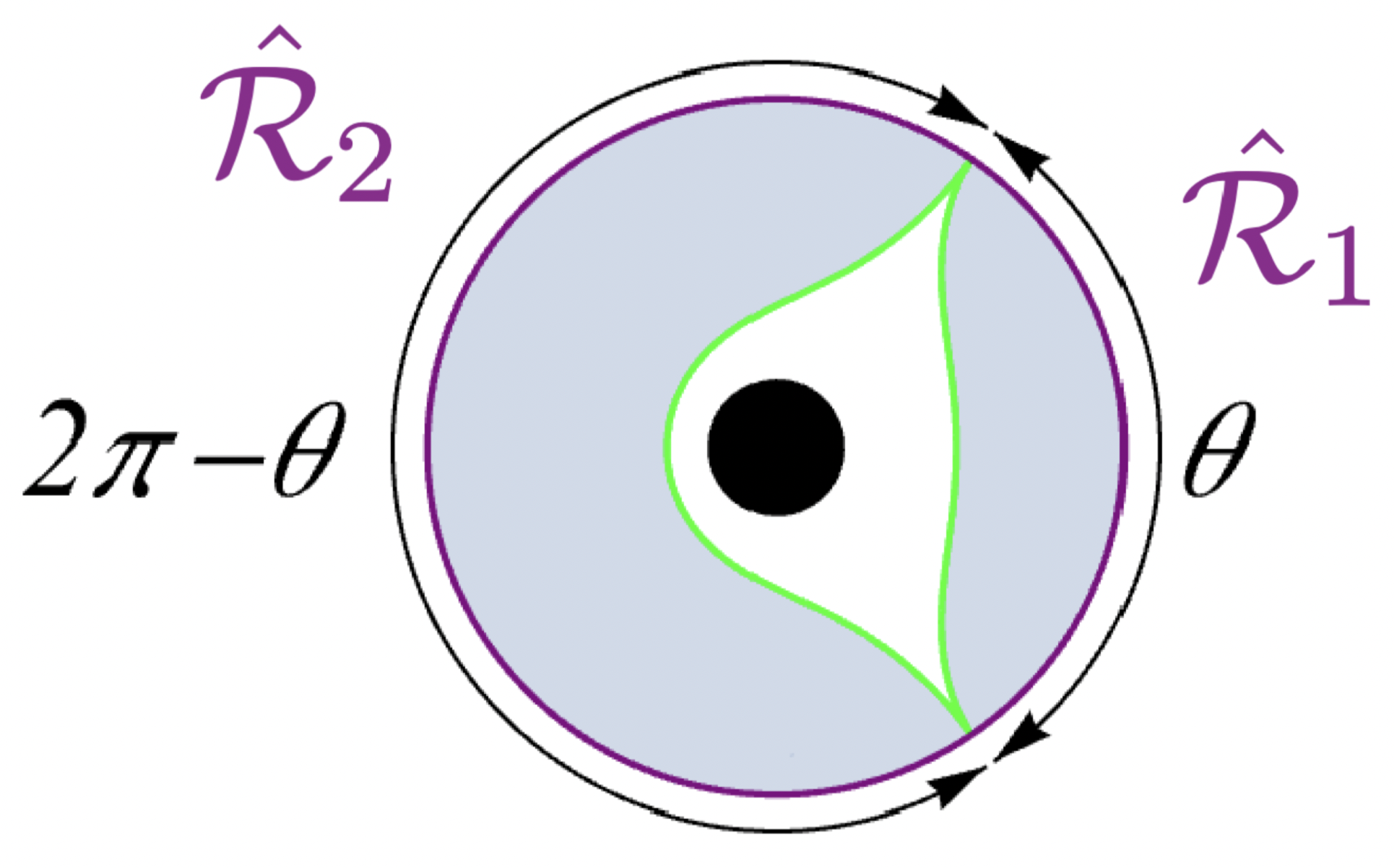}
\caption{When $\theta \geq 2\pi - \Delta \phi^*(M)$, there is a gap between the entanglement wedges of $\mh{R}_1$ and $\mh{R}_2$. Compare figure \ref{fig:BTZgeodesics}.}\label{fig:shadow}
\end{figure}

Accordingly, the first regime which we examine for scattering in the BTZ geometry is $\theta \geq 2\pi - \Delta \phi^*(M)$, for which we have
\begin{equation}
J_{12\to12}=\m{J}^+(\m{C}(\mh{V}_1))\cap\m{J}^+(\m{C}(\mh{V}_2))\cap\m{J}^-(\m{C}(\mh{R}_1))\cap\m{J}^-(\m{C}(\mh{R}_2)) \,.
\label{eq:special1}
\end{equation}
In this regime, the RT surfaces for $\mh{R}_1$ and $\mh{R}_2$ are separated, \ie $\gamma_{\mh{R}_1}$ and $\gamma_{\mh{R}_2}$ have no points in common (apart from the boundary endpoints at $r\to\infty$) --- see figure \ref{fig:shadow}.
Thus, even though $\mh{R}_1$ and $\mh{R}_2$ cover a constant-$t$ slice of the boundary, the union of their entanglement wedges does not cover the bulk constant-$t$ slice.
This gap between the entanglement wedges reflects, in part, the mixed nature of the thermal CFT state dual to the single-sided BTZ geometry.\footnote{When the boundary state is pure, the RT surfaces for complementary regions are always identical, and the corresponding entanglement wedges cover an entire bulk time slice \cite{Hubeny:2012wa}. This property is sometimes called entanglement wedge complementarity.}
We revisit this point in section \ref{sec:sec34}.

Recalling that the $\mh{V}_i$ and $\mh{R}_i$ are causal diamonds, we can rewrite eq.~\eqref{eq:special1} as
\begin{equation}
J_{12\to12} = \m{J}^+(c_1)\cap\m{J}^+(c_2)\cap\m{J}^-(r_1)\cap\m{J}^-(r_2) \,,
\end{equation}
where $c_i$ are the past-most points of $\mh{V}_i$ and $r_i$ are the future-most points of $\mh{R}_i$. 
Thus, $J_{12\to12}\neq\emptyset$ if and only if particles leaving from $c_1, c_2 \in \partial\m{M}$ can scatter to $r_1, r_2 \in \partial\m{M}$. 
An analogous situation was displayed in figure \ref{fig:phasetransition3dintro}.
Hence, in this regime, understanding the regions-based scattering process reduces to examining a points-based process. 
As we show in appendix \ref{sec:pointsappendix}, any such points-based process is fastest when the scattering point lies on the boundary, due to large gravitational time delays in the BTZ spacetime. 
This means $J_{12\to12}\neq\emptyset\implies\hat{J}_{12\to12}\neq\emptyset$, and so holographic scattering is impossible in this regime. From the $u<d$ inequality \eqref{eq:uineqbtz}, it is possible to have $u<d$ and $\theta \ge 2\pi - \Delta \phi^*(M)$ together. This means that unlike in the defect case, $u<d$ is not a sufficient condition for holographic scattering. However, we show below that it is still a necessary condition.

Consider the second regime, where $\theta < 2\pi - \Delta \phi^*(M)$.
In this case, by eq.~\eqref{eq:gammaBTZ}, $\gamma_{\mh{R}_2}$ is the union of $\gamma_{\mh{R}_1}$ with (the bifurcation surface of) the black hole horizon.
Thus, we can employ not only the simplification \eqref{eq:threeCs} but also the following analog of eq.~\eqref{eq:rtpast}:
\begin{equation}
\m{J}^-(\m{C}(\mh{R}_1))\cap\m{J}^-(\m{E}(\mh{R}_2))=\m{J}^-(\gamma_{\mh{R}_1})\,.
\label{eq:rtpastbtz}
\end{equation}
As with eq.~\eqref{eq:rtpast}, eq.~\eqref{eq:rtpastbtz} may be justified by the fact that the boundary of the causal past of a codimension-two spacelike surface is generated by null rays orthogonal to the surface.
The presence of the black hole horizon in  $\gamma_{\mh{R}_2}$ does not spoil the proof because the null sheet it contributes to $\partial\m{J}^-(\m{E}(\mh{R}_2))$ consists of past-directed null rays which fall behind or along the horizon and never intersect $\m{J}^-(\m{C}(\mh{R}_1))$.
We emphasize that this argument for eq.~\eqref{eq:rtpastbtz} only works because both entanglement wedges contain $\gamma_{\mh{R}_1}$, \ie there is no gap of the type of figure \ref{fig:shadow}.

The remainder of the analysis proceeds as in the defect case.
By the symmetric placement of $c_1$ and $c_2$ about the center of $\mh{R}_1$, we find $J_{12\to12}\neq\emptyset$ if and only if particles leaving from $c_1$ and $c_2$ are able to meet at $P_*$, where $P_*:=(t_*, r_*, \phi_*)$ is defined as the deepest (minimal-radius) point of $\gamma_{\mh{R}_1}$.
See figure \ref{fig:prettywedges_BTZ}.
Let $\Delta t$ be the coordinate time required for a null geodesic to reach $(r=r_*, \phi=0)$, which are coordinates of $P_*$, from the point $(r=\infty, \phi=\theta/2)$, which are coordinates of $c_2$. 
If $\Delta t \leq t_*-t_{c_2}=x$, then the scattering region is nonempty. 
Conversely, if $t_*>x$, then the scattering region is empty. 

\begin{figure}[htbp]
\centering
\includegraphics[width=.3\textwidth]{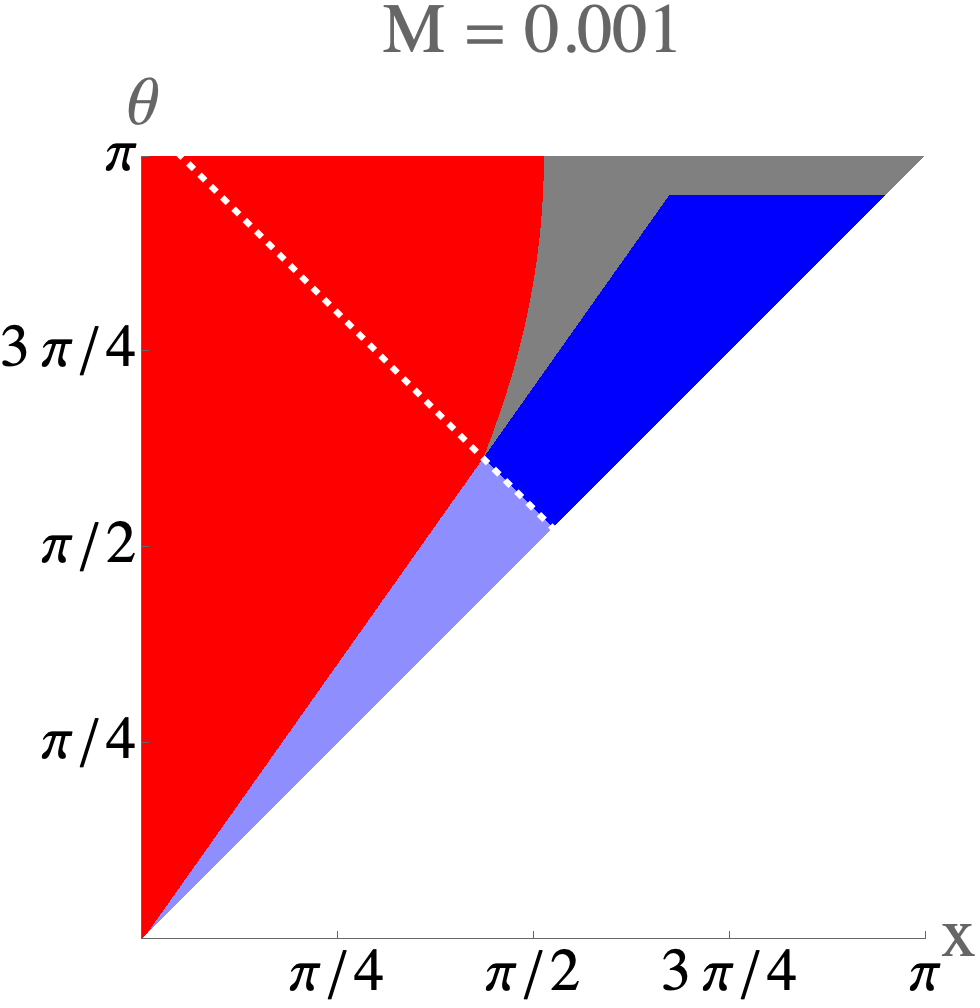}
\hfill
\includegraphics[width=.3\textwidth]{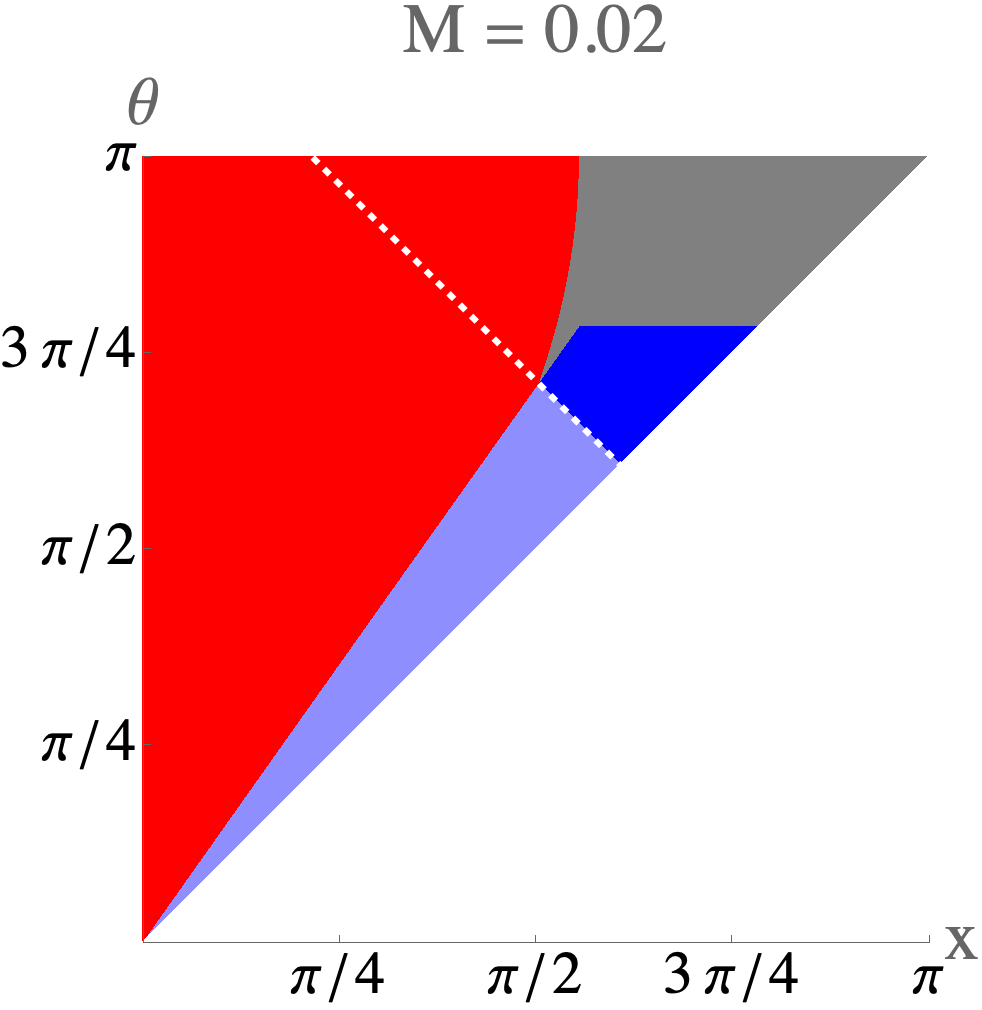}
\hfill
\includegraphics[width=.3\textwidth]{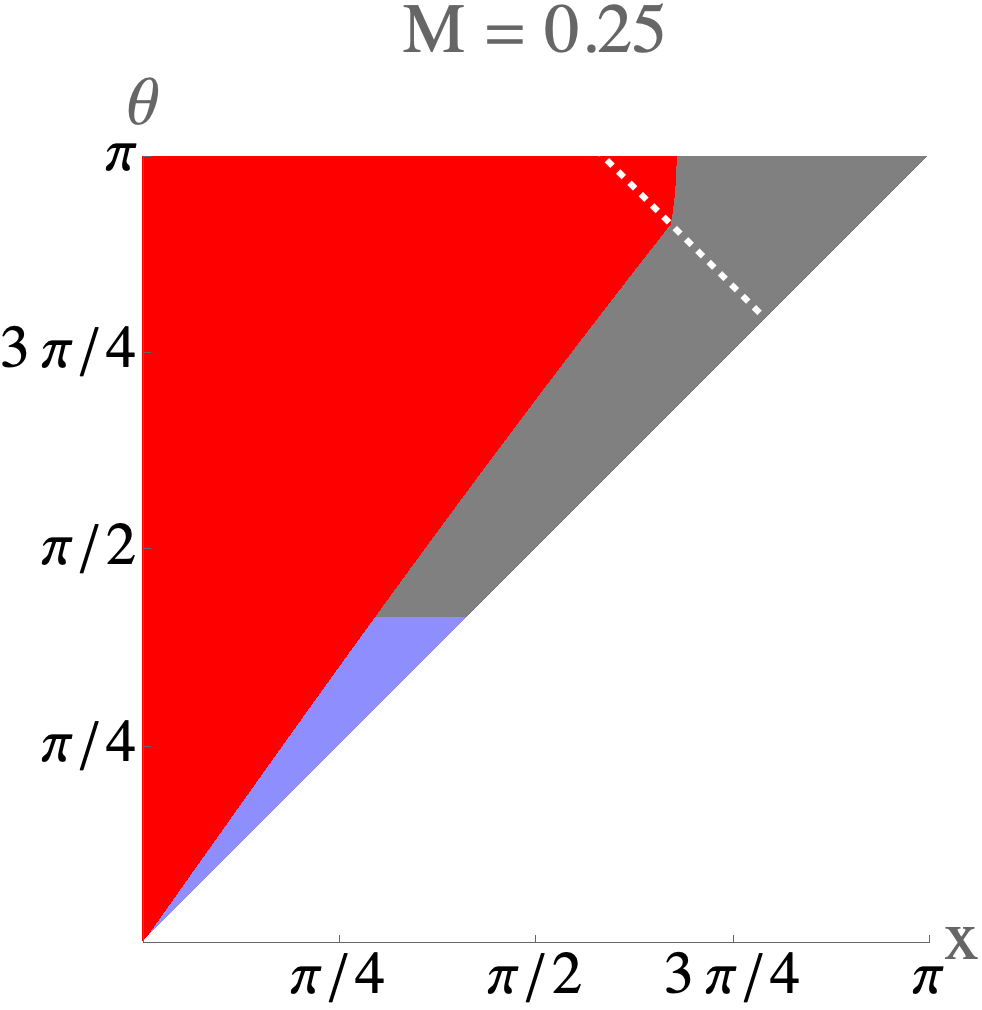}
\caption{Holographic scattering phase diagram in BTZ geometries with $M=0.001$, $0.02$, and $0.25$. The light and dark blue regions indicate the entanglement wedge is connected and has a corresponding holographic scattering process. Grey indicates the entanglement wedge is connected but there is no scattering. Condition 2 in eq.~\eqref{sleeper} is responsible for the horizontal border of the grey region, whereas condition 1 introduces a border similar to that of figure \ref{fig:tempregionsconical}. The red (disconnected) region expands as $M$ increases, as should be clear from comparing to the entanglement wedge phase diagram of figure \ref{fig:BTZphasediagram}.
\label{fig:btzregionsphase}}
\end{figure}

This computation is performed in appendix \ref{sec:regionsappendix}. The result is that holographic scattering for $\theta<2\pi-\Delta\phi^*$ occurs if and only if
\begin{equation}
\cosh^2 \left(\sqrt{M}\,\theta\, /\, 2\right) < \cosh \left(\sqrt{M} \, x\right) \,,
\label{eq:regionsineqbtz}
\end{equation}
which is equivalent to the $u<d$ inequality \eqref{eq:uineqbtz}.
Thus, a connected entanglement wedge has a corresponding holographic scattering process if and only if
\begin{equation}
\begin{aligned}
&(1)\ \ u<d\,,\quad {\rm and} \\
&(2)\ \ 
 \theta<2\pi-\Delta\phi^*\,.
 \label{sleeper}
\end{aligned}
\end{equation}
We illustrate this result in figure \ref{fig:btzregionsphase}.

It is natural to ask whether condition 2, like condition 1, can be interpreted as an ordering inequality among extremal surfaces.
The answer is no. 
None of the ordering inequalities reproduces the condition $\theta<2\pi - \Delta \phi^*$, even if we include the BTZ analogs of the non-minimal surfaces in figure \ref{fig:dcircsunusual}.
Nevertheless, condition 2 has a strong effect on holographic scattering.
In particular, in the limit $M\gg1$, which corresponds to the large-temperature limit of the dual CFT state, the condition $\theta<2\pi-\Delta\phi^*(M)$ is satisfied only for a vanishing range of $\theta$'s --- see eq.~\eqref{eq:btzthresh} and figure \ref{fig:btzregionsphase}. 
This means a vanishing portion of entanglement wedges will have a corresponding holographic scattering process, despite the persistence of $u<d$ configurations.

To summarize, $u<d$ is a necessary but not sufficient condition for holographic scattering in the BTZ spacetime.
There is an additional condition, namely $\theta<2\pi-\Delta \phi^*(M)$, which did not appear in the defect case, and which does not seem to permit an interpretation in terms of non-minimal extremal surfaces.
We discuss this additional condition in the following section.

\subsection{Removing the obstruction to holographic scattering}
\label{sec:sec34}

In the previous section, we found 
an obstruction to holographic scattering in the BTZ geometry related to the failure of $\m{E}(\mh{R}_1)$ and $\m{E}(\mh{R}_2)$ to be complementary in a bulk time slice, as shown in figure \ref{fig:shadow}.
Recall that when the boundary state is pure, entanglement wedges of complementary regions are always complementary.
Thus, we might expect that the obstruction to scattering for $\theta<2\pi-\Delta\phi^*(M)$ can be removed by purifying the boundary state.

To do this, recall that if we consider a single asymptotically AdS$_3$ boundary of the BTZ black hole, the dual CFT state is a (mixed) thermal state with temperature $T_\mt{BTZ}=\frac{\sqrt{M}}{2\pi}$.  
On the other hand, if we consider the fully extended geometry -- the eternal BTZ black hole -- we have a (pure) thermofield double (TFD) state, where the original CFT is entangled with a second copy of the CFT living on the second asymptotic boundary \cite{Maldacena:2001kr}.

Denote the second asymptotic boundary by $\mh{T}_2$, and define a new extended output region as $\mh{R}_2' =\mh{R}_2\cup\mh{T}_2$.
Since $\mh{R}_1$ and $\mh{R}_2'$ are complementary regions in the combined CFT and the global state is pure, we have $\gamma_{\mh{R}_1}=\gamma_{\mh{R}_2'}$ for all $\theta$.
Now, recall that our strategy in section \ref{sec:holscat} to determine when $J_{12\to12}\neq\emptyset$ was to examine if particles leaving from $c_1$ and $c_2$ could meet the minimal-radius point of the RT surface $\gamma_{\mh{R}_1}$.
If this point is also on the RT surface 
of the second output region, then holographic scattering is governed by the $u<d$ inequality.
This is the case here, so we indeed find that the condition $\theta<2\pi-\Delta \phi^*(M)$ is no longer required for holographic scattering.

\begin{figure}[htbp]
\centering
\begin{subfigure}[b]{0.3\textwidth}
    \centering
    \includegraphics[width=\textwidth]{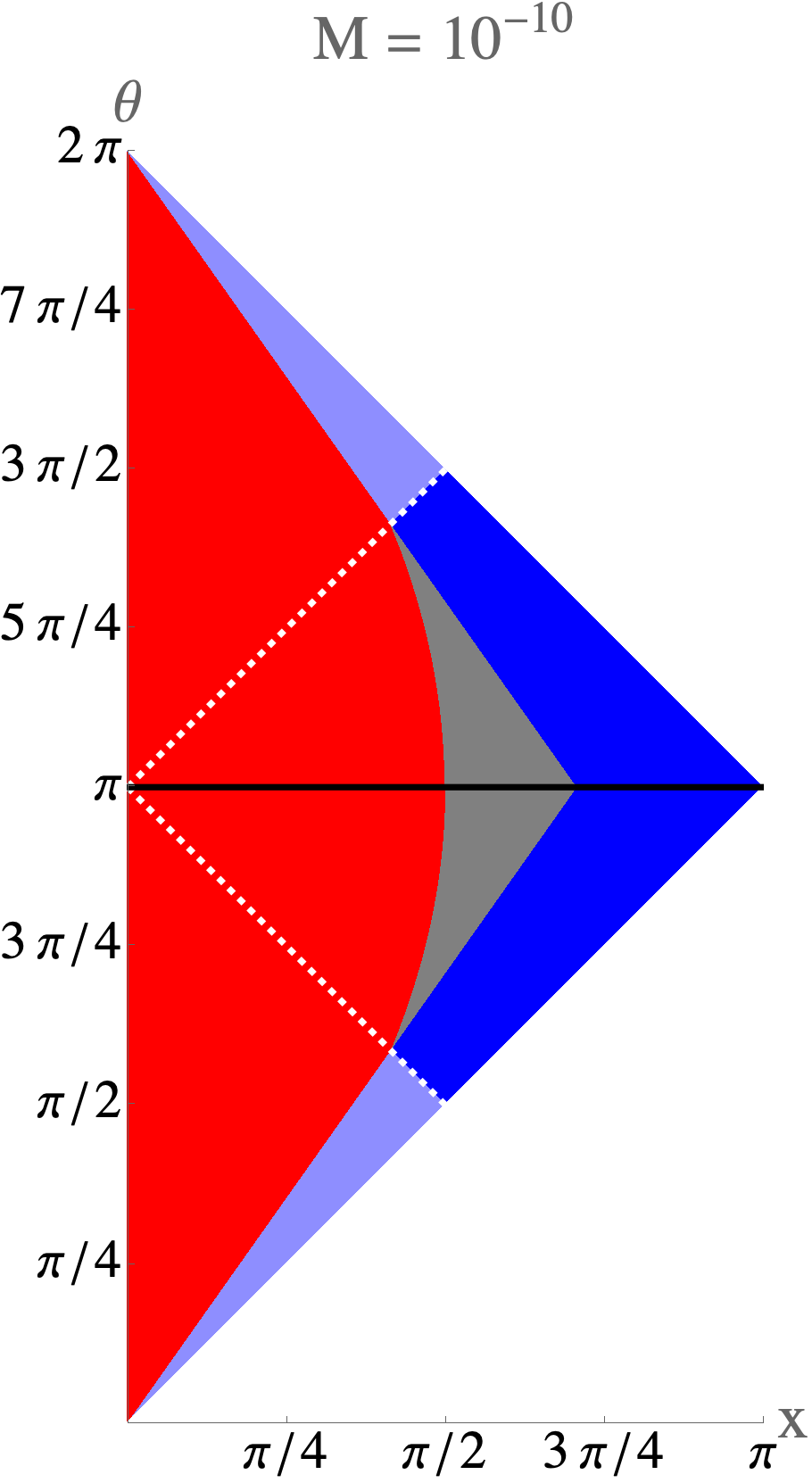}
    \caption{}
\end{subfigure}
\qquad\qquad\qquad\qquad
\begin{subfigure}[b]{0.3\textwidth}
    \centering
    \includegraphics[width=\textwidth]{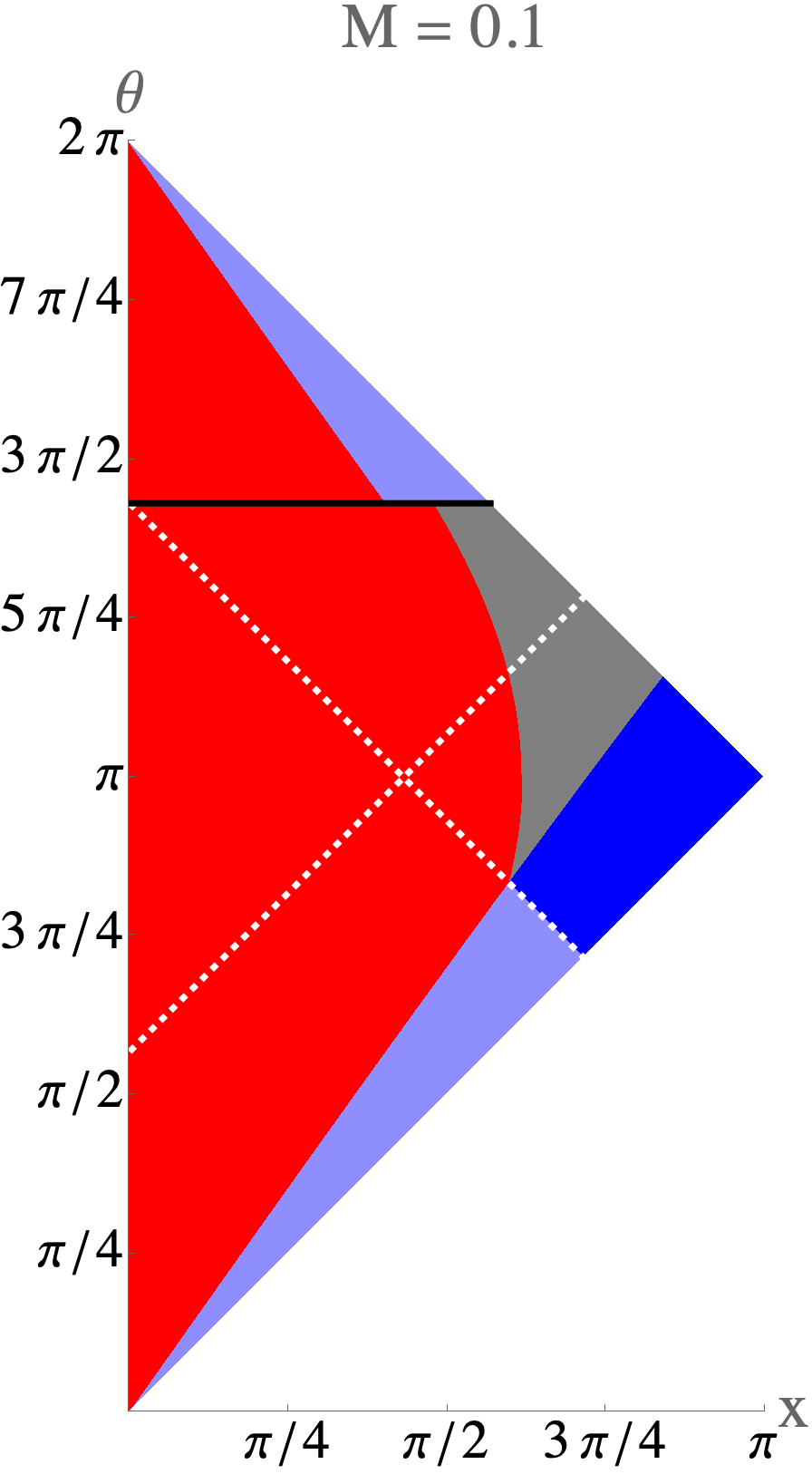}
    \caption{}
\end{subfigure}
\caption{Underneath the black line, \ie when $\theta<\Delta \phi^*(M)$, the holographic scattering inequality is $u<d$, analogously to figure \ref{fig:tempregionsconical}. Above the black line, the holographic scattering inequality is $u'<d$. The white lines indicate where $o=u$ and $o=u'$. Where there is holographic scattering, the $o$ configuration is denoted by dark blue and both the $u$ and $u'$ configurations are denoted by light blue.
In figure (a), we have $\Delta\phi^*(M)=\Delta\phi^*(10^{-10})\approx \pi$ and in figure (b), $\Delta\phi^*(M)=\Delta\phi^*(0.1)\approx 5.7 \, \pi /4$.
\label{fig:fullrange}}
\end{figure}

From this, we might expect that the holographic scattering phase diagram resembles that of figure \ref{fig:scatteringphasediagram}, where the only border of the grey region is at $u=d$.
Indeed, by the previous paragraph, this is the case for $\theta \le \pi$ --- see figure \ref{fig:fullrange}. 
However, note that by associating $\mh{T}_2$ with $\mh{R}_2$, we have broken the symmetry $\theta \mapsto 2\pi-\theta$, so that surprises may be lurking in the regime $\theta>\pi$.

Consider first when $\theta\in(\Delta \phi^*(M),2\pi)$.
In this case, $\gamma_{\mh{R}_1}=\gamma_{\mh{R}_2'}$ consists of the horizon together with the RT surface for an interval of width $\Delta \phi = 2\pi-\theta$.
Then, the holographic scattering computation proceeds as in section \ref{sec:sec33}, except with $\theta \mapsto 2\pi-\theta$.
The resulting holographic scattering inequality is equivalent to $u'<d$.
This is consistent with our findings at the end of section \ref{sec:sec32} that the defect scattering inequality is $u'<d$ for $\theta>\pi$.

Next, consider when $\theta\in(\pi, \Delta \phi^*(M))$.
In this case, $\gamma_{\mh{R}_1}=\gamma_{\mh{R}_2'}$ consists of a single surface: the solution eq.~\eqref{eq:gammaBTZ1} for an interval with $\Delta \phi = \theta$.
Then, the holographic scattering computation proceeds as in section \ref{sec:sec33}, with the only difference being that $\theta >\pi$.
A straightforward check shows that the result of appendix \ref{sec:regionsappendix} is automatically valid for values of $\theta>\pi$, \ie the inequality reads $u<d$.
These results are illustrated in figure \ref{fig:fullrange}.

We conclude that there is no obstruction to holographic scattering of the form $\theta> 2\pi-\Delta\phi^*(M)$ when considering the thermofield double state.
It may seem surprising that $\mh{T}_2$ has a role to play when, from a points scattering perspective, all of the input and output particles are confined to the opposite boundary.
Nevertheless, it is clear that correlations between $\mh{R}_1$ and $\mh{T}_2$ serve a resource that can be used to enable holographic scattering in this regime.
The remaining conditions $u<d$ and $u'<d$ reinforce the relationship uncovered in the defect case of section \ref{sec:sec32} between holographic scattering and RT candidates that are not necessarily minimal.
The physical meaning of this relationship is discussed in the next section.

\section{Discussion}
\label{sec:discuss}

In this note, we have studied two kinds of CFT correlations: those that have a corresponding holographic scattering process and those that do not. 
We found that holographic scattering requires the condition $u<d$ on the areas of extremal surfaces in both the heavy conical defect and BTZ geometries. 
Below, we interpret the $u<d$ condition as a statement about the entanglement structure of the CFT's internal degrees of freedom, in agreement with suggestions that sub-AdS scale physics is encoded in the entanglement between internally, rather than spatially, organized CFT degrees of freedom.
We then discuss physical interpretations for the additional constraint $\theta<2\pi-\Delta\phi^*(M)$ which appears for holographic scattering in the one-sided BTZ geometry, and touch upon the higher-dimensional case.
Lastly, we discuss the perspective that CFT correlations serve as a resource for nonlocal quantum computations, thus pointing to open questions at the intersection of quantum information and quantum gravity.

\subsection{\texorpdfstring{Boundary interpretation of $u<d$ condition}{Boundary interpretation of u<d condition}}

In this section, we discuss a boundary interpretation of the $u<d$ condition, which is necessary for bulk scattering in the heavy defect and BTZ geometries.
We can view this condition as pointing to a possible strengthening of the connected wedge theorem, since that theorem implies only $I(\mh{V}_1:\mh{V}_2)=O(1/G_N)$, \textit{i.e.}~$\min(o,u)<d$, a weaker condition than $u<d$.

Recall that bulk scattering processes have been identified as natural settings to probe sub-AdS scale locality in the AdS/CFT correspondence \cite{susskind1999holography,Heemskerk_2009}. 
One puzzle illustrating this connection is discussed in \cite{Heemskerk_2009}: consider two particles falling into the bulk of an AdS spacetime. 
Suppose the particles come near to each other but miss, and emerge near the boundary again later. 
This corresponds in the boundary to two excitations that smear over the boundary and overlap, but miraculously fail to interact and then coalesce again into localized excitations. 
Mysteriously, if we shift the excitations slightly such that they interact in the bulk, the boundary excitations now interact as well, even while the excitations appear similarly smeared over the boundary. 
The puzzle is to understand what is the difference, from a boundary perspective, between these two settings. 

For particles that miss each other by  distances larger than the AdS scale $\ell_{AdS}$, this phenomenon can be explained by energy localization. The two particles are at different depths in the bulk and so associated with different boundary energy scales, and excitations at different energy scales can indeed fail to interact in this way \cite{Heemskerk_2009}.  
Below the AdS scale, however, this explanation breaks down, and another boundary mechanism is needed to explain bulk locality.

One suggestion has been that the organization of the internal degrees of freedom in the boundary theory gives rise to sub-AdS scale bulk locality. 
To support this viewpoint, we can consider the BFSS model \cite{banks1999m}, which describes flat space in terms of a ($0+1$)-dimensional matrix model. 
Here, it is the organization of the degrees of freedom within the matrices appearing in the BFSS model that describe a theory with sub-AdS scale locality (since effectively we have $\ell_{AdS}\rightarrow \infty$ in flat space). 
The authors of \cite{susskind1999holography} make a similar argument within AdS/CFT: considering $\mathcal{N}=4$ SYM with $SU(N)$ gauge symmetry, we can note that an $\ell_{AdS}$-sized patch of the bulk is described by $N^2$ degrees of freedom, which is just few enough that it can be described by spatially homogeneous excitations of the boundary matrix degrees of freedom.
Because of the connection between entanglement and geometry observed when considering spatially organized degrees of freedom, it was suggested that entanglement among internal degrees of freedom play a role in the emergence of sub-AdS scale bulk physics \cite{anous2020areas}.

Inspired by these ideas, we will argue that the $u<d$ condition means that not only $\mh{V}_1$ and $\mh{V}_2$ are strongly correlated but also the entanglement must be organized within the internal degrees of freedom in a particular way. 
To explain this claim, we first focus on the conical defect geometries of section \ref{sec:sec21} with $M=-\frac{1}{n^2}$ for integer $n$. 
These geometries are $\mathbb{Z}_n$ orbifolds of the pure AdS$_3$ vacuum geometry.
Following \cite{Balasubramanian_2005}, we correspondingly interpret the boundary CFT as an orbifold of a second CFT, which we denote CFT$'$ and which lives on the boundary of the AdS$_3$ covering geometry. 
As discussed in \cite{Entwinement_2015}, in this covering theory we have ``ungauged'' a discrete $\mathbb{Z}_n$ gauge symmetry and physical quantities appear as $\mathbb{Z}_n$-invariant quantities in the ungauged theory.
Further, we note that CFT$'$ has a central charge $c'=c/n$.\footnote{The bulk version of this statement is $G_N'=n \,G_N$. One can verify with these choices, together with a suitable cutoff prescription, that the entanglement entropy of a small interval $\mh{V}_1$ in CFT matches the entanglement entropy of the corresponding region $\mathbf{V}_1:=\mh{V}_1^1\cup...\cup\mh{V}_1^n$ in CFT$'$. Similar statements hold for the discussion near eq.~\eqref{eq:unwrapBTZ} regarding covering spaces of the BTZ geometry. \label{footy}}
This setting provides an interesting way of studying the internal degrees of freedom in the original CFT. 
In particular, the $O(c)$ internal degrees of freedom in the CFT are now split up into $n$ boundary subregions in CFT$'$, each with $O(c/n)$ internal degrees of freedom.
This allows the RT formula to probe the structure of entanglement among internal degrees of freedom. 

\begin{figure}[htbp]
\centering
\begin{subfigure}[b]{0.4\textwidth}
    \centering
    \includegraphics[width=\textwidth]{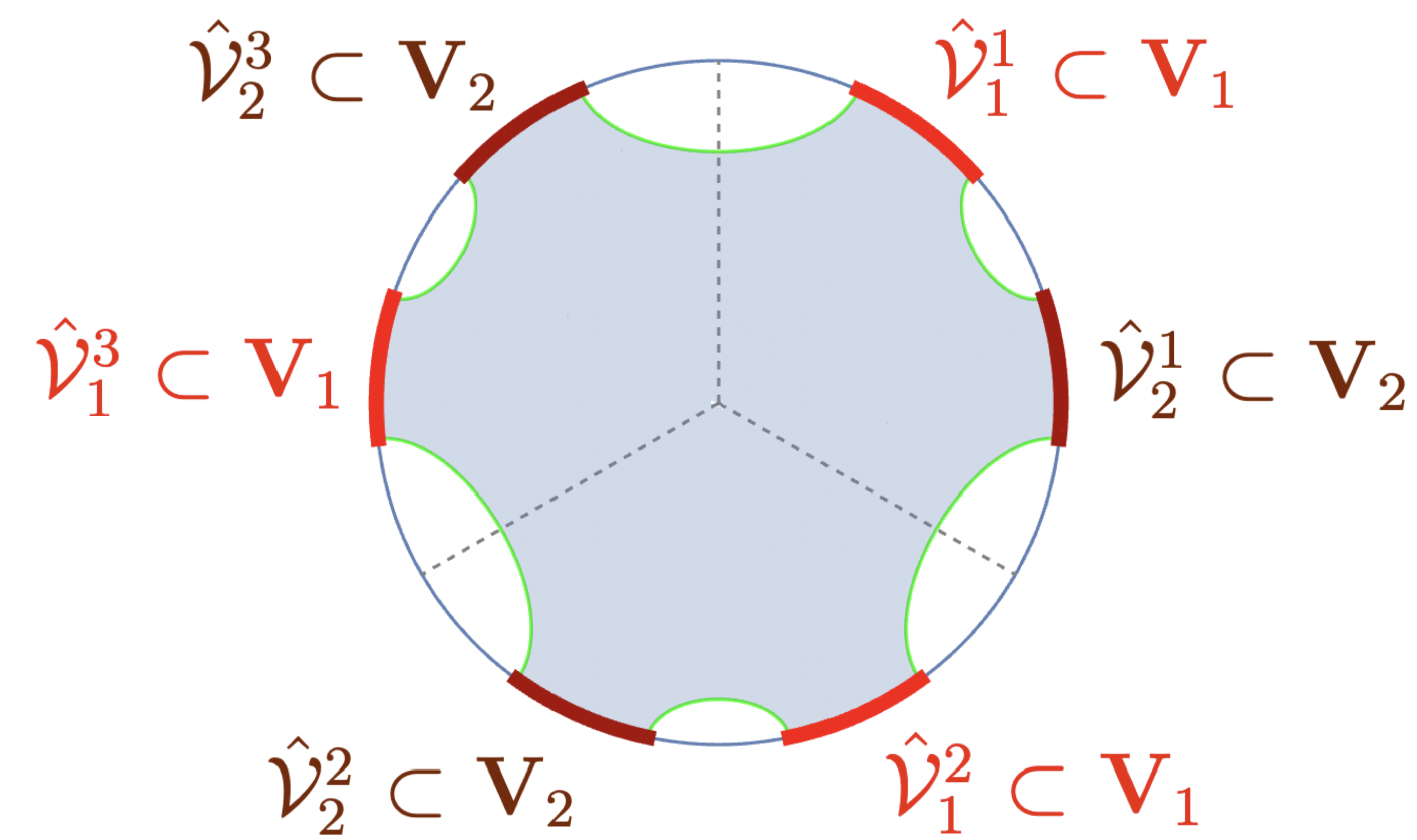}
    \caption{}
\end{subfigure}
\hfill
\begin{subfigure}[b]{0.25\textwidth}
    \centering
    \includegraphics[width=\textwidth]{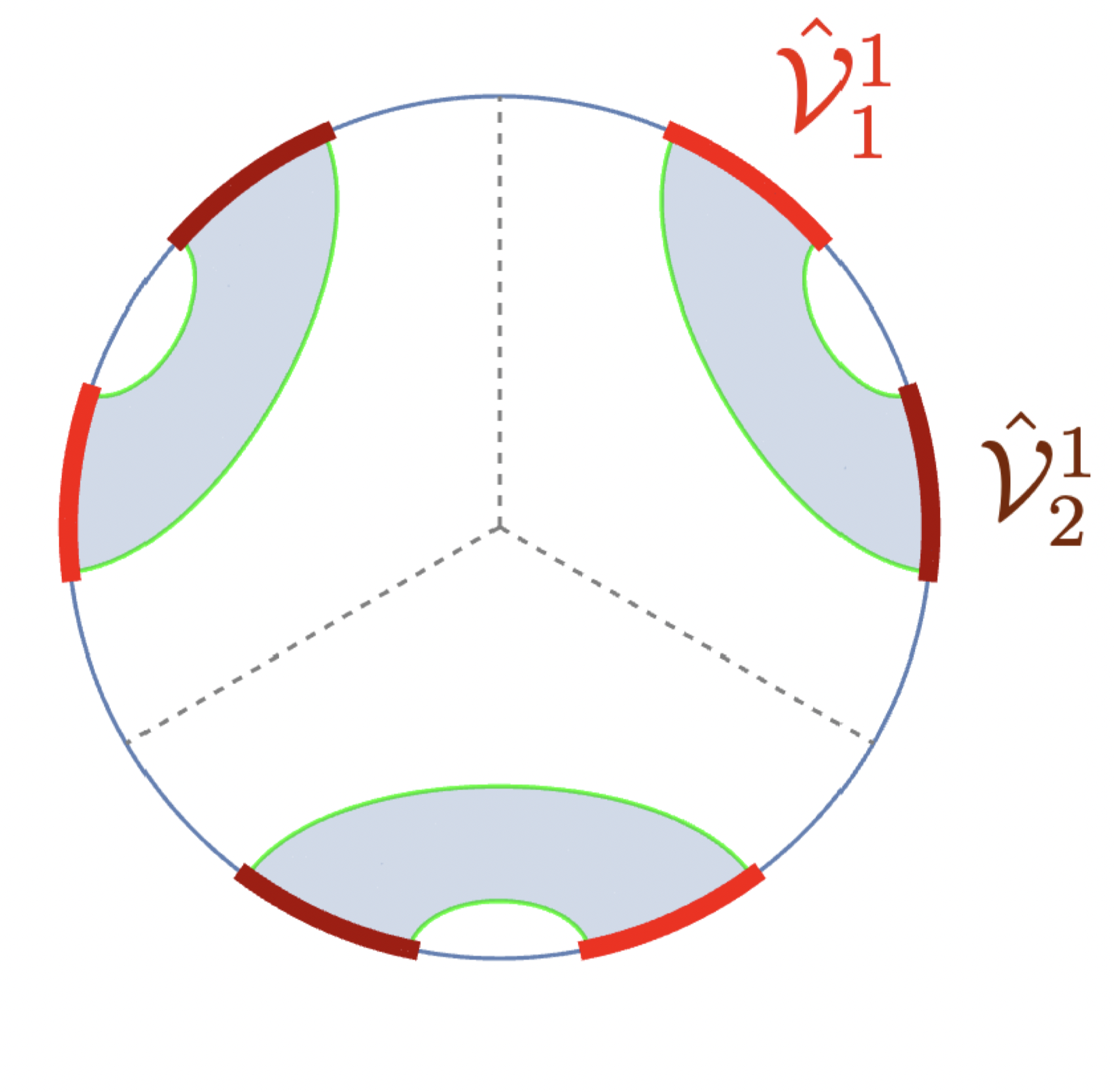}
    \caption{}
    \label{fig:ulessdcovering}
\end{subfigure}
\hfill
\begin{subfigure}[b]{0.25\textwidth}
    \centering
    \includegraphics[width=\textwidth]{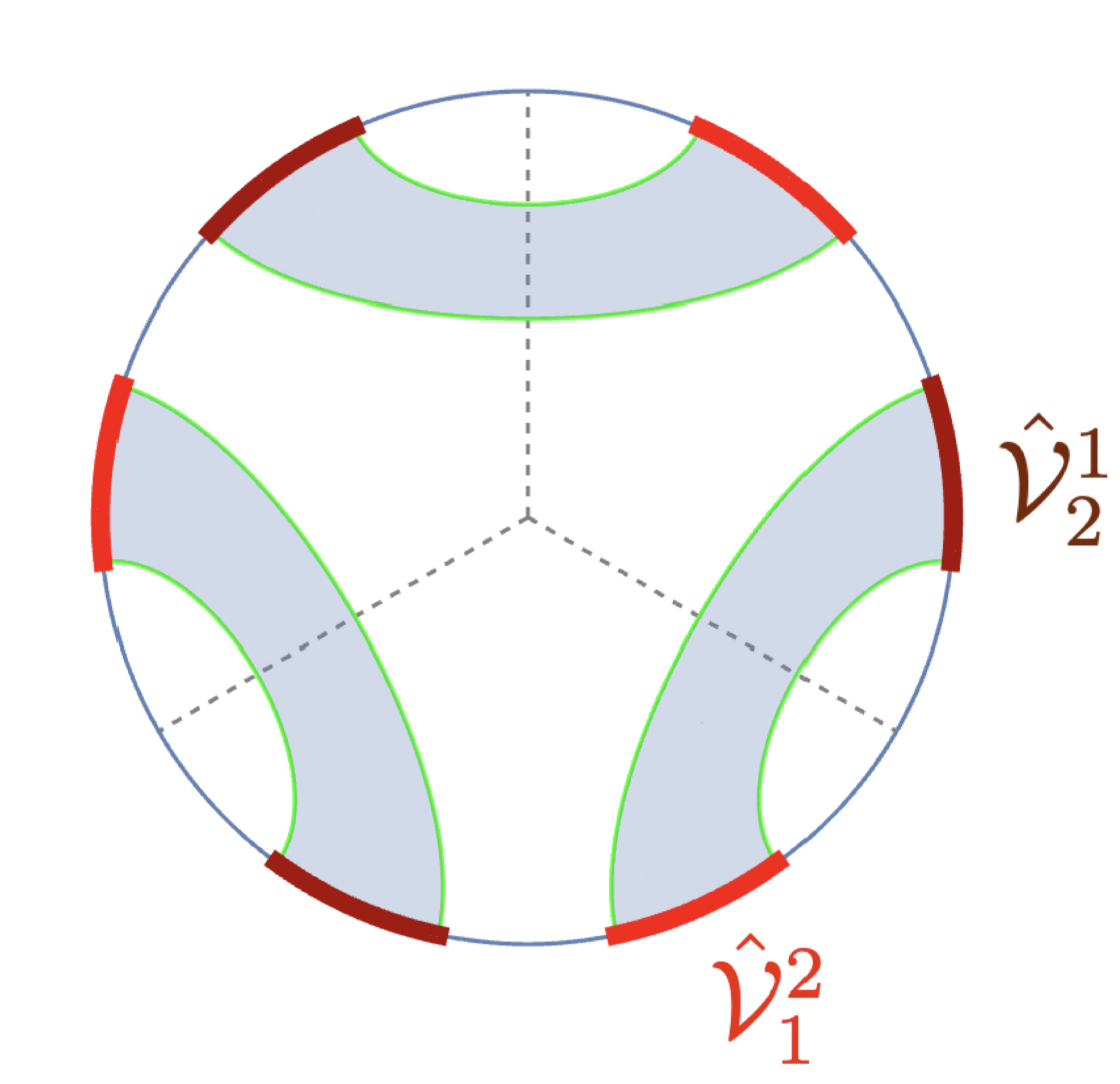}
    \caption{}
    \label{fig:uprimecovering}
\end{subfigure}
\caption{
Covering space of the AdS$_3$/$\mathbb{Z}_3$ orbifold geometry. In the covering space, there are $n=3$ copies of the input regions $\mh{V}_1$ and $\mh{V}_2$. We illustrate a few candidates for a connected entanglement wedge in this covering space. Each of these candidates is $\mathbb{Z}_3$ symmetric and so can also be realized in the orbifold geometry. Respectively, panels (a), (b) and (c) correspond to the $o$ wedge in figure \ref{fig:dcircs_o},  the $u$ wedge in figure \ref{fig:dcircs_u}, and the $u'$ wedge in figure \ref{fig:dcircs_h}. Note that in (b) and (c), the entanglement wedge has three disconnected components corresponding to localized correlations in CFT$'$.}
\label{fig:entwine}
\end{figure}

Now return to the connected wedge theorem and the $u<d$ condition. 
The connected wedge theorem applied to boundary subregions $\mh{V}_1$ and $\mh{V}_2$
gives that when bulk scattering occurs, $I(\mh{V}_1:\mh{V}_2)=O(1/G_N)$, corresponding to a connected wedge in the bulk. 
To describe this connected wedge in the covering geometry, we first note that the CFT regions $\mh{V}_1$ and $\mh{V}_2$ are identified with sets of regions in CFT$'$: $\mh{V}_1\mapsto \mh{V}_1^1\cup...\cup\mh{V}_1^n=:\mathbf{V}_1$, $\mh{V}_2\mapsto \mh{V}_2^1\cup...\cup\mh{V}_2^n=:\mathbf{V}_2$. 
The CFT condition $I(\mh{V}_1:\mh{V}_2)=O(1/G_N)$ becomes, in CFT$'$, the statement that
\begin{align}
    I(\mathbf{V}_1:\mathbf{V}_2)=O(1/G_N) \,.
\end{align}
In bulk language, this says the entanglement wedge of $\mathbf{V}_1$ and $\mathbf{V}_2$ is connected. 
There are many ways in which this can be realized, and we illustrate a few examples in figure \ref{fig:entwine}. 
Note that all three configurations are $\mathbb{Z}_3$ symmetric, but a feature which distinguishes the second and third examples from the first is that their entanglement  wedges have (three) disconnected components.

\begin{figure}
\centering
\begin{subfigure}[t]{0.44\textwidth}
    \includegraphics[width=\textwidth]{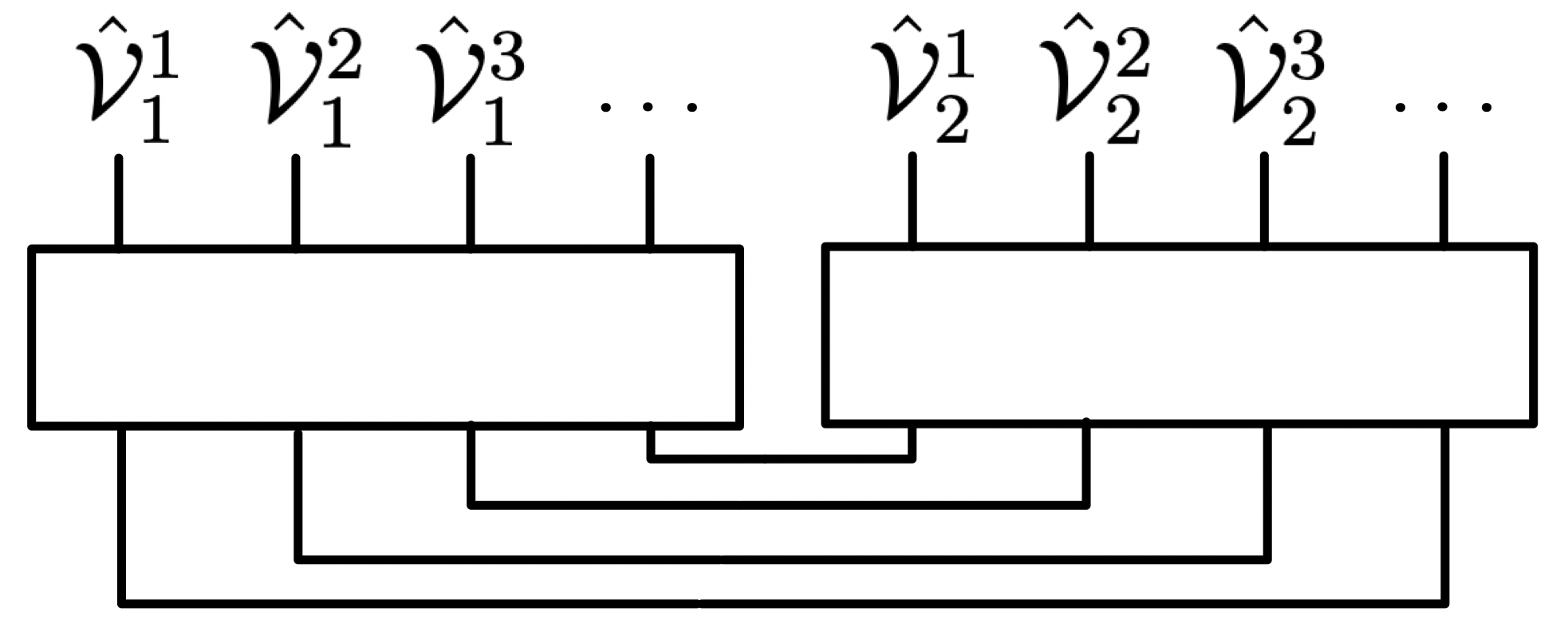}
    \caption{}
    \label{fig:alltoall}
\end{subfigure}
\hfill
\begin{subfigure}[t]{0.54\textwidth}
    \includegraphics[width=\textwidth]{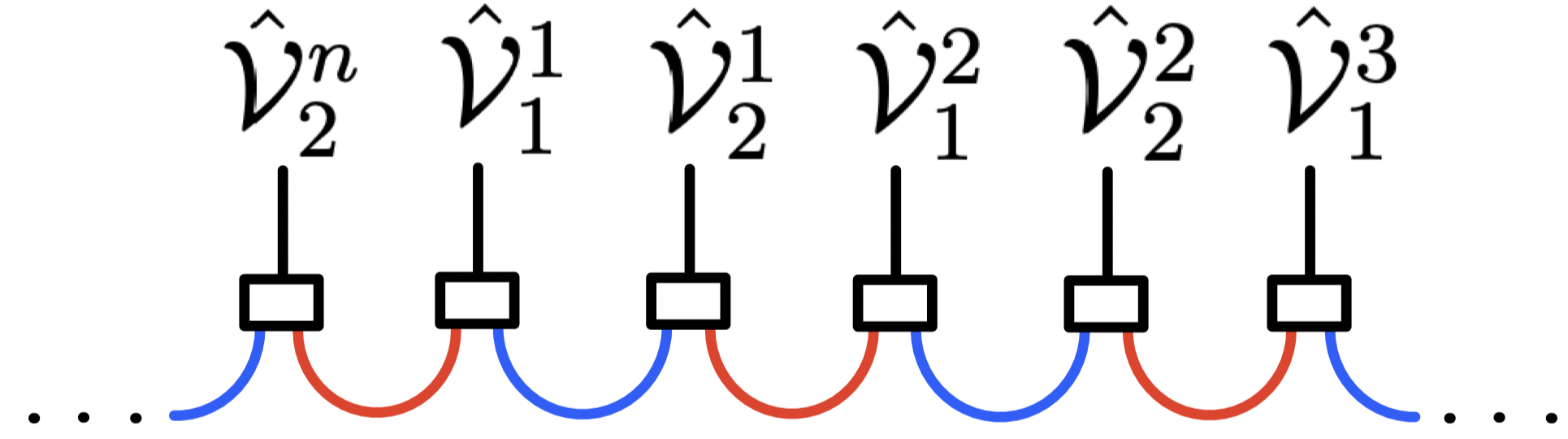}
    \caption{}
    \label{fig:pairwise}
\end{subfigure}
\caption{(a) When $\m{E}(\mh{V}_1\cup \mh{V}_2)$ is connected in the defect geometry, $\mathbf{V}_1=\mh{V}_1^1\cup\mh{V}_1^2\cup...\cup\mh{V}_1^n$ and $\mathbf{V}_2=\mh{V}_2^1\cup\mh{V}_2^2\cup...\cup\mh{V}_2^n$ share strong correlation, but individual pairs $\mh{V}_1^i$ and $\mh{V}_2^i$ need not. (b) When $\m{E}(\mh{V}_1\cup \mh{V}_2)$ has a holographic scattering process, we have $u<d$. Thus, by eq.~\eqref{rocket8}, each corresponding pair of subsystems $\mh{V}_1^i$ and $\mh{V}_2^i$ in the covering space shares strong correlation, as indicated by the blue legs. As discussed below eq. ~\eqref{rocket1}, it is also possible to have $I(\m{V}_{1}^i:\m{V}_2^{i-1})=O(1/G_N)$ --- see the red legs.
\label{fig:organizedentanglement}}
\end{figure}

In the covering geometry, the $u<d$ condition amounts to the statement that \emph{each pair} of the input regions shares a connected entanglement wedge (see figure \ref{fig:ulessdcovering}). 
In terms of the boundary correlations in CFT$'$, this configuration yields
\begin{align}
    \forall \, i\,\,\text{with}\,\, 1\leq i\leq n\,,\,\,\,\,\,\,\,\,\,I(\mh{V}_1^i:\mh{V}_2^i) = O(1/G_N) \,.
    \label{rocket8}
\end{align}
This correlation structure is illustrated by the blue legs in figure \ref{fig:pairwise}.
It is a stronger condition than the full sets $\mathbf{V}_1$ and $\mathbf{V}_2$ being strongly correlated with one another, as illustrated in figure \ref{fig:alltoall}.
From the perspective of the original CFT, eq.~\eqref{rocket8} means $\mh{V}_1$ and $\mh{V}_2$ share strong correlations between corresponding subsets of their internal degrees of freedom.

By symmetry, we expect a similar conclusion to apply when $\theta>\pi$.
Recall from the end of section \ref{sec:sec32} that for $\theta>\pi$ the holographic scattering condition is $u'<d$ instead of $u<d$. 
In the covering space, this translates to the condition
\begin{equation}
I(\mh{V}_1^i:\mh{V}_2^{i-1}) = O(1/G_N) \, \label{rocket1},
\end{equation}
which corresponds to the RT surfaces in figure \ref{fig:uprimecovering} and the red legs in figure \ref{fig:pairwise}.
Thus, holographic scattering again requires $O(1/G_N)$ correlations for certain nearest-neighbor subsets of degrees of freedom in CFT$'$.

We can ask whether other pairs of boundary intervals in the covering space can share large correlations, \ie  $I(\mh{V}_a^i:\mh{V}_b^k) = O(1/G_N)$ for $i \neq k, k-1,$ or $k+1$.\footnote{By leaving the subscripts unspecified here, we are including possibilities where, \eg $I(\mh{V}_1^i:\mh{V}_1^{i-1}) = O(1/G_N)$. See further comments below.} A careful analysis shows the answer is no; the only possibilities are eq.~\eqref{rocket8} and eq.~\eqref{rocket1}. 
Now, the latter arises in the regime $u'<d$, \ie
\begin{equation}
\begin{aligned}
&\sin^2\frac{\sqrt{|M|}\,x}{2}>\sin\frac{\sqrt{|M|}\,(2\pi-\theta-x)}{2}\sin\frac{\sqrt{|M|}\,\left(2\pi-\theta+x\right)}{2} \,.\\
\end{aligned}
\label{eq:uprimeineq}
\end{equation}
Comparing to the $u<d$ inequality \eqref{eq:uineq}, we see that there exist choices of $x$ and $\theta$ for which $\mh{V}_1^i$ shares strong correlations with both $\mh{V}_2^{i}$ and $\mh{V}_2^{i-1}$.

It follows that figure \ref{fig:pairwise} shows the maximal possible bipartite correlation structure of the regions under consideration.
That is, the intervals in CFT$'$ carry at most nearest-neighbour correlations.
This figure is reminiscent of the tensor network describing a matrix product state, \eg see \cite{Orus:2018dya,Banuls:2022vxp}.
In particular, when the nodes are isometric tensors, we expect the resulting state to have at most nearest-neighbor correlations.
As an aside, note that if we had considered configurations where the two intervals were different sizes, we would have found that next-to-nearest-neighbor correlations of the form $I(\mh{V}_1^i:\mh{V}_1^{i-1}) = O(1/G_N)$ are also possible.
However, this would be the furthest possible extent of $O(1/G_N)$ correlations amongst intervals in CFT$'$.

To summarize, the conical defect geometries with opening angle $2\pi/n$ suggest the $u<d$ condition for bulk scattering is related to the entanglement structure among the internal degrees of freedom of the boundary CFT. 
In particular, we saw that holographic scattering required $O(1/G_N)$ correlations for nearest-neighbour subsets of degrees of freedom in the covering theory (CFT$'$), as depicted by the blue legs in figure \ref{fig:pairwise}, or red legs in the case of $\theta>\pi$.

With the goal of extending our results beyond the orbifolds of AdS$_3$, note that we can apply the same ``ungauging'' procedure to more general geometries.
For example, given a defect geometry with opening angle $\alpha$, we 
can consider it to be a $\mathbb{Z}_n$ orbifold of a defect geometry with opening angle $n\,\alpha$.\footnote{We restrict ourselves to $n\,\alpha\le2\pi$ (which implies $\alpha \le \pi$, \ie $M \ge -1/4$). A covering geometry with $n \, \alpha>2\pi$ would have a conical excess in the bulk and an energy which is less than that of the AdS$_3$ vacuum. This, together with various other arguments (\eg see figure 1 of \cite{Basile:2023ycy}), indicate that such conical excess geometries do not correspond to  physical states.} 
That is, the covering geometry would be constructed by gluing together $n$ copies of the original defect geometry. The $\mathbb{Z}_n$ orbifolds considered previously correspond to the special case where $n\,\alpha=2\pi$ and the covering geometry is the smooth AdS$_3$ vacuum solution. 
We are now generalizing the construction to allow the covering geometry to have a conical defect. 

The ``ungauging" procedure still  unpacks some of the internal degrees of freedom into spatially organized degrees of freedom. 
Hence, the interpretation of $u<d$ in terms of nearest-neighbor correlations still applies.
In particular, this suggests that holographic scattering in generic defect geometries (\ie with $\alpha<\pi$ not necessarily equal to $2\pi/n$) again requires a specific correlation structure among the internal degrees of freedom of the boundary CFT.

Similar reasoning applies for the BTZ black hole. Every BTZ geometry can be understood as a $\mathbb{Z}_n$ orbifold of another BTZ geometry, where the covering geometry is obtained by extending the periodicity of the angular coordinate to $\delta\phi=2\pi n$.
In other words, just as we did above, we are constructing a covering geometry by gluing together $n$ copies of the original geometry.\footnote{We note that the horizon entropy is unchanged in the covering space, since as commented in footnote \ref{footy}, $G_N'=n\,G_N$.
Hence, both the area and Newton's constant increase by a factor of $n$, and we have $S'_{BH}=\frac{{\rm Area}(r=\sqrt{M})}{4\,G'_N}=\frac{\pi\sqrt{M}}{2\,G_N}$ as for the original black hole.} 
If we describe the covering geometry with the metric \eqref{eq:mumetric} with coordinates $(t,r,\phi)$ and mass $M$, we can choose new coordinates $(\tilde{t},\tilde{r},\tilde{\phi})$ and mass $\tilde{M}$ such that the angular periodicity is restored to $\delta\tilde\phi=2\pi$. This follows with the definition
\begin{equation}
\tilde{r} = n\, r \,,\qquad 
\tilde{t} = \frac{t}{n} \,,\qquad 
\tilde{\phi} = \frac{\phi}{n} ,\qquad 
\tilde{M} = M\, n^2\,.
\label{eq:unwrapBTZ}
\end{equation}
The metric in the $(\tilde{t},\tilde{r},\tilde{\phi})$ coordinates has the same form as eq.~\eqref{eq:mumetric}, but now with mass parameter $\tilde{M}$. The  original geometry where $\delta\phi =2\pi$ corresponds to a wedge of width $\Delta \tilde{\phi} = \frac{2\pi}{n}$ in this description of the covering geometry.
Unlike in the defect case, there are no restrictions on $n$; it can be taken arbitrarily large.
For any fixed choice of $n$, however, the interpretation of the $u<d$ condition in terms of nearest-neighbor correlations will apply.
This again suggests a connection between holographic scattering and the organization of internal degrees of freedom for the BTZ black hole.

An extension of this work would be to verify that holographic scattering requires a condition analogous to $u<d$ in other geometries, such as rotating conical defects, rotating black holes, or shockwave geometries.
In some of these cases, we could again convert internal degrees of freedom to spatially organized degrees of freedom by considering the geometry as the orbifold of a covering space.
Finding a $u<d$ constraint would lead to the same conclusion that holographic scattering requires specific correlations among the internal degrees of freedom.
Unfortunately, it remains unclear how the above observations could be adapted to cases that can not arise as the orbifold of a physical covering geometry, such as conical defects with $\alpha>\pi$.

Another generalization would be to consider input regions $\mh{V}_1$ and $\mh{V}_2$ with differing sizes $x_1$ and $x_2$.
Since the distinction between entanglement and causal wedges played a central role in deriving holographic scattering inequalities in section \ref{sec:holscat}, it would be interesting to study cases where $x_1$ is so large that $\m{E}(\mh{V}_1)\neq\m{C}(\mh{V}_1)$.
In particular, we would like to know whether a condition like $u<d$ still holds and what is the covering space interpretation.

\subsection{Obstruction to holographic scattering in BTZ geometry}
\label{sec:BTZinterp}

In this section, we interpret the constraint $\theta<2\pi-\Delta \phi^*(M)$  noted in eq.~\eqref{sleeper} for holographic scattering in the BTZ geometry.
To minimize clutter, let us denote $\Delta\phi^*(M)$ as simply $\Delta\phi^*$.

\begin{figure}[b]
\centering
\includegraphics[width=.47\textwidth]{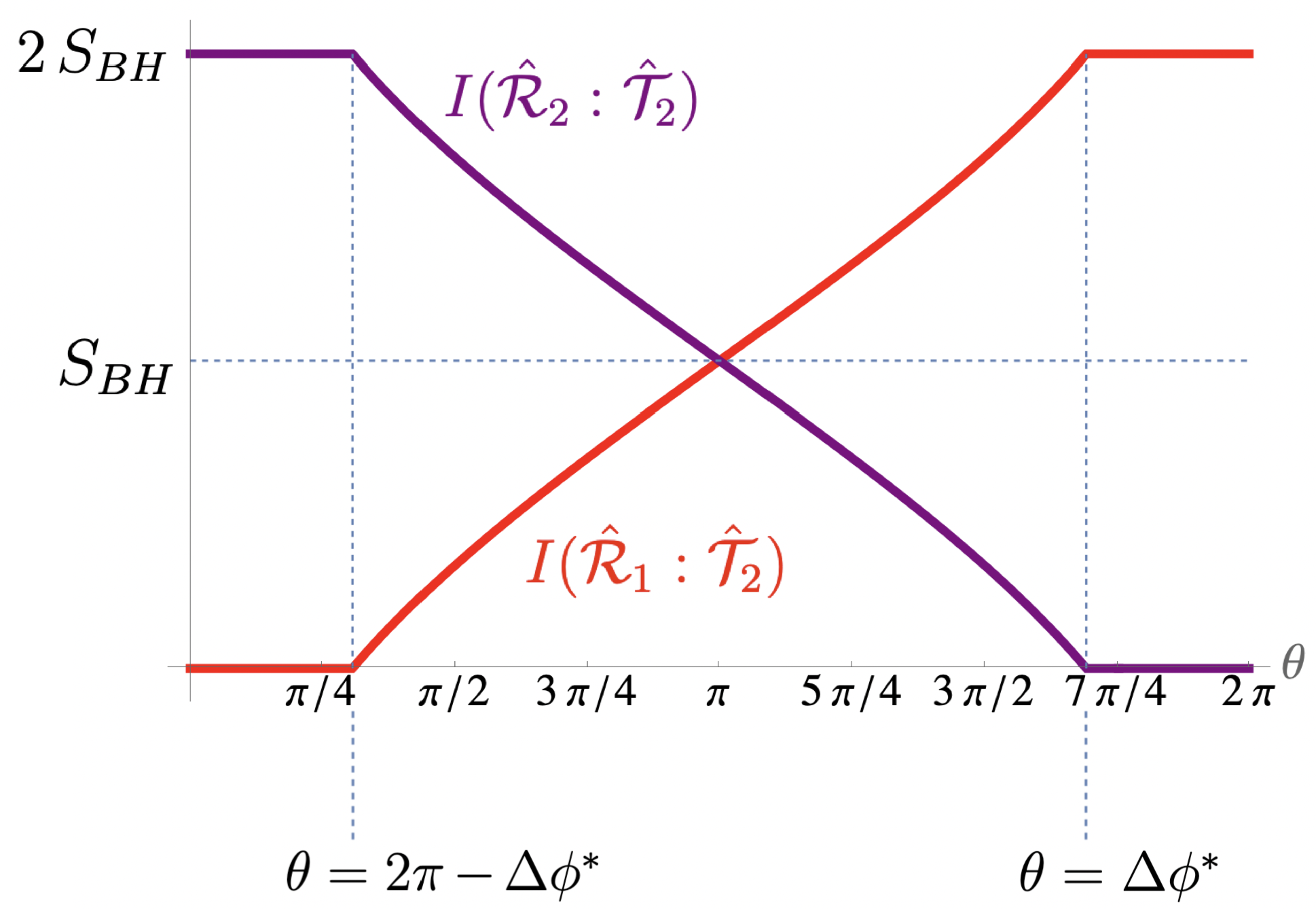}
\hfill
\includegraphics[width=.49\textwidth]{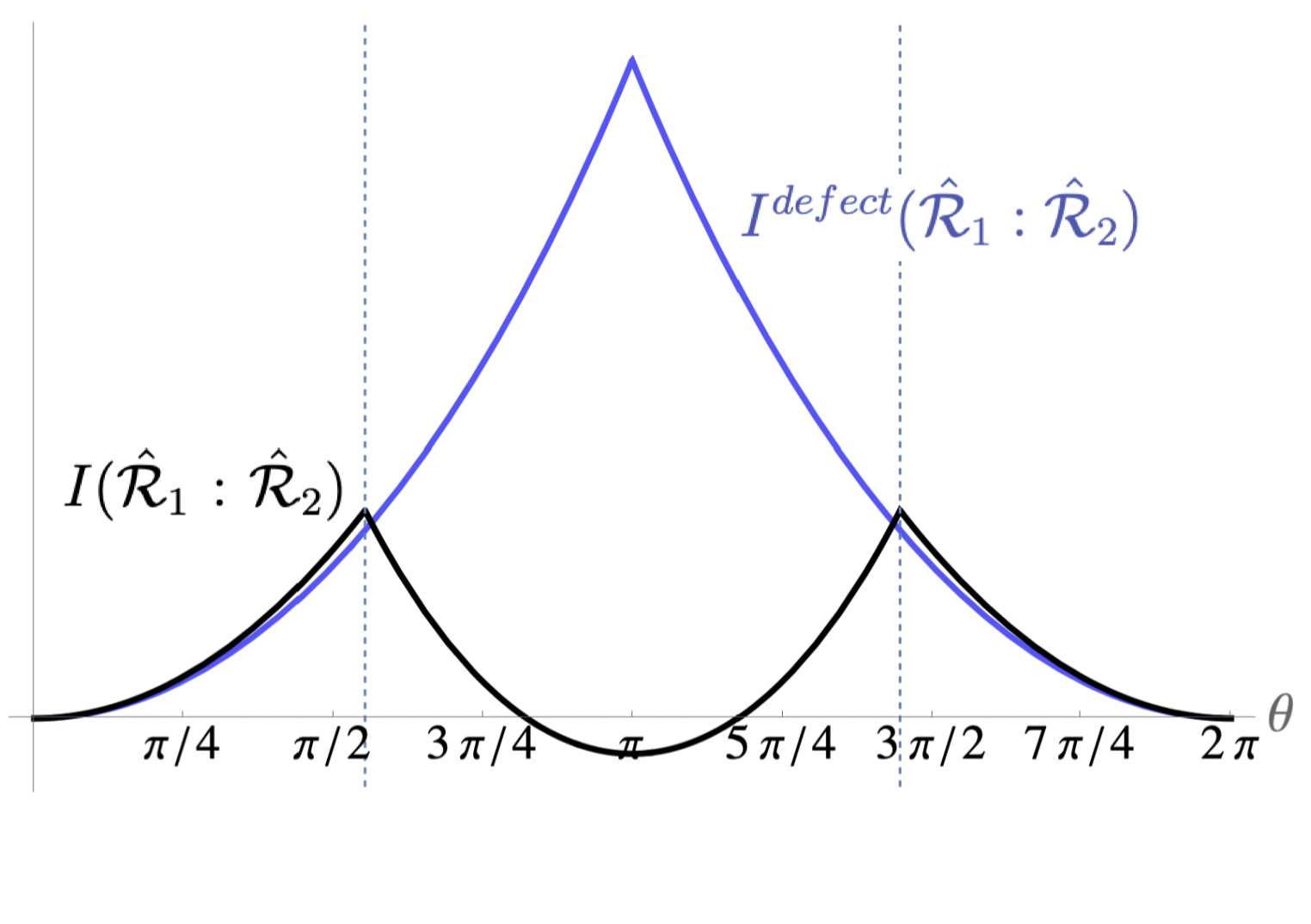}
\caption{The graphs of $I(\mh{R}_1:\mh{T}_2)$ (left, purple), $I(\mh{R}_2:\mh{T}_2)$ (left, red), and $I(\mh{R}_1:\mh{R}_2)$ (right, black) have kinks at $\theta = 2\pi- \Delta \phi^*$ and $\theta = \Delta \phi^*$.
This indicates a change of character in the correlations among the $\mh{R}_i$ and $\mh{T}_2$.
In the left diagram, $M=1$.
In the right diagram, $M=0.1$.
Note, the mutual information $I(\mh{R}_1:\mh{R}_2)$ is technically divergent, so we have plotted the vacuum-subtracted quantity. 
The analogous quantity for a very heavy defect ($M = - 10^{-10}$) is shown in blue for reference.
\label{fig:IRiR}}
\end{figure}

Recall that in the boundary, we have a thermal CFT state with temperature $T_\mt{BTZ}=\frac{\sqrt{M}}{2\pi}$. 
As discussed in section \ref{sec:sec34}, the thermal state can be purified by considering the TFD state on two copies of the CFT; call them $\mh{T}_1$ and $\mh{T}_2$. 
Since the $\theta$ parameter appearing in the obstruction inequality controls the size of the $\mh{R}_i$ regions, it is natural to study the correlations that each of the $\mh{R}_i$ shares with $\mh{T}_2$, the second copy of the CFT.
The correlations obey
\begin{equation}
    I(\mh{R}_1:\mh{T}_2)+I(\mh{R}_2:\mh{T}_2) = I(\mh{T}_1:\mh{T}_2) = 2 S_{BH}\,,
\end{equation}
where $S_{BH}$ is the Bekenstein-Hawking entropy of the black hole horizon.
Further, as shown in figure \ref{fig:IRiR}, for $\theta<2\pi-\Delta \phi^*$, $\mh{T}_2$ is correlated only with the bigger region, $\mh{R}_2$, and not the smaller region, $\mh{R}_1$.
Similarly, for $\theta>\Delta \phi^*$, $\mh{T}_2$ is correlated only with $\mh{R}_1$.\footnote{The latter statement follows from $\theta \mapsto 2\pi-\theta$ symmetry. In the following paragraphs, to avoid redundancy, we discuss only the cases with $\theta< \pi$.}
On the other hand, with $\theta \in (2\pi-\Delta \phi^*, \Delta \phi^*)$, we find $I(\mh{R}_1:\mh{T}_2)$ increases with increasing $\theta$, while $I(\mh{R}_2:\mh{T}_2)$ decreases -- see the kinks in the various mutual information curves of figure \ref{fig:IRiR} at $\theta = 2\pi -\Delta \phi^*, \Delta \phi^*$.
Thus, at $\theta = 2\pi-\Delta \phi^*(M)$, the structure of correlations among $\mh{R}_1$, $\mh{R}_2$, and $\mh{T}_2$ undergoes a transition, corresponding to a transition to a new set of RT surfaces in the bulk (\textit{c.p.} figure \ref{fig:shadow}).

What does this transition correspond to in the boundary?
Note that in the thermal state, long-range correlations (\ie over distances larger than $\sim 1/T_\mt{BTZ}$) are suppressed.
In the TFD setting, the diminishing of long-range correlations within the single boundary corresponds to the formation of strong correlations of low-energy modes between the two copies of the CFT.
Thus, it makes sense that when one of the regions, say $\mh{R}_1$, becomes small, losing access to long-range modes in $\mh{T}_1$, it stops being correlated with $\mh{T}_2$.
We will further argue that the entanglement of low-energy modes between $\mh{T}_1$ and $\mh{T}_2$ disrupts precisely those correlations between $\mh{R}_1$ and $\mh{R}_2$ which are necessary for holographic scattering.

Our argument goes as follows:
imagine beginning with small $\theta$ and expanding $\mh{R}_1$.
Then, the minimal-radius point $P$ on $\gamma_{\mh{R}_1}$ moves to smaller and smaller radius, \ie towards the horizon. 
Given the key role noted for this point in holographic scattering, we expect that (at least some of) the correlations required for scattering are moving to correlations amongst lower energy modes in $\mh{R}_1$ and $\mh{R}_2$.
However, as we observed above, the TFD state enjoys strong correlations between the low-energy modes of the two CFTs. 
When $\theta$ approaches $2\pi-\Delta\phi^*(M)$, it is plausible that the energy of the modes involved in the desired scattering correlations approaches that for the thermal correlations.
Beyond this point, the thermal correlations in the TFD state would be able to disrupt those necessary for scattering.
In particular, by the monogamy of entanglement, we expect that entanglement of low-energy modes between $\mh{R}_1$ and $\mh{R}_2$ is diminished as these modes become entangled with the purifying system.
This is manifested in the bulk via the separation of $\gamma_{\mh{R}_1}$ and $\gamma_{\mh{R}_2}$ presented in figure \ref{fig:shadow}.

Therefore, the behavior of low-energy modes may provide physical insight into the $\theta < 2\pi - \Delta \phi^*$ constraint on holographic scattering in the thermal state.
In future work, it would be desirable to relate the $2\pi-\Delta\phi^*$ threshold more directly to the length scale associated with suppression of correlations in the thermal state.

\subsection{Higher dimensions}
\label{sec:higherdDiscuss}

It is not known whether the connected wedge theorem holds for higher dimensions, \ie $d>2$. However, it is still possible to explore the relationship between entanglement wedge connectivity and bulk-only scattering.
Such explorations may result in counterexamples to, or evidence for, a higher-dimensional version of the theorem.
In particular, we can ask, does scattering also imply a relationship among non-minimal surfaces in the higher-dimensional context?
Note that pure AdS$_4$ does not admit points-type causal discrepancies for two-to-two scattering \cite{maldacena2015looking,May:2019odp}, so it is particularly natural in the higher-dimensional context to focus on regions-type casual discrepancies (as we have done in this work for the three-dimensional case).
Higher dimensional geometries that are likely to yield interesting relationships among non-minimal surfaces and scattering include AdS$_4$ Schwarschild and the AdS soliton \cite{Horowitz_1998,May:2019odp}.

A simple starting point is to work with AdS$_{4}$ in Poincar\'e coordinates, which we study in appendix \ref{sec:higherdappendix}. 
There, we find no counterexamples to the connected wedge theorem; within the class of examples we study, whenever bulk scattering occurs there is also a large mutual information between the input regions $\mh{V}_i$. 
To compare to our more detailed claim that the $u<d$ condition is necessary for scattering, we can observe simply that since there is no massive object in the bulk, and hence no $o$-type wedge configuration.
Therefore, bulk scattering is indeed implying $u<d$ as it does in AdS$_{3}$. 
A difference arises, however, when we consider the further observation that in the conical defect and two-sided BTZ geometries, the $u<d$ condition is strong enough to imply bulk scattering. 
We do not find an analogue of this in Poincar\'e AdS$_4$. In appendix \ref{sec:higherdappendix}, we construct examples where the relevant wedge is connected, and hence $u<d$, but no bulk scattering occurs.
Indeed, in any asymptotically AdS$_4$ geometry, the relevant RT surfaces at small $x$ and $\theta$ live in a region that approximates pure AdS$_4$, so it is possible that \textit{all} asymptotically AdS$_4$ geometries permit cases with $u<d$ and no scattering.

It would be interesting to study this effect further in AdS$_4$ or general dimensions, and in particular to ask if there are additional conditions on boundary correlation that are strong enough to imply bulk scattering. 

\subsection{CFT correlations as a resource}

As mentioned in the introduction, the distinction between CFT correlations that do \textit{vs.} those that do not support bulk scattering is interesting not only for understanding how boundary entanglement gives rise to geometry and causal processes in the bulk, but also from a quantum information theorist's perspective. 
In the boundary picture, bulk scattering is described by a non-local quantum computation, a subject of independent interest in quantum cryptography \cite{kent2011quantum,buhrman2014position,allerstorfer2023relating}. 
In \cite{May:2019odp,Maycomplexity_2022,dolev2022holography,may2023nonlocal}, it was argued that the AdS/CFT setting provides a more powerful method for implementing these non-local computations than protocols that have so far been understood in the quantum information literature. 
In particular, holographic protocols for non-local computation were argued to implement much higher complexity operations while using low entanglement than is possible in existing, explicit protocols. 

Our work provides a step towards understanding how these highly efficient holographic protocols work by clarifying which CFT states are useful resources for this quantum information processing task.
In particular, our work suggests that states satisfying the $u<d$ criterion enable efficient holographic protocols. 
Further, we observed that applying local CFT operators does not suffice to access bulk-scattering in many cases.
This suggests higher complexity, more delocalized, operations are needed to realize efficient protocols for the corresponding CFT states.

\begin{acknowledgments}
We are happy to thank Eric Chitambar, Dongjin Lee, Luis Lehner, Beni Yoshida, Jessica Yeh, and Jacob Abajian for fruitful discussions and useful comments.  Research at Perimeter Institute is supported in part by the Government of Canada through the Department of Innovation, Science and Economic Development Canada and by the Province of Ontario through the Ministry of Colleges and Universities. RCM is supported in part by a Discovery Grant from the Natural Sciences and Engineering Research Council of Canada, and by funding from the BMO Financial Group.
\end{acknowledgments}

\appendix

\section{The points theorem}
\label{sec:pointsappendix}

In this appendix, we repeat the holographic scattering analysis of section \ref{sec:holscat} using a ``points-based'' version of the connected wedge theorem, which we call the points theorem. In fact, the points theorem is a special case of the regions-based theorem used in the main text. The statement of the points theorem is exactly the same as the statement of the regions theorem (as in section \ref{sec:sec31}), except that $\mh{C}_i$ and $\mh{R}_i$ are restricted to be points $c_i$ and $r_i$ rather than extended regions. In this case, we have $\m{E}(\{c_1\})=\{c_1\}$, etc., so that the bulk inputs and outputs are identified with the boundary inputs and outputs.

As mentioned in the introduction, there are two physical reasons to consider the points formulation. 
Firstly, the points-based theorem can be easily argued for at the level of quantum information without assuming entanglement wedge reconstruction \cite{May:2019yxi}. 
Secondly, points-based scattering is more directly connected to the behaviour of boundary four-point functions.

We will find that a large class of connected entanglement wedges do not have a corresponding points-type scattering process despite having a corresponding regions-type scattering process.

\subsection{Results from the points theorem}
\label{sec:pointsresults}

We study when connected entanglement wedges $\m{E}(A\cup B)$ in the defect and BTZ spacetimes have or do not have a corresponding points-type holographic scattering process.

\begin{figure}[htbp]
\centering
\includegraphics[width=.5\textwidth]{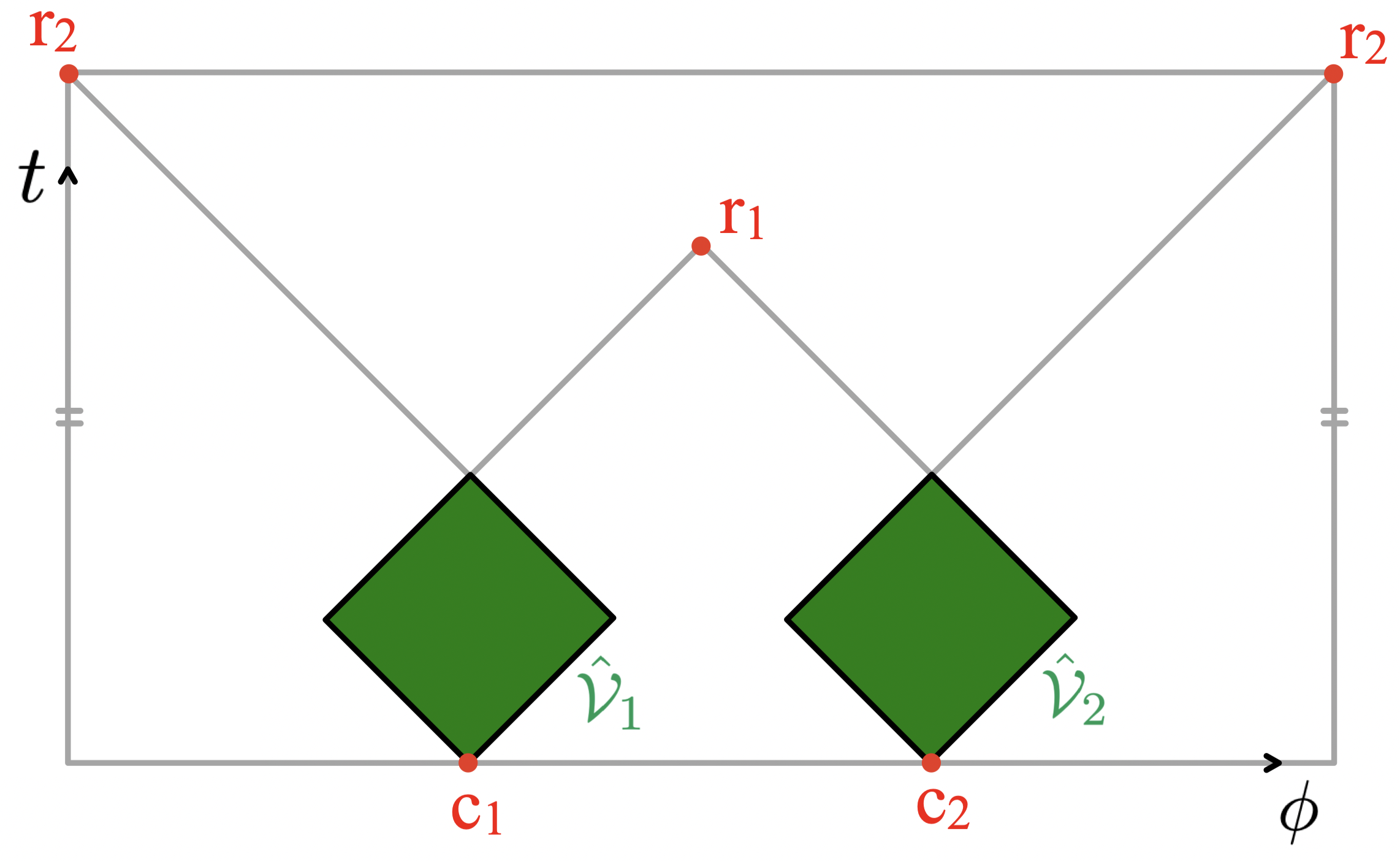}
\caption{An example boundary quantum task for the points-based theorem. The input and output data are encoded into points $c_i$ and $r_i$ rather than regions $\mh{C}_i$ and $\mh{R}_i$. 
\label{fig:pointsbdrysetup}}
\end{figure}

As in section \ref{sec:geometryandwedges}, we begin with two intervals $A$ and $B$. As in our previous analysis, we then identify the boundary decision regions as $\m{D}(A) = \mh{V}_1$ and $\m{D}(B) = \mh{V}_2$. 
Now, for the points-based analysis and following \cite{May:2019odp}, we place the $c_i$ at the past-most points of the $r_i$ and place the points $r_i$ in such a way as to satisfy the definition eq.~\eqref{eq:decisionregions} of the decision regions. That is, the points $r_i$ are positioned such that their past boundary light cones define the future boundaries of the two regions $\mh{V}_i$.
An example setup is shown in figure \ref{fig:pointsbdrysetup}.

As in the regions case, the choice of parameters $x$ and $\theta$ describing the regions $A$ and $B$ guarantees that boundary scattering is causally forbidden, \textit{i.e.}~$\hat{J}_{12\to12}=\emptyset$. Therefore, it remains to determine what parameter values of $x$ and $\theta$ lead to ${J}_{12\to12}\neq\emptyset$.

Physically, the condition ${J}_{12\to12}\neq\emptyset$ means that a bulk scattering process exists such that system 1 leaves from $\phi=\phi_{c_1}$ and returns to $\phi_{r_1}$ in time $\Delta T_{c_1,c_2\to r_1}$ and system 2 leaves from $\phi_{c_2}$ and returns to $\phi_{r_2}$ in time $\Delta T_{c_1,c_2\to r_2}$ such that $\Delta T_{c_1,c_2\to r_1}\leq t_{r_1}$ and $\Delta T_{c_1,c_2\to r_2}\leq t_{r_2}$. Here, $t_{r_1}$ and $t_{r_2}$, the time coordinates of $r_1$ and $r_2$ on the boundary, are fixed by $x$ and $\theta$ as follows:
\begin{equation}
t_{r_1}=\frac{\theta}{2}+x \,, \qquad \qquad
t_{r_2}=\pi+x-\frac{\theta}{2} \,,
\end{equation}
where the origin of time is taken to be $t_{c_1}=t_{c_2}$. 
See figure \ref{fig:phasetransition3dintro}.

Such a scattering process is possible if and only if a scattering process with only null geodesic trajectories is possible. This fact may be argued from the time-translation and rotational symmetries of the metrics considered in section \ref{sec:sec21}, and intuitively comes from the statement that nothing travels faster than light. 

\begin{figure}[htbp]
\centering
\includegraphics[width=.32\textwidth]{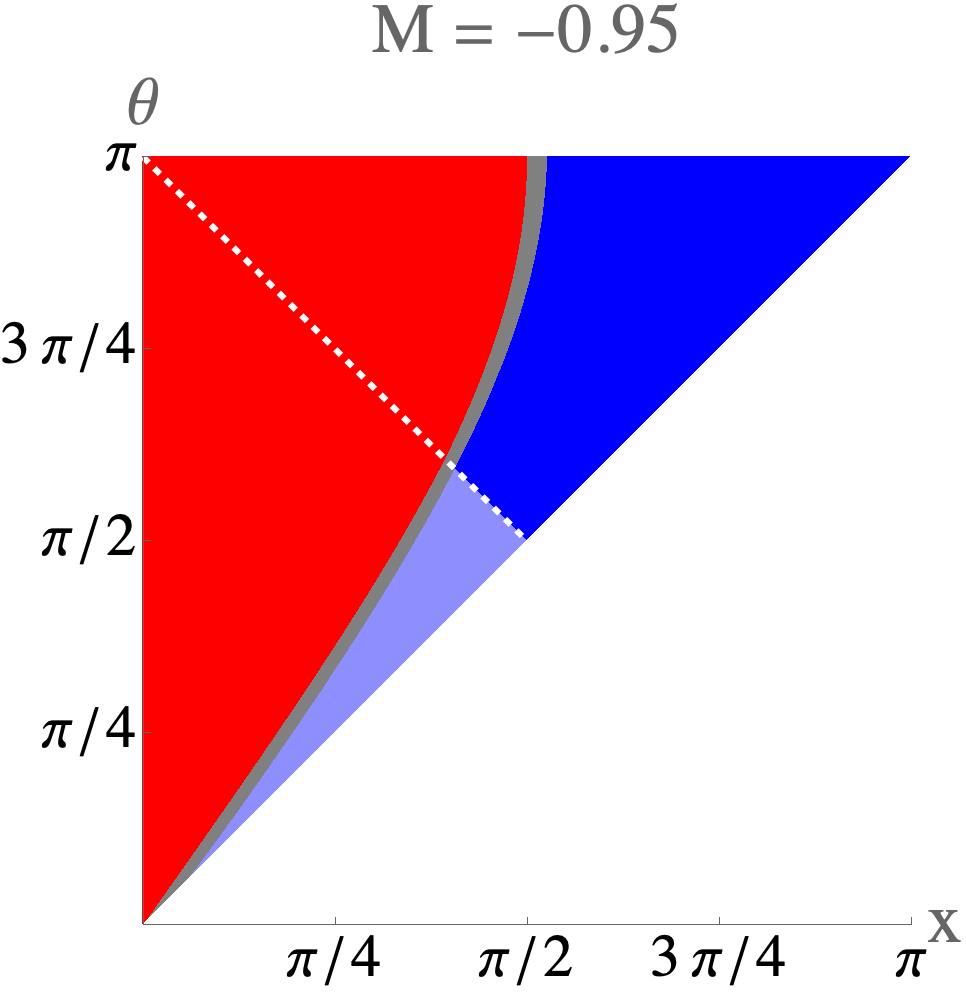}
\hfill
\includegraphics[width=.32\textwidth]{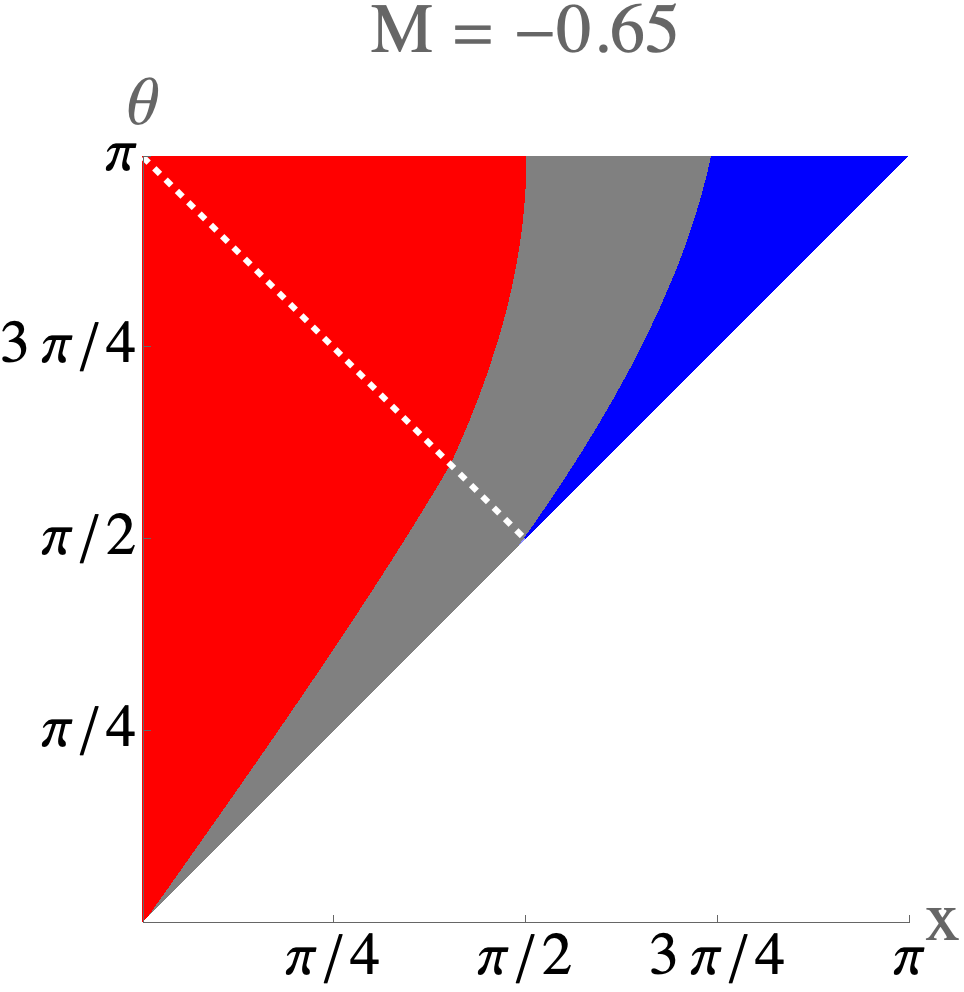}
\hfill
\includegraphics[width=.32\textwidth]{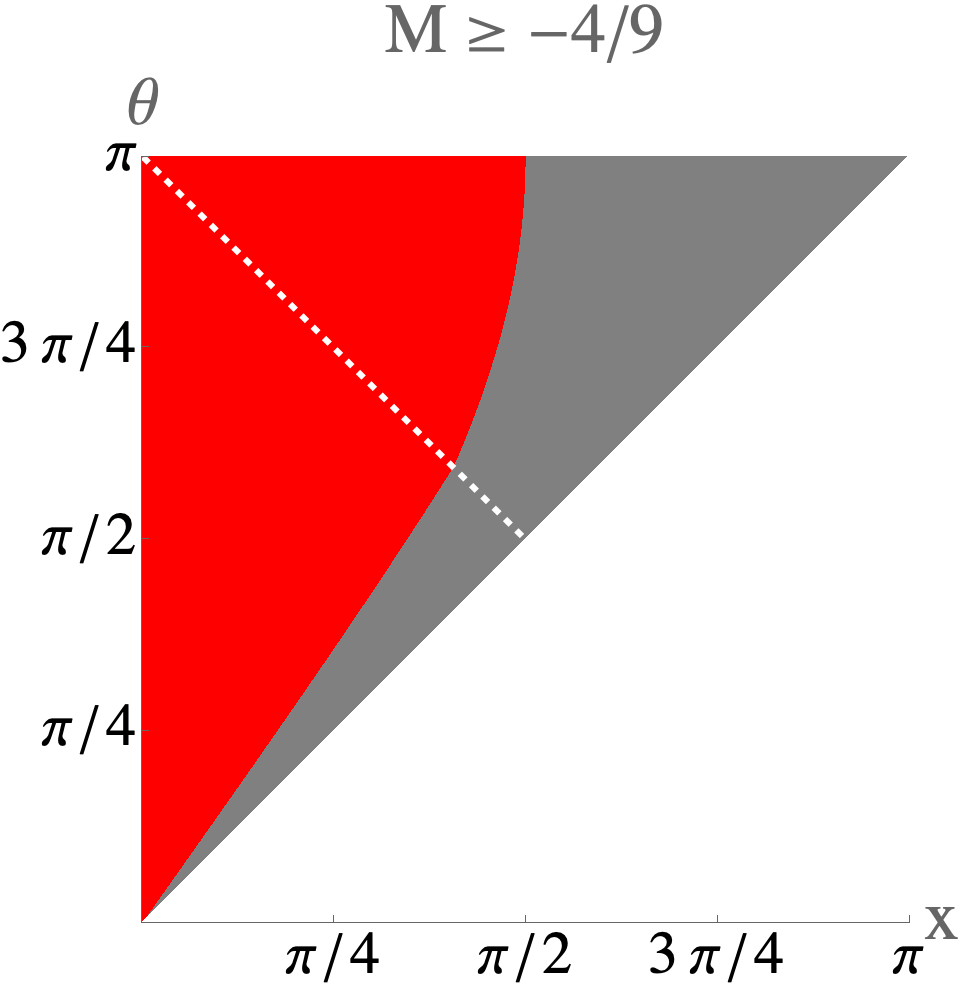}
\caption{Points-based holographic scattering phase diagrams in defect spacetimes with (from left to right) $M=-0.95$, $-0.65$, and $-4/9\simeq -0.444$. Contrast the regions-based phase diagram of figure \ref{fig:tempregionsconical}.
\label{fig:scatteringphasediagram}}
\end{figure}

To allow for a self-contained discussion, the details of the scattering problem are left to section \ref{sec:pointsdetails}. The result of the computation is the phase diagram \ref{fig:scatteringphasediagram} showing whether holographic scattering exists for each $x$ and $\theta$. As in the main text, the grey regions indicate $x$ and $\theta$ values for which $\m{E}(A\cup B)$ is connected but has no corresponding holographic scattering process, whereas the blue regions indicate cases of holographic scattering.

Note that in pure AdS ($M=-1$), every connected entanglement wedge has a corresponding holographic scattering process. As $M$ increases, fewer entanglement wedges correspond to a holographic scattering process, similarly to what we found in the regions-based discussion of section \ref{sec:sec32}. A striking difference with the regions case is that for $M\geq-\frac{4}{9}$, no connected entanglement wedges whatsoever have a corresponding holographic scattering process. 

As detailed in section \ref{sec:pointsdetails}, this effect persists in the BTZ case; namely, for points-type processes, scattering is impossible in the bulk for all $\theta$ and $x$ for which scattering is impossible in the boundary. Therefore, no connected entanglement wedges have a corresponding holographic scattering process. 
Thus, we found a huge class of connected entanglement wedges that do not have a corresponding points-type scattering process despite having a corresponding regions-type scattering process. 
This is consistent with the regions formulation of the theorem being more general than the points formulation.

\begin{figure}[htbp]
\centering
\begin{subfigure}[t]{0.35\textwidth}
    \centering
    \includegraphics[width=\textwidth]{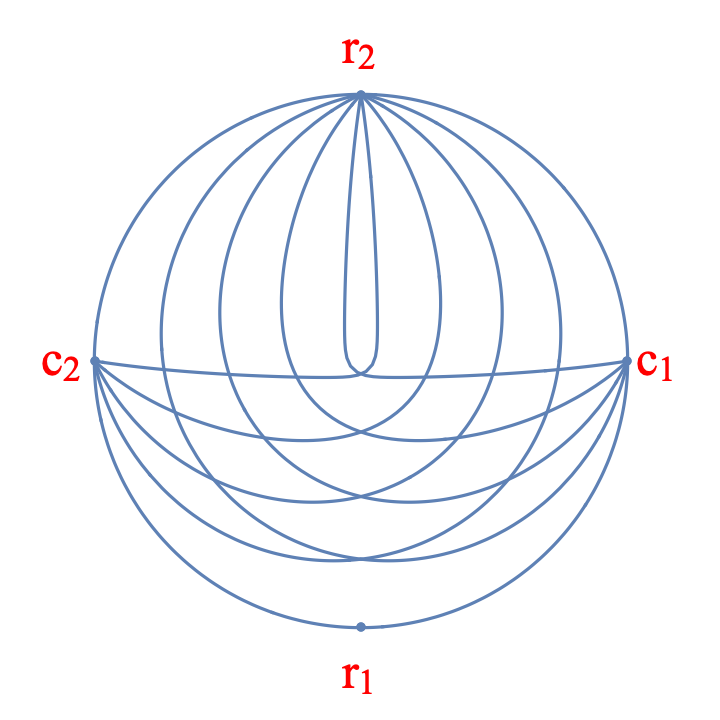}
    \caption{}
\end{subfigure}
\qquad \qquad
\begin{subfigure}[t]{0.35\textwidth}
    \centering
    \includegraphics[width=\textwidth]{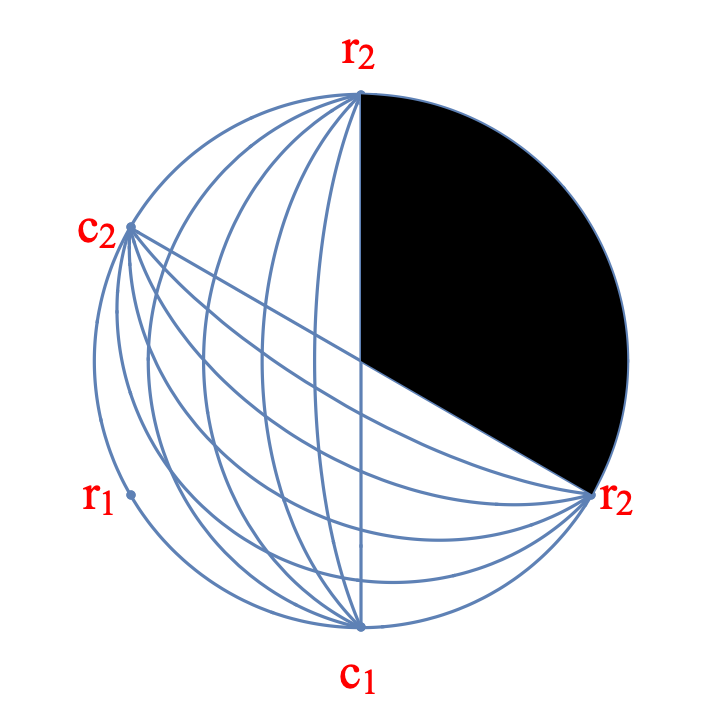}
    \caption{}
\end{subfigure}
\caption{Scattering problem with $\theta=\pi$ in the $M=-\frac{4}{9}$ geometry. The blue curves show possible null trajectories, with time direction suppressed. In figure (a), we employ the usual defect coordinates of eq.~\eqref{eq:mumetric}. In figure (b), we embed the defect geometry into pure AdS$_3$ via eq.~\eqref{eq:defecttopure}, revealing that $r_2$ is antipodal to both $c_1$ and $c_2$, and therefore a continuous family of curves with time delay $\pi$ passes from between antipodal points such as $c_1$ and $r_2$.
\label{fig:doublyantipodal}}
\end{figure}

Note, there is a geometric reason why $M=-\frac{4}{9}$ provides a threshold between the spacetimes in which holographic scattering is permitted vs. entirely forbidden. 
Recall from section \ref{sec:sec21} that the conical defect spacetime can be obtained from the AdS$_3$ spacetime by removing a wedge of angular extent $2\pi\left(1-\sqrt{|M|}\right)$. In the $M=-4/9$ case, one removes a wedge of $2\pi/3$ radians, as shown in figure \ref{fig:doublyantipodal}. 
Figure \ref{fig:doublyantipodal} also shows that, when $\theta=\pi$, $r_2$ is antipodal to both $c_1$ and $c_2$ from the perspective of pure AdS$_3$ (before removing the wedge). 
By symmetry, the analogous statement holds for $r_1$. Now, if $N$ and $S$ are antipodal points in the AdS$_3$ boundary, then every null geodesic emanating from $N$ arrives at $S$ in precisely $\pi$ of pure AdS$_3$ coordinate time.
From this, and the fact that $c_1$ and $c_2$ are fixed to lie in the $t=0$ slice, it follows that no matter where one picks the scattering point $(r=r_p, \phi=\phi_p)$ for null geodesics to meet, the scattering time always equals $\pi$. 
Thus, a degeneracy arises in which not only the minimal boundary scattering time and the minimal bulk scattering time are equal, but in fact the scattering time is the same no matter where one places the bulk scattering point. 
When $M$ increases beyond $-4/9$, the degeneracy breaks to a single optimal scattering point on the boundary, explaining the geometric reason for the $-4/9$ threshold.\footnote{Finding a scattering degeneracy for the extreme case of $\theta=\pi$ is key for understanding the importance of the $M=-4/9$ threshold. However, note that such a scattering degeneracy occurs in every defect spacetime when the regions are separated by $\theta=2\pi\left(1/\sqrt{|M|}-1\right)$. 
Indeed, this is the $\theta$-value in figure \ref{fig:scatteringphasediagram} where the blue and grey regions together meet the line $\theta=x$.}

\subsection{Details of the points computation}
\label{sec:pointsdetails}

In this section, we compute the main result of section \ref{sec:pointsresults}, namely the holographic scattering inequality in conical defect AdS$_3$. This computation generalizes the $M=-1$ computation in appendix C of \cite{May:2019yxi}. We conclude with a discussion of the BTZ case.

Recall from section \ref{sec:pointsresults} that a connected entanglement wedge between $A$ and $B$ has a corresponding holographic scattering process if and only if a scattering process consisting of exclusively null trajectories exists within the time constraint. Accordingly, consider null geodesics in the defect geometry. They are described by
\begin{equation}
\label{eq:sol}
\begin{aligned}
r(\lambda) &= \frac{1}{\sqrt{1-l^2}}\sqrt{|M|l^2+\lambda^2(1-l^2)^2} \,,\\
t(\lambda) &= \frac{1}{\sqrt{|M|}}\arctan\left(\frac{1-l^2}{\sqrt{|M|}}\lambda\right)+\frac{\pi}{2\sqrt{|M|}} + t_{-\infty} \,,\\
\phi(\lambda) &= \frac{1}{\sqrt{|M|}}\arctan\left(\frac{1-l^2}{l\sqrt{|M|}}\lambda\right)+\frac{\pi}{2\sqrt{|M|}} \, \text{sgn}(l) + \phi_{-\infty} \,.
\end{aligned}
\end{equation}
where $l$ is a constant of motion with the interpretation of angular momentum.

Note that as $\lambda\to-\infty$, we have $r(\lambda)\to\infty$. In words, $r$ decreases towards its minimal value as $\lambda$ increases to $0$. Therefore, the solution \eqref{eq:sol} captures the behavior of a massless test particle with angular momentum $l$ that begins its trajectory at $(t,\phi)=(t_{-\infty}, \phi_{-\infty})$ on the asymptotic boundary. In practice, the subscript $-\infty$ will be replaced with one of the $c_i$.

We can solve for the angular momentum $l$ by demanding that the particle passes through some bulk point $P$. In practice, $P=(t_p,r_p,\phi_p)$ will be the interaction point for the scattering. We find
\begin{equation}
\label{eq:angularmom}
\begin{aligned}
l^2 &= \frac{1}{1+\frac{|M|}{r_p^2}\csc^2\left(\sqrt{|M|}\,\Delta\phi\right)} \,,
\end{aligned}
\end{equation}
\noindent
where $\Delta\phi:=|\phi_{\infty}-\phi_p|$.

The time it takes the particle to travel from $(t,\phi)=(t_{-\infty}, \phi_{-\infty})$ on the boundary to $P$ in the bulk is then given by 
\begin{equation}
\label{eq:traveltime}
\begin{aligned}
\Delta T_{-\infty}:=|t(\lambda_p)-t_{-\infty}|=\frac{1}{\sqrt{|M|}}\left(\frac{\pi}{2}-\arctan\frac{\cos\left(\sqrt{|M|}\,\Delta\phi\right)}{\sqrt{\sin^2\left(\sqrt{|M|}\,\Delta\phi\right)+\frac{|M|}{r_p^2}}}\right) \,.
\end{aligned}
\end{equation}

By the time-reflection symmetry of the metric, this equation applies not only for a particle that falls towards $P$ from the boundary, but also a particle that shoots away from $P$ towards the boundary. In the former context, the subscript $-\infty$ will be replaced with one of the $c_i$; in the latter, we write $r_i$.

As discussed in section \ref{sec:pointsresults}, holographic scattering is possible if and only if the bulk scattering time from $c_1$ and $c_2$ to $r_1$ can be less than or equal to the boundary scattering time $t_{r_1}$, and likewise for $r_2$. We write
\begin{equation}
\label{eq:probtosolve}
\begin{aligned}
\Delta T_{c_1,c_2\to r_1}&\leq t_{r_1}=\frac{\theta}{2}+x \,,\\
\Delta T_{c_1,c_2\to r_2}&\leq t_{r_2}=\pi+x-\frac{\theta}{2} \,,
\end{aligned}
\end{equation}
\noindent
where $\Delta T_{c_1,c_2\to r_1} = \max(\Delta T_{c_1},\Delta T_{c_2}) + \Delta T_{r_1}$ and $\Delta T_{c_1,c_2\to r_2} = \max(\Delta T_{c_1},\Delta T_{c_2}) + \Delta T_{r_2}$. Since $t_{c_1}=t_{c_2}=0$, we have the convenient relation $\Delta T_{c_1}=\Delta T_{c_2}$. Also, $\phi_p=\phi_{r_1}$, since from figure \ref{fig:pointsbdrysetup} it is clear that $\phi_{r_1}$ lies halfway between $\phi_{c_1}$ and $\phi_{c_2}$. We can now write the $\Delta T$'s explicitly:

\begin{equation}
\label{eq:deltats}
\begin{aligned}
\Delta T_{c_1}=\Delta T_{c_2}&=\frac{1}{\sqrt{|M|}}\left(\frac{\pi}{2}-\arctan\frac{\cos\left(\sqrt{|M|}\, \theta \, / \,2\right)}{\sqrt{\sin^2\left(\sqrt{|M|}\, \theta\,/\,2\right)+\frac{|M|}{r_p^2}}}\right) \,,\\
\Delta T_{r_1}&=\frac{1}{\sqrt{|M|}}\left(\frac{\pi}{2}-\arctan\frac{r_p}{\sqrt{|M|}}\right) \,,\\
\Delta T_{r_2}&=\frac{1}{\sqrt{|M|}}\left(\frac{\pi}{2}-\arctan\frac{\cos\left(\sqrt{|M|}\,\pi\right)}{\sqrt{\sin^2\left(\sqrt{|M|}\,\pi\right)+\frac{|M|}{r_p^2}}}\right) \,,
\end{aligned}
\end{equation}

\noindent
and eq.\ \eqref{eq:probtosolve} becomes
\begin{equation}
\label{eq:ineqs}
\begin{aligned}
f_M(x, \theta,r_p)&:=\frac{\theta}{2}+x-(\Delta T_{c_1}+\Delta T_{r_1})&\geq0 \,,\\
g_M(x, \theta,r_p)&:=\pi+x-\frac{\theta}{2}-(\Delta T_{c_1}+\Delta T_{r_2})&\geq 0 \,.
\end{aligned}
\end{equation}

One can check that $f_M$ is monotonically increasing in $r_p$ and asymptotes to $x$ as $r_p\to\infty$. 
Therefore, there are always values of $r_p$ for which $f_M\geq0$.  On the other hand, $g_M$ can be either monotonically increasing or decreasing in $r_p$, depending on $x$ and $\theta$. 
Since $g_M$ asymptotes to a negative number, namely $x-\theta$, it is impossible to find values of $r_p$ where $g_M\geq0$ in the regime where $g_M$ is monotonically increasing. 
Fortunately, in the case where $g_M$ is monotonically decreasing towards $x-\theta$, one can check that there are always values of $r_p$ for which $g_M\geq0$.

Now, in an interval of $r_p$ values where both $f_M\geq0$ and $g_M\geq0$, we have that $f_M$ starts at the $(f_M=0)$-axis and monotonically increases in $r_p$, and $g_M$ starts out large and monotonically decreases in $r_p$ until it reaches the $(g_M=0)$-axis. Therefore, the existence of an interval of $r_p$ values where both $f_M\geq0$ and $g_M\geq0$ implies the existence of some $r_p^*$ where $f_M(r_p^*)=g_M(r_p^*)$. 

Therefore, we may reduce the two inequalities \eqref{eq:ineqs} to a single inequality that depends only on $x$ and $\theta$ by solving for $r_p^*$ and plugging it back into one of the inequalities. For concreteness, the solution for $r_p^*$ is the following:
\begin{equation}
\label{eq:rpstar}
\begin{aligned}
r_p^*=\frac{\sqrt{|M|}\,\sin\left(\sqrt{|M|}\,(\pi-\theta)\right)}{\cos\left(\sqrt{|M|}\,(\pi-\theta)\right)-\cos\left(\sqrt{|M|}\,\pi\right)} \,.
\end{aligned}
\end{equation}

By solving the resulting inequality numerically, we obtain the phase diagrams of figure \ref{fig:scatteringphasediagram}, which specify the ranges of $x$ and $\theta$ for which holographic scattering exists.

The numerics also verify that for $M \ge -4/9$, holographic scattering is impossible for all choices of $x$ and $\theta$, as argued from a geometric perspective in section \ref{sec:pointsresults}.
Further, as claimed in section \ref{sec:pointsresults}, this effect persists in the black hole case ($M\geq0$).
Let us verify this claim by studying null geodesics in the BTZ spacetime:
\begin{equation}
\label{eq:btzsol}
\begin{aligned}
r(\lambda) &= \frac{1}{\sqrt{1-l^2}}\sqrt{-M\,l^2+\lambda^2(1-l^2)^2} \,,\\
\phi(\lambda) &= \phi_{-\infty}-\frac{1}{\sqrt{M}}\,\text{arctanh}\left(\frac{l\sqrt{M}}{\lambda(1-l^2)}\right) \,,\\
t(\lambda) &= t_{-\infty}-\frac{1}{\sqrt{M}}\,\text{arctanh}\left(\frac{\sqrt{M}}{\lambda(1-l^2)}\right) \,.
\end{aligned}
\end{equation}

As $\lambda\to-\infty$, we have $r(\lambda)\to\infty$. Therefore, as with eq.\ \eqref{eq:sol}, eq.\ \eqref{eq:btzsol} captures the behavior of a massless test particle with angular momentum $l$ that begins its trajectory at $(t,\phi)=(t_{-\infty}, \phi_{-\infty})$ on the asymptotic boundary.

We again solve for the angular momentum $l$ by demanding that the particle passes through $(t_p,r_p,\phi_p)$. We find
\begin{equation}
\label{eq:angularmombtz}
\begin{aligned}
l^2 &= \frac{1}{1+\frac{M}{r_p^2}\text{csch}^2\left(\sqrt{M}\,\Delta\phi\right)} \,,
\end{aligned}
\end{equation}
\noindent
where $\Delta\phi:=|\phi_{\infty}-\phi_p|$.

We now solve for the time that it takes the particle to travel from $\left(t,\phi\right)=\left(t_{-\infty}, \phi_{-\infty}\right)$ on the boundary to $P$ in the bulk:
\begin{equation}
\label{eq:traveltimeBTZ}
\begin{aligned}
\Delta T_{-\infty}:=|t(\lambda_p)-t_{-\infty}|=\frac{1}{\sqrt{M}}\text{arctanh}\frac{\sqrt{\sinh^2\left(\sqrt{M}\,\Delta\phi\right)+\frac{M}{r_p^2}}}{\cosh\left(\sqrt{M}\,\Delta\phi\right)} \,.
\end{aligned}
\end{equation}

Crucially, \eqref{eq:traveltimeBTZ} is monotonically decreasing in $r_p$ for any choice of $M>0$ and $\Delta\phi$. 
Therefore, we could choose some $\phi_p$ for the scattering process and calculate $\Delta T_{c_1,c_2\to r_1}$ and $\Delta T_{c_1,c_2\to r_2}$ as a function of $r_p$. 
However, since both of these quantities are minimized when $r_p$ is taken to infinity, clearly they are always greater than the boundary scattering times $t_{r_1}$ and $t_{r_2}$. 
Therefore, scattering is impossible in the bulk for all $\theta$ and $x$ for which scattering is impossible in the boundary.\footnote{Note that eq.~\eqref{eq:traveltime} is monotonically decreasing in $r_p$ for any choice of $M>-\frac{1}{4}$, so the arguments here also apply to conical defect geometries with $M>-\frac{1}{4}$.\label{foot:fourthfootnote}}
It follows that there is no points-type holographic scattering for any $\m{E}(A\cup B)$ in the BTZ spacetime.
\section{Lightcones and wedges}
\label{sec:lightconeappendix}

In this appendix, we justify several claims in the main text regarding lightcones, entanglement wedges, and causal wedges in the defect and BTZ geometries.

\subsection{Causal vs.~entanglement wedges in defect geometry}
\label{sec:defectwedgesappendix}

We begin by examining the entanglement and causal wedges for a single interval on the boundary of the conical defect spacetime.
Above eq.~\eqref{eq:threeCs}, we claimed that the entanglement wedge $\m{E}(\mh{S})$ for a causal diamond $\mh{S}$ equals its causal wedge $\m{C}(\mh{S})$ if and only if the width of $\mh{S}$ is less than $\pi$.
In this section, we justify this claim.
Note, in the main text and throughout this appendix, we assume $\mh{S}=\m{D}(A)$, where $A$ is a region of width $\Delta \phi$ lying within the $t=0$ slice of $\hat{\partial}\m{M}$.

The entanglement wedge of $\mh{S}$, $\m{E}(\mh{S})$, is defined as the causal development of any spacelike codimension-1 surface $m$ in the bulk satisfying the homology condition $\partial{m}=\gamma \cup A$, where $\gamma$ is the RT surface for $A$. The causal wedge $\m{C}(\mh{S})$ of $\mh{S}$ is defined as $\m{J}^+(\mh{S})\cap\m{J}^-(\mh{S})$. Letting $s_-$ and $s_+$ be the past-most and future-most points of $\mh{S}$, respectively, we find $\m{C}(\mh{S})=\m{J}^+(s_-)\cap\m{J}^-(s_+)$.

In general, $\m{C}(\mh{S})\subseteq\m{E}(\mh{S})$ \cite{Hubeny_2012, Wall_2014, Engelhardt_2015}. 
To check whether we have strict inclusion or equality, we compare the $t=0$ slice of $\m{C}(\mh{S})$ with the $t=0$ slice of $\m{E}(\mh{S})$. 
By time-reflection symmetry, it suffices to check whether the boundary of $\m{J}^+(s_-) \cap \{t=0\}$ coincides with the RT surface.
We accordingly compute the surface $t=t(r,\phi)$ describing the boundary of $\m{J}^+(s_-)$, \ie the future lightcone of $s_-$. 

\begin{figure}[htbp]
\centering
\includegraphics[width=.22\textwidth]{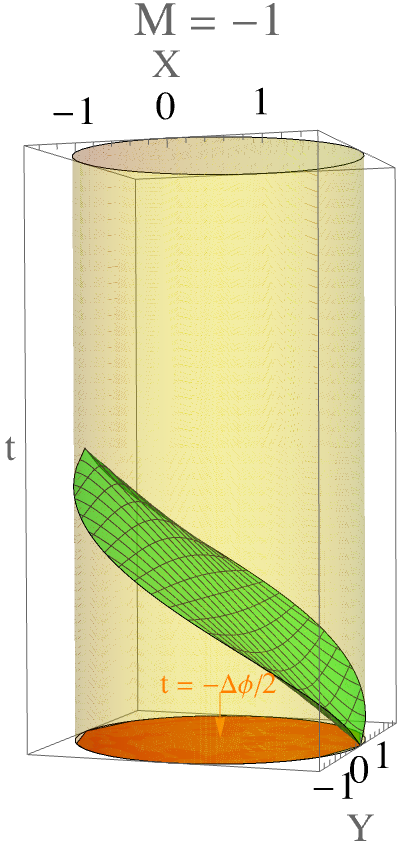}
\hfill
\includegraphics[width=.22\textwidth]{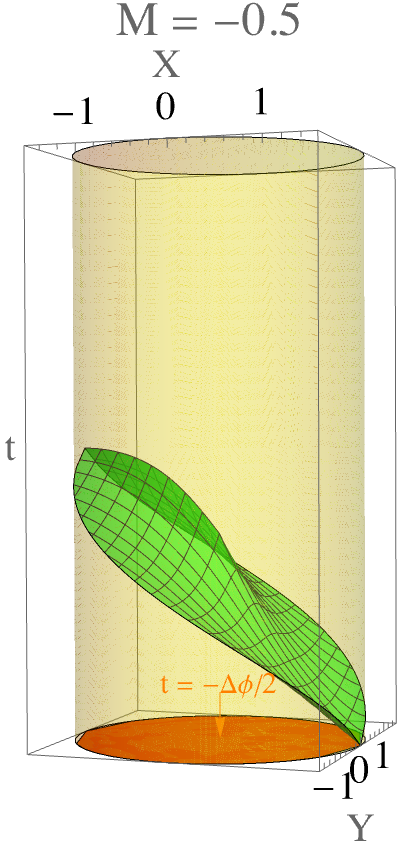}
\hfill
\includegraphics[width=.22\textwidth]{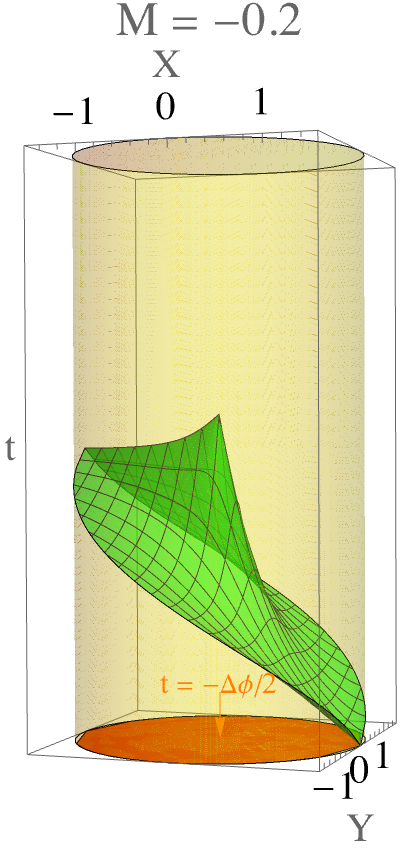}
\hfill
\includegraphics[width=.22\textwidth]{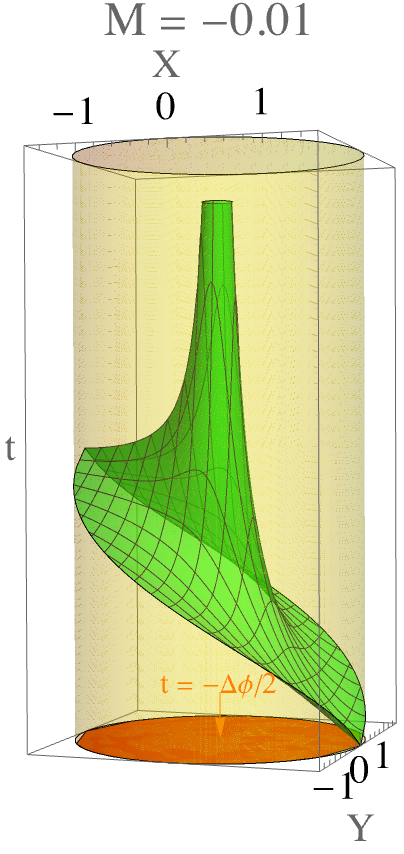}
\caption{The lightsheet \eqref{eq:conicallightsheet} of future-directed null rays emanating from a point $s_-$ on the conformal boundary of conical defect AdS$_3$. For reference, the leftmost diagram shows the case of pure AdS$_3$ ($M=-1$).
\label{fig:conicallightcones}}
\end{figure}

To do this, begin with eq.~\eqref{eq:sol} of appendix \ref{sec:pointsdetails}, which describes a general null geodesic in the conical defect geometry. 
In that equation, $t_{-\infty}$ and $\phi_{-\infty}$ describe the location of the null ray as $r\to\infty$, which in this case should be $s_-$. 
Centering $A$ at $(\phi=0,t=0)$, we obtain $\phi_{-\infty}=0$ and $t_{-\infty}=-\Delta\phi/2$. 
Eliminating $l$ and $\lambda$ in eqs.~\eqref{eq:sol}, we obtain the lightsheet
\begin{equation}
\label{eq:conicallightsheet}
t(r,\phi)=-\frac{1}{\sqrt{|M|}}\arctan\left(\cos\left(\sqrt{|M|} \,\phi\right)\sqrt{\frac{r^2}{|M|+r^2\sin^2\left(\sqrt{|M|}\,\phi\right)}}\right)+\frac{\pi}{2\sqrt{|M|}}+t_{-\infty} \,,
\end{equation}
\noindent
where even though we substituted $\phi_{-\infty}=0$, we have chosen to write $t_{-\infty}$ instead of $-\Delta\phi/2$ for clarity. This surface is plotted in figure \ref{fig:conicallightcones} for various values of $M<0$. 

\begin{figure}[htbp]
\centering
(a)
\includegraphics[width=.27\textwidth]{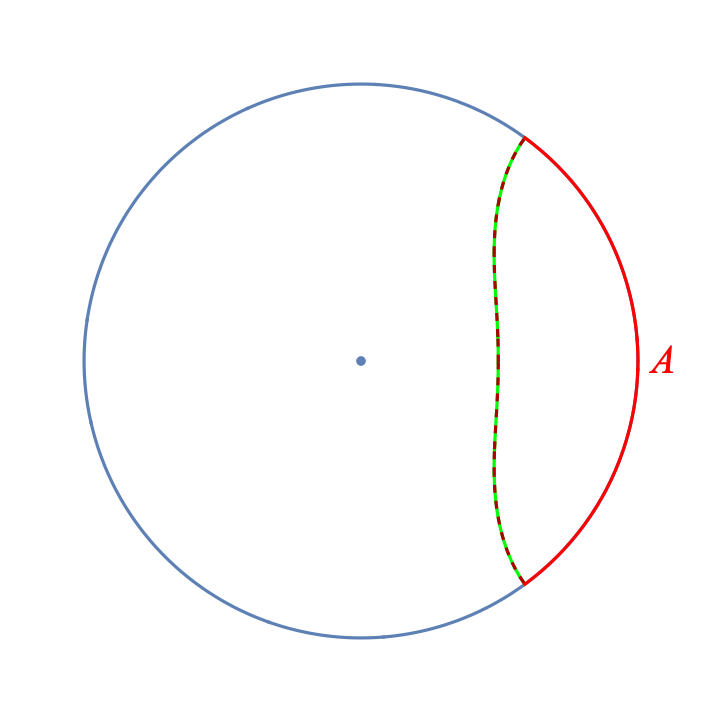}
(b)
\includegraphics[width=.27\textwidth]{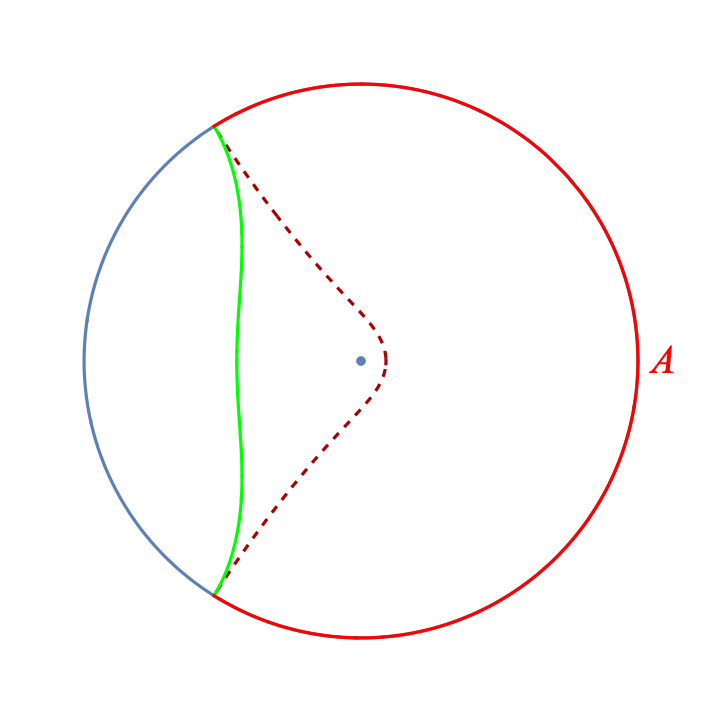}
(c)
\includegraphics[width=.27\textwidth]{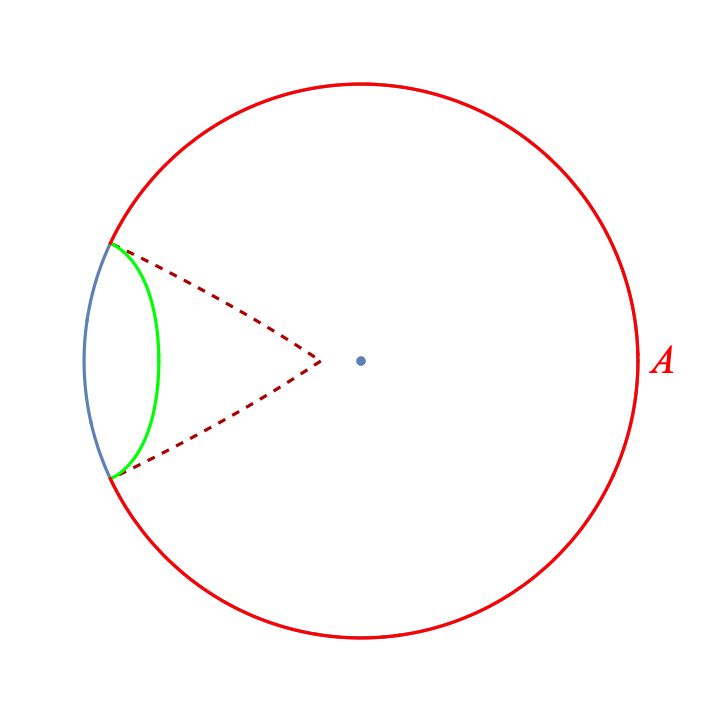}
\caption{Causal wedge boundary (red, dashed) versus entanglement wedge boundary (green) for the conical defect spacetime with $M=-0.4$. (a) When $\Delta\phi<\pi$, the entanglement wedge and causal wedge coincide. (b) When $\pi<\Delta\phi<\pi/\sqrt{|M|}$, the causal wedge boundary coincides with the non-minimal RT candidate of section \ref{sec:sec32}. (c) When  $\Delta\phi>\pi/\sqrt{|M|}$, the causal wedge has a kink.
\label{fig:circCW}}
\end{figure}

The intersection of the surface \eqref{eq:conicallightsheet} with the $t=0$ slice in the bulk reads
\begin{equation}
\label{eq:cw_at_tzero}
r(\phi) = 
\frac{\sqrt{|M|}\,\sec\!\left(\sqrt{|M|}\, \phi\right)}{\sqrt{\tan^2\!\left(\sqrt{|M|} \, \Delta\phi\, /\, 2\right)-\tan^2\!\left(\sqrt{|M|}\,\phi\right)} }\,,
\end{equation}
\noindent
where the range of $\phi$ is given by
\begin{equation}
\phi\in
\begin{cases}
\left(-\frac{\pi}{2\sqrt{|M|}},\frac{\pi}{2\sqrt{|M|}}\right) &\Delta\phi<\frac{\pi}{\sqrt{|M|}} \,, \\
\left[-\pi,-\frac{\pi}{2\sqrt{|M|}}\right)\cup\left(\frac{\pi}{2\sqrt{|M|}},\pi\right] &\Delta\phi>\frac{\pi}{\sqrt{|M|}} \,.
\end{cases}
\end{equation}
Examples of this curve are illustrated by the dashed red lines of figure \ref{fig:circCW}.

Consulting the RT surface \eqref{eq:sol1conical} for $A$, we find the causal wedge boundary coincides with the RT surface if and only if $A$ has width $\Delta\phi<\pi$. We conclude that $\m{E}(\mh{S})=\m{C}(\mh{S})$ if and only if $\Delta\phi<\pi$, as claimed.

\subsection{Causal vs.~entanglement wedges in BTZ geometry}

\begin{figure}[htbp]
\centering
\qquad
\includegraphics[width=.2\textwidth]{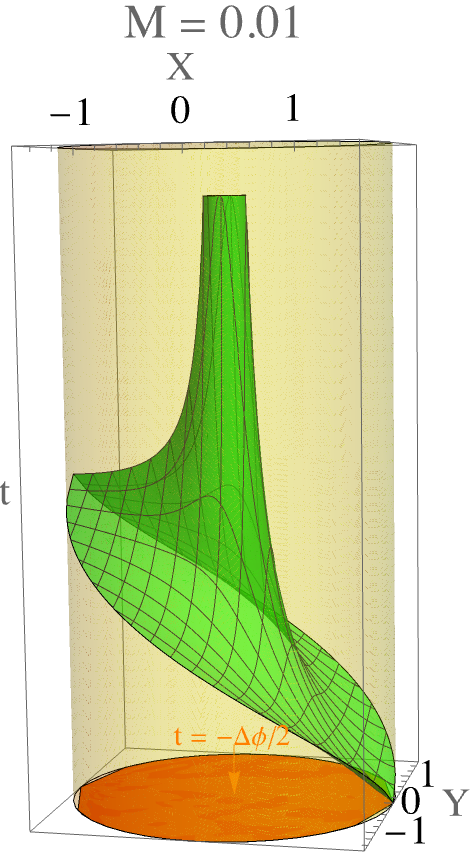}
\hfill
\includegraphics[width=.2\textwidth]{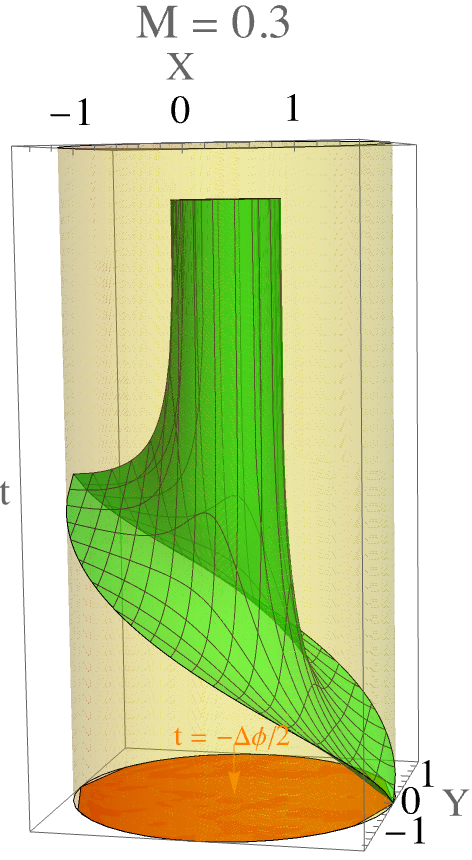}
\hfill
\includegraphics[width=.2\textwidth]{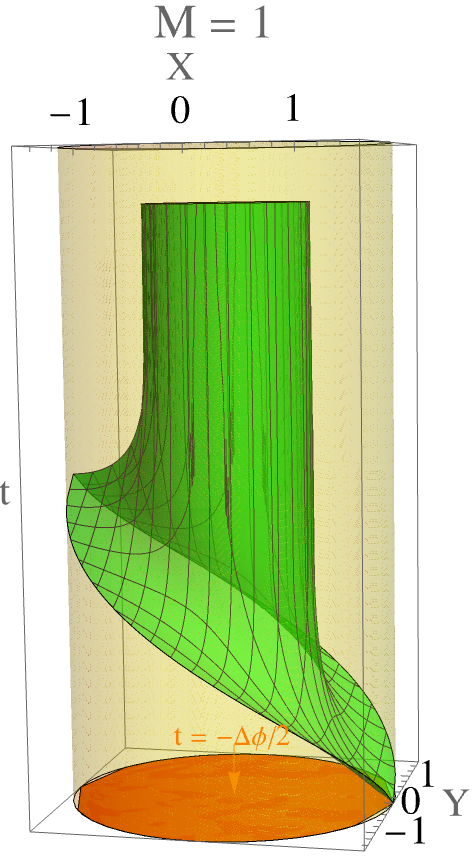}
\qquad
\caption{The lightsheet \eqref{eq:btzlightsheet} of future-directed null rays emanating from a point $s_-$ on the conformal boundary of various BTZ spacetimes.
\label{fig:btzlightcones}}
\end{figure}

In this section, we study when entanglement and causal wedges of a single boundary interval coincide in the BTZ spacetime.
Using eq.~\eqref{eq:btzsol}, the future lightcone of $s_-$ is described by
\begin{equation}
\label{eq:btzlightsheet}
t(r,\phi)=\frac{1}{\sqrt{M}}\, \mathrm{arctanh}\frac{\sqrt{M\,\mathrm{csch}^2\!\left(\sqrt{M}\phi\right)+r^2}}{r \, \mathrm{coth}\!\left(\sqrt{M}\phi\right)}+t_{-\infty} \,,
\end{equation}
\noindent
and examples are shown in figure \ref{fig:btzlightcones}. 
As in the previous subsection, we have imposed $\phi_{-\infty}=0$ and we can further set $t_{-\infty}=-\Delta\phi/2$.

\begin{figure}[htbp]
\centering
(a)
\includegraphics[width=.27\textwidth]{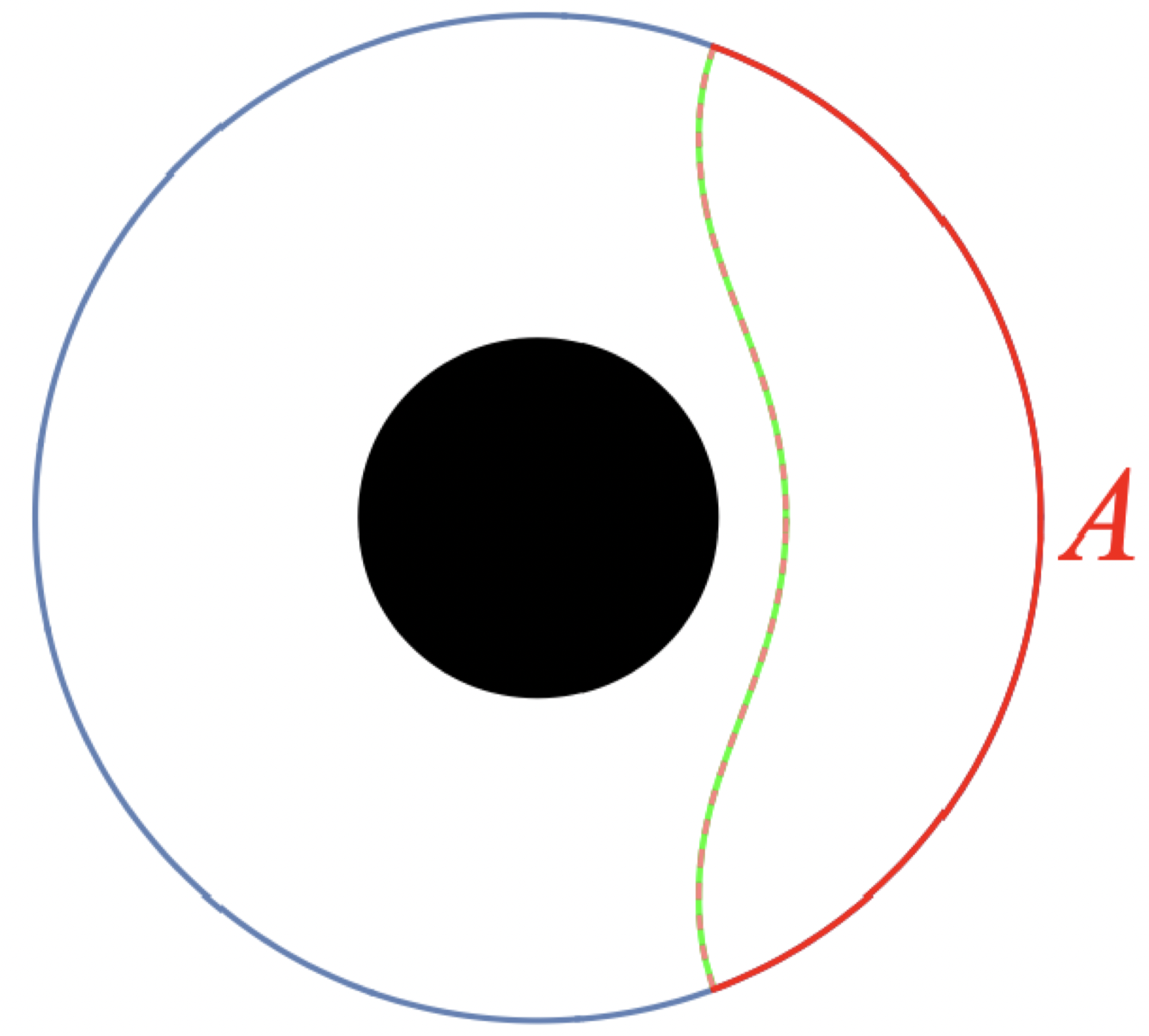}
\qquad\qquad
(b)
\includegraphics[width=.27\textwidth]{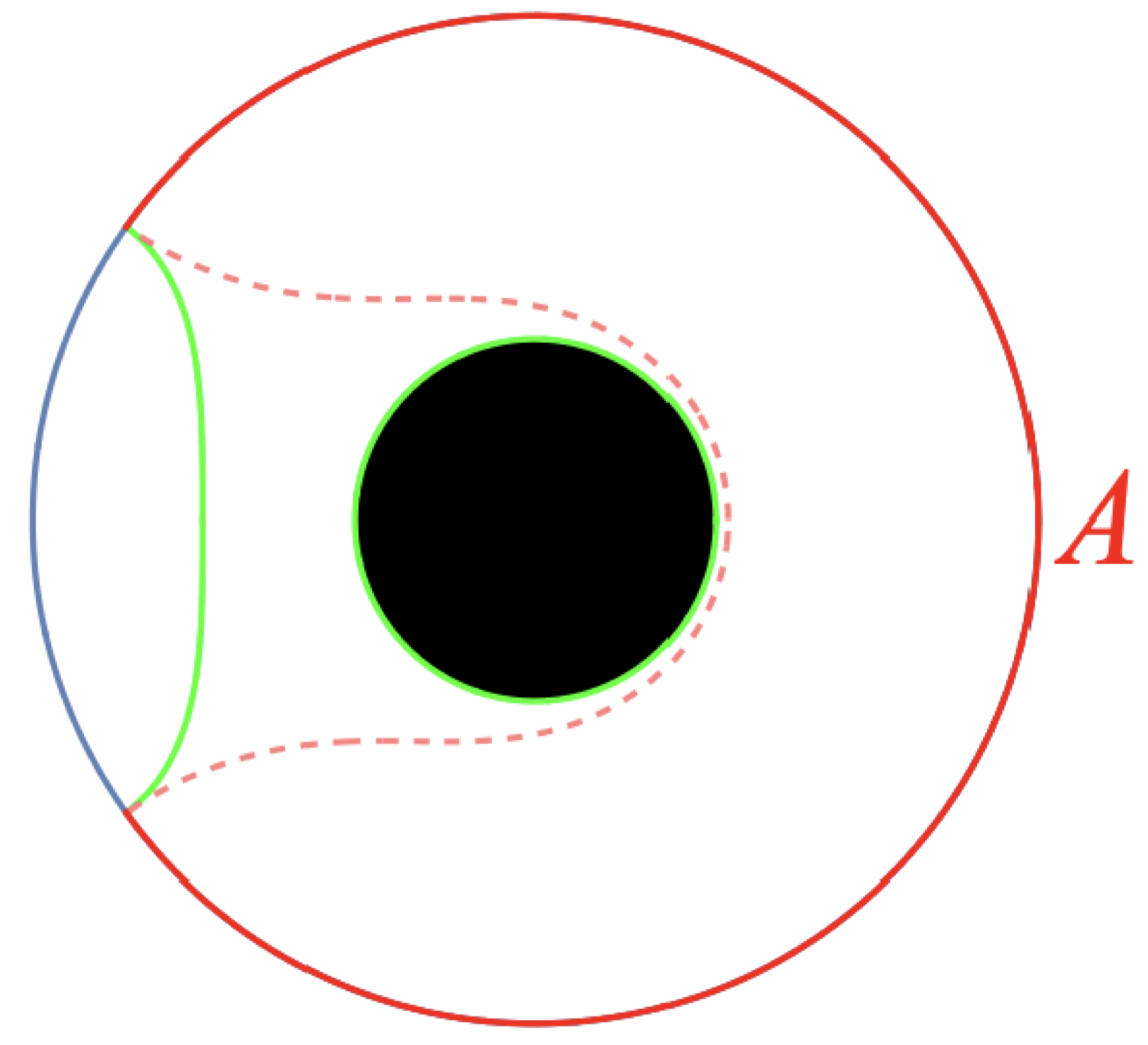}
\caption{The boundaries of the causal wedge (red, dashed) and the entanglement wedge (green) for the BTZ spacetime with $M=0.4$, for which $\Delta\phi^*(M)\approx 5$. (a) When $\Delta\phi<\Delta\phi^*(M)$, the entanglement and causal wedges coincide. (b) When $\Delta\phi>\Delta\phi^*(M)$, the causal wedge boundary coincides with the non-minimal RT candidate of section \ref{sec:sec33}.
\label{fig:circCWBH}}
\end{figure}

Restricting to $t=0$ yields the curve
\begin{equation}
\begin{aligned}
\label{eq:btz_cw_at_tzero}
r(\phi) &= \frac{\sqrt{M}\,\mathrm{sech}\!\left(\sqrt{M}\,\phi\right)}{\sqrt{\tanh ^2\!\left(\sqrt{M} \, \Delta\phi \, /\, 2\right)-\tanh ^2\!\left(\sqrt{M}\, \phi\right)}}\,,
\end{aligned}
\end{equation}
\noindent
which is depicted in figure \ref{fig:circCWBH}. 

We conclude, by comparison with eq.~\eqref{eq:gammaBTZ}, that the curve \eqref{eq:btz_cw_at_tzero} is the RT surface if and only if $A$ has width $\Delta\phi<\Delta\phi^*(M)$.

\subsection{Symmetry argument for the meeting of two particles}
\label{sec:meetingatP}

In this section, we prove the following claim of section \ref{sec:sec32}: particles leaving from $c_1$ and $c_2$ are able to meet at $\gamma_{\m{R}}$ if and only if they can meet at the minimal-radius point of $\gamma_{\m{R}}$, denoted $P_*$. The ``if'' direction is trivial because $P_*\in \gamma_{\m{R}}$. To prove the ``only if'' direction, we proceed by way of contradiction. 

Assume the particles can meet at $P\in \gamma_{\m{R}}$ such that $P\neq P_*$, yet they cannot meet at $P_*$. By symmetry, $c_1$ can signal to both $P$ and the reflection of $P$ about $\phi=0$, but it cannot signal to $P_*$. It follows that $\partial \m{J}^+(c_1)$ crosses $\gamma_{\m{R}}$ at least twice in the slice $t=t_{\gamma_{\m{R}}}$. Let $\mh{B}:=\mh{J}^+(c_1)\cap \mh{J}^-(b_1)$ where $b_1$ is the reflection of $c_1$ about $t=t_{\gamma_{\m{R}}}$. By construction, $\mh{B}$ is a causal diamond with the property $\m{J}^+(c_1)|_{_{t=t_{\gamma_{\m{R}}}}}=\m{C}(\mh{B})|_{_{t=t_{\gamma_{\m{R}}}}}$. By the monotonicity of $r'(\phi)$ in eqs.~\eqref{eq:sol1conical} and \eqref{eq:cw_at_tzero}, we conclude that $\partial \m{J}^+(c_1)=\partial \m{C}(\mh{B})$ crosses $\gamma_{\m{R}}$ exactly twice. It follows that $\mh{B} \supset \mh{R}_1$. Then, $\m{C}(\mh{B}) \supset \m{C}(\mh{R}_1)=\m{E}(\mh{R}_1)$, where the first relation follows from the nesting property of causal wedges \cite{Hubeny_2013}, and the second follows from $\theta<\pi$. Putting these together, we have $\m{C}(\mh{B})\supset\gamma_{\m{R}}$. This contradicts $\partial \m{J}^+(c_1) = \partial \m{C}(\mh{B})$ crossing $\gamma_{\m{R}}$.
\section{Holographic scattering inequalities from null geodesics}
\label{sec:regionsappendix}

In section \ref{sec:sec32}, we explained that holographic scattering exists in the defect spacetime if and only if the coordinate time $\Delta t$ that a null geodesic takes to travel from $(r, \phi)=(\infty, \frac{\theta}{2})$, the coordinates of $c_2$, to $(r, \phi)=(r_p^*, 0)$, the coordinates of the center $P^*$ of $\gamma_{\mh{R}_1}$, is less than $x$, the actual time separation between $c_1$ and $P^*$.

To determine the time $\Delta t$ from $(t,r, \phi)=(0,\infty, \frac{\theta}{2})$ to $(t, r, \phi)=(\Delta t,r_p^*, 0)$, let us first determine $r_p^*$, the radial coordinate for the turning point of $\gamma_{\m{R}_1}$. From eq.~\eqref{eq:gammaconicalAdS} with $\Delta\phi=\theta$, we have
\begin{equation}
\begin{aligned}
r(\phi) &= 
\frac{\sqrt{|M|}\,\sec\left(\sqrt{|M|}\,\phi\right)}{\sqrt{\tan^2\left(\sqrt{|M|}\,\theta \,/\,2\right)-\tan^2\left(\sqrt{|M|}\,\phi\right)}}\,,\qquad\text{ for }\phi\in\left(-\frac{\theta}{2},\, \frac{\theta}{2}\right) \,,
\end{aligned}
\end{equation}
\noindent
and by the symmetry about $\phi=0$, we obtain $r_p^*=r(0)=\sqrt{|M|}\cot\frac{\sqrt{|M|}\theta}{2}$.

To solve for $\Delta t$, treat eq. \eqref{eq:sol} with $t_{-\infty}=0$, $\phi_{-\infty}=\theta/2$ as a system of three equations for $\lambda$, $l$, and $\Delta t$. This yields
\begin{equation}
\begin{aligned}
\cot^2\frac{\sqrt{|M|}\,\theta}{2} &= \cot^2\left(\sqrt{|M|}\,\Delta t\right)\left(1+\sec^2\frac{\sqrt{|M|}\,\theta}{2}\right) \,.
\end{aligned}
\end{equation}
The condition $\Delta t<x$ then translates to 
\begin{equation}
\begin{aligned}
\cot^2\frac{\sqrt{|M|}\,\theta}{2} > \cot^2\left(\sqrt{|M|}\,x\right)\left(1+\sec^2\frac{\sqrt{|M|}\,\theta}{2}\right) \,,
\label{eq:conicalregionsineq}
\end{aligned}
\end{equation}
\noindent
which upon simplification yields $\cos^2\left(\sqrt{|M|}\,\theta \,/\,2\right) > \cos \left(\sqrt{|M|} \, x\right)$, \ie eq.~\eqref{eq:regionsineqconical} in the main text.

We now turn to the analogous computation in the BTZ case (section \ref{sec:sec33}) and compute the time $\Delta t$ along a null geodesic from $(t,r, \phi)=(0, \infty, \frac{\theta}{2})$ to $(t, r, \phi)=(\Delta t,r_p^*, 0)$, where $P^*$ is defined as the turning point of $\gamma_{\hat{\m{R}}_1}$ in the BTZ spacetime.

From eq. \ref{eq:gammaBTZ1} with $\Delta\phi=\theta$, we have the following curve for $\gamma_{\hat{\m{R}}_1}$:
\begin{equation}
\begin{aligned}
r(\phi) &= 
\frac{\sqrt{M}\left|\,\mathrm{sech}\left(\sqrt{M}\,\phi\right)\right|}{\sqrt{\tanh^2\left(\sqrt{M}\,\theta\,/\,2\right)-\tanh^2\left(\sqrt{M}\,\phi\right)}}\,, \qquad \text{ for }\phi\in\left(-\frac{\theta}{2},\,\frac{\theta}{2}\right) \,.
\end{aligned}
\end{equation}
\noindent
By the symmetry about $\phi=0$, we obtain $r_p^*=r(0)=\sqrt{M}\coth\frac{\sqrt{M}\theta}{2}$.

To solve for $\Delta t$, treat eq.~\eqref{eq:btzsol} with $t_{-\infty}=0$, $\phi_{-\infty}=\theta/2$ and the endpoint boundary condition $(t,r,\phi)=(\Delta t, r_p^*, 0)$ as a system of three equations for $\lambda$, $l$, and $\Delta t$. This yields
\begin{equation}
\begin{aligned}
\coth^2\left(\sqrt{M}\,\theta\,/\,2\right) = \coth^2\left(\sqrt{M}\,\Delta t\right)\left(1+\mathrm{sech}^2\left(\sqrt{M}\,\theta\,/\,2\right)\right) \,.
\end{aligned}
\end{equation}
The condition $x>\Delta t$ then translates to 
\begin{equation}
\begin{aligned}
\coth^2\left(\sqrt{M}\,\theta\,/\,2\right) > \coth^2\left(\sqrt{M}\,x\right)\left(1+\mathrm{sech}^2\left(\sqrt{M}\,\theta\,/\,2\right)\right) \,,
\label{eq:btzregionsineq}
\end{aligned}
\end{equation}
\noindent
which upon simplification yields $\cosh^2\left(\sqrt{M} \, \theta \, / \, 2\right) < \cosh \left(\sqrt{M}\, x\right)$, \textit{i.e.}~eq.~\eqref{eq:regionsineqbtz} in the main text.
\section{\texorpdfstring{Holographic scattering in pure AdS$_4$}{Holographic scattering in pure AdS4}}
\label{sec:higherdappendix}

In section \ref{sec:higherdDiscuss}, we claimed that Poincar\'e AdS$_4$ admits connected wedges with no corresponding scattering process.
In this section, we justify this claim.

In Poincar\'e coordinates, the pure AdS$_4$ metric reads
\begin{equation}
ds^2=r^2(-dt^2+dx_1^2+dx_2^2)+\frac{dr^2}{r^2}\,.
\label{eq:poincaremetric}
\end{equation}
At the boundary ($r\rightarrow \infty$), consider strips lying in the $t=0$ slice with infinite extent along the $x_2$ direction.
To emulate the Poincar\'e AdS$_3$ setup of \cite{May:2021nrl}, we choose the strips to have width $x$ and separation $\theta$ (of their midpoints) along the $x_1$ direction.
For the regions not to overlap, we demand $\theta > x$.
From \cite{Ben_Ami_2014}, the entanglement wedge inequality reads
\begin{equation}
\theta < \frac{1+\sqrt{5}}{2}\,x = \phi_g x\,,
\label{eq:stripwedgesconnected}
\end{equation}
where $\phi_g$ is the golden ratio.

To compute the holographic scattering inequality, define $\hat{\m{V}}_1$ and $\hat{\m{V}}_2$ as the causal developments of the strip intervals discussed above.
Similarly to the case of figure \ref{fig:C1C2plot_bdry} and surrounding discussion, it is optimal to define $\hat{\m{R}}_1$ and $\hat{\m{R}}_2$ as causal developments of complementary strips in the $t=\frac{x}{2}$ slice, where $\hat{\m{R}}_1$ has width $\theta$ and the complementary strip $\hat{\m{R}}_2$ has two components of infinite width extending out to $x_1\to\pm \infty$, respectively. 
Both $\hat{\m{R}}_1$ and $\hat{\m{R}}_2$ also have infinite extent in the $x_2$ direction.

\begin{figure}[tbp]
\centering
\includegraphics[width=.49\textwidth]{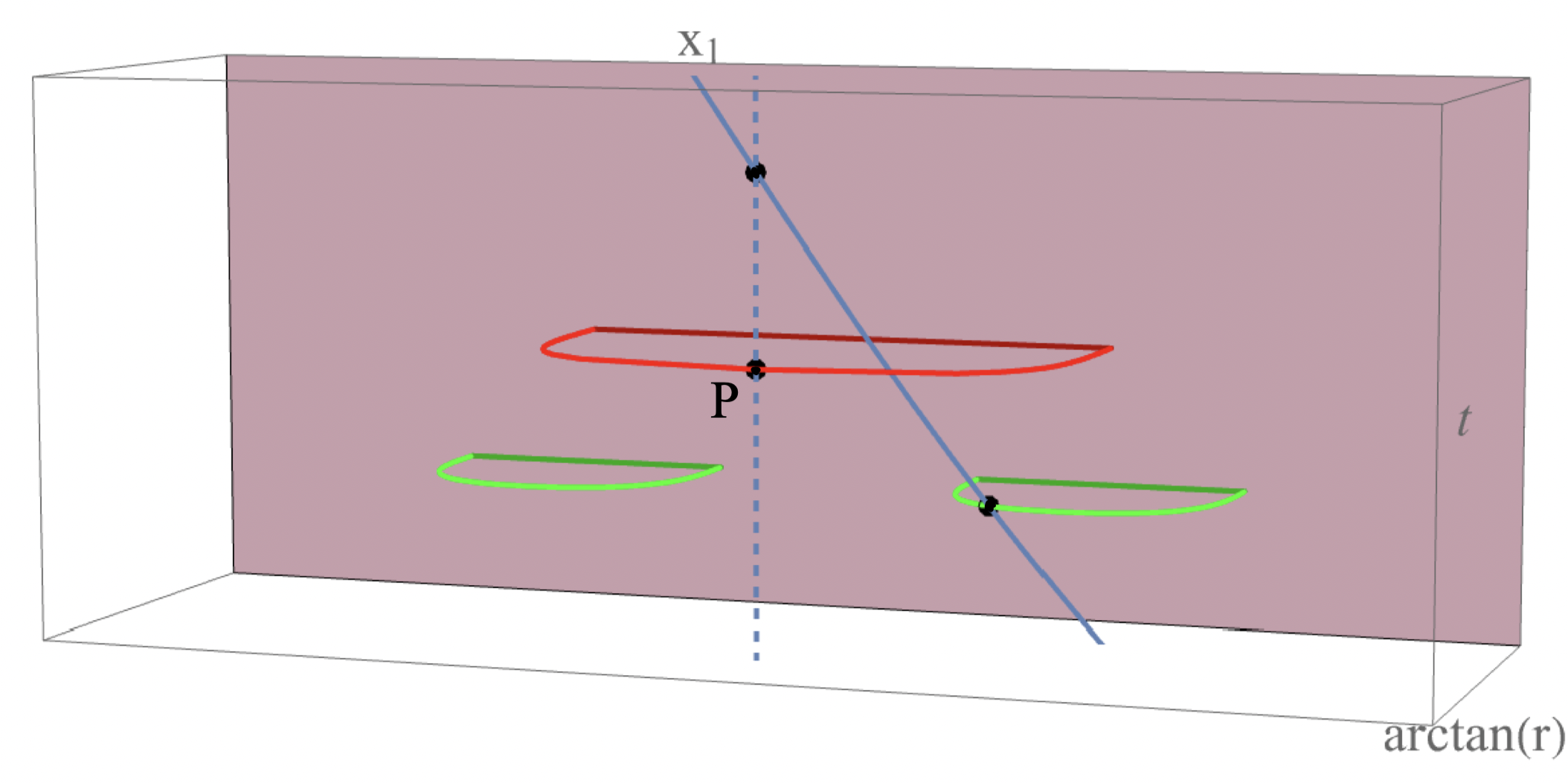}
\hfill
\includegraphics[width=.49\textwidth]{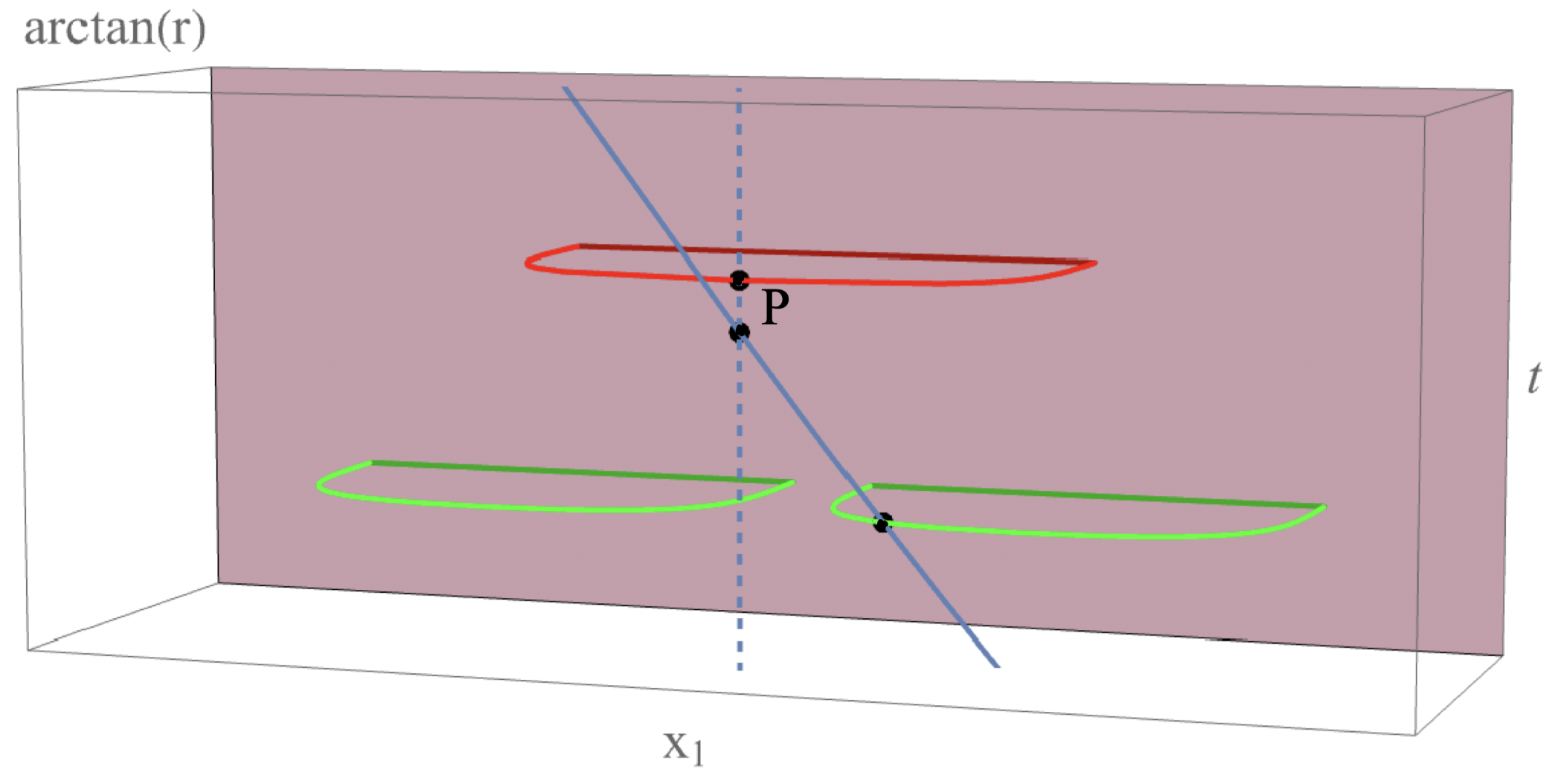}
\caption{Holographic scattering setup in Poincar\'e AdS$_4$ with the $x_2$ direction suppressed.
The purple plane represents the conformal boundary at $r\rightarrow \infty$.
The causal development of the green strip regions in the boundary are $\hat{\m{V}}_1$ and $\hat{\m{V}}_2$; the causal development of the red strip region in the boundary is $\hat{\m{R}}_1$.
The red and green geodesics hanging into the bulk are the corresponding RT surfaces: $\gamma_{\hat{\m{V}}_1}$, $\gamma_{\hat{\m{V}}_2}$, and $\gamma_{\hat{\m{R}}}$.
The center of $\gamma_{\hat{\m{R}}}$ is called $P$.
In both diagrams, the fastest null ray from $\gamma_{\hat{\m{V}}_2}$ to the blue, dashed line defined by $(r,p) = (r_p, x_p)$ is shown in blue.
In the left diagram, this null ray reaches $(r_p, x_p)$ to the future of $P$, so there is no holographic scattering.
In the right diagram, the null ray reaches $(r_p, x_p)$ to the past of $t=t_p$, so there is holographic scattering.
\label{fig:AdS4Scat}}
\end{figure}

Using the same reasoning as in section \ref{sec:sec32}, we can reduce the problem of showing $J_{12\rightarrow12}\neq\emptyset$ to the problem of showing that massless particles leaving from $\m{E}(\hat{\m{V}}_1)$ and $\m{E}(\hat{\m{V}}_2)$ can meet at $\gamma_{\m{R}}:=\gamma_{\hat{\m{R}}_1}=\gamma_{\hat{\m{R}}_2}$. 
In section \ref{sec:sec32}, the entanglement and causal wedges were equal for $\hat{\m{V}}_1$ and $\hat{\m{V}}_2$, so the massless particles leaving from $\m{E}(\hat{\m{V}}_1)$ and $\m{E}(\hat{\m{V}}_2)$ could be equivalently understood as null particles leaving from the past-most tips of $\hat{\m{V}}_1$ and $\hat{\m{V}}_2$. 
In the present case, the entanglement and causal wedges are distinct for the strip regions \cite{Hubeny_2012}, and so we must minimize the scattering time over all possible starting points in $\m{E}(\hat{\m{V}}_i)$.
By arguments about symmetry and promptness \cite{Witten_2020}, this amounts to determining whether a null geodesic departing orthogonally from $\gamma_{\hat{\m{V}}_2}$ (or, equivalently, $\gamma_{\hat{\m{V}}_1}$) can reach the line defined by $(r, x_1)=(r_p, x_p)$ before $t=t_p$, where $P:=(t_p, r_p, x_p)$ is the minimum-$r$ point of $\gamma_{\hat{\m{R}}_1}$.
Note, since it would cost extra time for the light ray to travel in the $x_2$ direction, we can simply set $x_2$ to an arbitrary constant throughout this discussion.
A diagram of the setup is shown in figure \ref{fig:AdS4Scat}.

Let us derive the condition for the null geodesic to be orthogonal to $\gamma_{\mh{V}_2}$.
The surface $\gamma_{\mh{V}_2}$ is determined by a profile function $x_1(r) = \frac{\theta}{2} \pm x_1^{RT}(r)$, where \cite{Fonda_2015}
\begin{equation}
\frac{d x_1^{RT}}{dr}=\frac{1}{r^2 \sqrt{\left(\frac{r}{r_{min}}\right)^4-1}}\,,
\label{eq:diffeq}
\end{equation}
and the conserved quantity $r_{min}$ is determined by setting  $\lim_{r\to\infty}x_1^{RT}(r) = x/2$.
This yields
\begin{equation}
\begin{aligned}
x_1^{RT}(r)&=\frac{\sqrt{\pi}\,\Gamma(\frac{3}{4})}{r_{min} \Gamma(\frac{1}{4})} - \frac{r_{min}^2}{3r^3}\,_2F_1\left(\frac{1}{2},\frac{3}{4},\frac{7}{4},\frac{r_{min}^4}{r^4}\right),\\
r_{min}&=r_{min}(x):=\sqrt{\frac{2}{\pi}}\cdot\frac{1}{x}\Gamma\left(\frac{3}{4}\right)^2\,.
\label{eq:rtcurve}
\end{aligned}
\end{equation}

Since the tangent vector to this curve (in the $t=const.$, $y=const.$ slice) is given by
\begin{equation}
    (r'(r), x_1'(r)) = \left(1, \pm \frac{d x_1^{RT}}{dr}\right)\,,
\end{equation}
the null ray leaving the RT surface orthogonally has, up to an overall rescaling, the following tangent vector at $\gamma_{R}$:
\begin{equation}
    X:=(X^t, X^r, X^{x_1}, X^{x_2}):= \left(- \textrm{sign}(l) \sqrt{1+r^4 \left(\frac{d x_1^{RT}}{dr}\right)^2}, \pm r^4 \frac{d x_1^{RT}}{dr}, -1, 0\right)\,,
    \label{eq:slope}
\end{equation}
where the meaning of $\textrm{sign}(l)$ will become apparent below.

Next, we solve for null geodesics in Poincar\'e AdS$_4$, parameterized by $\lambda$.
From the Killing symmetries of the metric, we obtain the following conserved quantities:
\begin{equation}
l=r^2 \frac{dx_1}{d\lambda} \qquad\qquad e = r^2 \frac{dt}{d\lambda}\,.
\end{equation}
Rescaling $\lambda$ such that $e=1$, the differential equation for $r(\lambda)$ becomes
\begin{equation}
\left(\frac{dr}{d\lambda}\right)^2=1-l^2.
\end{equation}
Choosing $\lambda<0$, the full solution reads
\begin{equation}
\begin{aligned}
r(\lambda)&=-\sqrt{1-l^2}\,\lambda\\
x_1(\lambda)&=-\frac{l}{(1-l^2)\lambda}+x_{-\infty}\\
t(\lambda)&=-\frac{1}{(1-l^2)\lambda}+t_{-\infty}.
\label{eq:ads4lightrays}
\end{aligned}
\end{equation}

To determine $x_{-\infty}$, $t_{-\infty}$, and $l$, we use the boundary conditions on the null geodesic. 
As discussed above, the curve can start anywhere on $\gamma_{\hat{\m{V}}_2}$, but with a specified slope given by eq.~\eqref{eq:slope}, and it must end somewhere on the line $(r,x_1)=(r_p, x_p)$.
We pick our origin such that $t=0$ at $\gamma_{\hat{\m{V}}_2}$ and $x_1=0$ at $P$ (\ie $x_p=0$). Then, the future-most endpoint of the geodesic is
\begin{equation}
(t, r, x_1)=(t_2, r_p, 0),
\label{eq:endpt}
\end{equation}
where $t_2$ is so far unknown, and $r_p=r_{min}(\theta)$ with $r_{min}$ defined in eq.~\eqref{eq:rtcurve}.
Notice that the holographic scattering inequality now simply reads $t_2<t_p=\frac{x}{2}$.

The starting point for the geodesic is
\begin{equation}
(t, r, x_1)=\left(0, \, r_1,\, \frac{\theta}{2} - x_1^{RT}(r_1)\right)\,,
\label{eq:startpt}
\end{equation}
where  $r_1$ is so far unknown, and we have chosen the minus sign in $x_1(r)$ because this gives the branch of $\gamma_{\hat{\m{V}}_2}$ that lies closer to $P^*$.

The slope condition \eqref{eq:slope} implies
\begin{equation}
    l = \frac{dx_1/d\lambda}{dt/d\lambda} = \frac{X^{x_1}}{X^{t}} = \pm \frac{1}{\sqrt{1+r^4 \left(\frac{d x_1^{RT}}{dr}\right)^2}}\Biggr|_{r=r_1}\,.
    \label{eq:slopeCondition}
\end{equation}
Note, we should impose $l<0$ for the null ray, because we want $x_1$ to decrease from $\frac{\theta}{2} - x_1^{RT}(r_1)$ towards zero as $r$ decreases from $r_1$ to $r_p$.

By imposing conditions \eqref{eq:endpt}--\eqref{eq:slopeCondition}, we can solve numerically for $t_2=t_2(x,\theta)$ and find $t_2<\frac{x}{2}$ if and only if $\theta \lessapprox 1.288 \, x$. 
Thus, the holographic scattering inequality reads
\begin{equation}
    \theta \lessapprox \frac{4}{5}\phi_g \,x \,.
\end{equation}

By comparing with eq.~\eqref{eq:stripwedgesconnected}, we conclude that for $\theta$ between $\frac{4}{5}\phi_g \,x$ and $\phi_g \,x$, the entanglement wedge is connected but there is no holographic scattering.

As mentioned in section \ref{sec:higherdDiscuss}, this implies that the $u<d$ condition is sufficient but not necessary for holographic scattering in AdS$_4$, at least within the class of examples we have studied.
Indeed, we expect that this argument can be straightforwardly lifted from Poincar\'e AdS$_4$ to global AdS$_4$, where the role of the infinite strip regions is played by annuli on the sphere at $r\rightarrow \infty$.

\bibliographystyle{jhep}
\bibliography{biblio}

\providecommand{\href}[2]{#2}\begingroup\raggedright\begin{thebibliography}{10}

\bibitem{May:2019odp}
A.~May, G.~Penington and J.~Sorce, \emph{{Holographic scattering requires a
  connected entanglement wedge}},
  \href{https://doi.org/10.1007/JHEP08(2020)132}{\emph{JHEP} {\bfseries 08}
  (2020) 132} [\href{https://arxiv.org/abs/1912.05649}{{\ttfamily
  1912.05649}}].

\bibitem{Mark_BuildingUpSpacetime}
M.~Van~Raamsdonk, \emph{{Building up spacetime with quantum entanglement}},
  \href{https://doi.org/10.1142/S0218271810018529}{\emph{Gen. Rel. Grav.}
  {\bfseries 42} (2010) 2323}
  [\href{https://arxiv.org/abs/1005.3035}{{\ttfamily 1005.3035}}].

\bibitem{CoolHorizons}
J.~Maldacena and L.~Susskind, \emph{{Cool horizons for entangled black holes}},
  \href{https://doi.org/10.1002/prop.201300020}{\emph{Fortsch. Phys.}
  {\bfseries 61} (2013) 781} [\href{https://arxiv.org/abs/1306.0533}{{\ttfamily
  1306.0533}}].

\bibitem{Faulkner_2014}
T.~Faulkner, M.~Guica, T.~Hartman, R.C.~Myers and M.~Van~Raamsdonk,
  \emph{{Gravitation from Entanglement in Holographic CFTs}},
  \href{https://doi.org/10.1007/JHEP03(2014)051}{\emph{JHEP} {\bfseries 03}
  (2014) 051} [\href{https://arxiv.org/abs/1312.7856}{{\ttfamily 1312.7856}}].

\bibitem{swingle2014universality}
B.~Swingle and M.~Van~Raamsdonk, \emph{{Universality of Gravity from
  Entanglement}},  \href{https://arxiv.org/abs/1405.2933}{{\ttfamily
  1405.2933}}.

\bibitem{Jacobson_1995}
T.~Jacobson, \emph{{Thermodynamics of space-time: The Einstein equation of
  state}}, \href{https://doi.org/10.1103/PhysRevLett.75.1260}{\emph{Phys. Rev.
  Lett.} {\bfseries 75} (1995) 1260}
  [\href{https://arxiv.org/abs/gr-qc/9504004}{{\ttfamily gr-qc/9504004}}].

\bibitem{Jacobson_2016}
T.~Jacobson, \emph{{Entanglement Equilibrium and the Einstein Equation}},
  \href{https://doi.org/10.1103/PhysRevLett.116.201101}{\emph{Phys. Rev. Lett.}
  {\bfseries 116} (2016) 201101}
  [\href{https://arxiv.org/abs/1505.04753}{{\ttfamily 1505.04753}}].

\bibitem{Donnelly_2008}
W.~Donnelly, \emph{{Entanglement entropy in loop quantum gravity}},
  \href{https://doi.org/10.1103/PhysRevD.77.104006}{\emph{Phys. Rev. D}
  {\bfseries 77} (2008) 104006}
  [\href{https://arxiv.org/abs/0802.0880}{{\ttfamily 0802.0880}}].

\bibitem{Donnelly_2012}
W.~Donnelly, \emph{{Decomposition of entanglement entropy in lattice gauge
  theory}}, \href{https://doi.org/10.1103/PhysRevD.85.085004}{\emph{Phys. Rev.
  D} {\bfseries 85} (2012) 085004}
  [\href{https://arxiv.org/abs/1109.0036}{{\ttfamily 1109.0036}}].

\bibitem{Bianchi_2023}
E.~Bianchi and E.R.~Livine, \emph{{Loop Quantum Gravity and Quantum
  Information}},  \href{https://arxiv.org/abs/2302.05922}{{\ttfamily
  2302.05922}}.

\bibitem{Rob_OnArch}
E.~Bianchi and R.C.~Myers, \emph{{On the Architecture of Spacetime Geometry}},
  \href{https://doi.org/10.1088/0264-9381/31/21/214002}{\emph{Class. Quant.
  Grav.} {\bfseries 31} (2014) 214002}
  [\href{https://arxiv.org/abs/1212.5183}{{\ttfamily 1212.5183}}].

\bibitem{Cooperman_2014}
J.H.~Cooperman and M.A.~Luty, \emph{{Renormalization of Entanglement Entropy
  and the Gravitational Effective Action}},
  \href{https://doi.org/10.1007/JHEP12(2014)045}{\emph{JHEP} {\bfseries 12}
  (2014) 045} [\href{https://arxiv.org/abs/1302.1878}{{\ttfamily 1302.1878}}].

\bibitem{Ryu06_1}
S.~Ryu and T.~Takayanagi, \emph{{Aspects of Holographic Entanglement Entropy}},
  \href{https://doi.org/10.1088/1126-6708/2006/08/045}{\emph{JHEP} {\bfseries
  08} (2006) 045} [\href{https://arxiv.org/abs/hep-th/0605073}{{\ttfamily
  hep-th/0605073}}].

\bibitem{Ryu06_2}
S.~Ryu and T.~Takayanagi, \emph{{Holographic derivation of entanglement entropy
  from AdS/CFT}},
  \href{https://doi.org/10.1103/PhysRevLett.96.181602}{\emph{Phys. Rev. Lett.}
  {\bfseries 96} (2006) 181602}
  [\href{https://arxiv.org/abs/hep-th/0603001}{{\ttfamily hep-th/0603001}}].

\bibitem{Hubeny_2007}
V.E.~Hubeny, M.~Rangamani and T.~Takayanagi, \emph{{A Covariant holographic
  entanglement entropy proposal}},
  \href{https://doi.org/10.1088/1126-6708/2007/07/062}{\emph{JHEP} {\bfseries
  07} (2007) 062} [\href{https://arxiv.org/abs/0705.0016}{{\ttfamily
  0705.0016}}].

\bibitem{Faulkner_2013}
T.~Faulkner, A.~Lewkowycz and J.~Maldacena, \emph{{Quantum corrections to
  holographic entanglement entropy}},
  \href{https://doi.org/10.1007/JHEP11(2013)074}{\emph{JHEP} {\bfseries 11}
  (2013) 074} [\href{https://arxiv.org/abs/1307.2892}{{\ttfamily 1307.2892}}].

\bibitem{Dong_2014}
X.~Dong, \emph{{Holographic Entanglement Entropy for General Higher Derivative
  Gravity}}, \href{https://doi.org/10.1007/JHEP01(2014)044}{\emph{JHEP}
  {\bfseries 01} (2014) 044} [\href{https://arxiv.org/abs/1310.5713}{{\ttfamily
  1310.5713}}].

\bibitem{May:2019yxi}
A.~May, \emph{{Quantum tasks in holography}},
  \href{https://doi.org/10.1007/JHEP10(2019)233}{\emph{JHEP} {\bfseries 10}
  (2019) 233} [\href{https://arxiv.org/abs/1902.06845}{{\ttfamily
  1902.06845}}].

\bibitem{May:2021nrl}
A.~May, \emph{{Holographic quantum tasks with input and output regions}},
  \href{https://doi.org/10.1007/JHEP08(2021)055}{\emph{JHEP} {\bfseries 08}
  (2021) 055} [\href{https://arxiv.org/abs/2101.08855}{{\ttfamily
  2101.08855}}].

\bibitem{May:2022clu}
A.~May, J.~Sorce and B.~Yoshida, \emph{{The connected wedge theorem and its
  consequences}}, \href{https://doi.org/10.1007/JHEP11(2022)153}{\emph{JHEP}
  {\bfseries 11} (2022) 153}
  [\href{https://arxiv.org/abs/2210.00018}{{\ttfamily 2210.00018}}].

\bibitem{Gary_2009}
M.~Gary, S.B.~Giddings and J.~Penedones, \emph{{Local bulk S-matrix elements
  and CFT singularities}},
  \href{https://doi.org/10.1103/PhysRevD.80.085005}{\emph{Phys. Rev. D}
  {\bfseries 80} (2009) 085005}
  [\href{https://arxiv.org/abs/0903.4437}{{\ttfamily 0903.4437}}].

\bibitem{Heemskerk_2009}
I.~Heemskerk, J.~Penedones, J.~Polchinski and J.~Sully, \emph{{Holography from
  Conformal Field Theory}},
  \href{https://doi.org/10.1088/1126-6708/2009/10/079}{\emph{JHEP} {\bfseries
  10} (2009) 079} [\href{https://arxiv.org/abs/0907.0151}{{\ttfamily
  0907.0151}}].

\bibitem{Penedones_2011}
J.~Penedones, \emph{{Writing CFT correlation functions as AdS scattering
  amplitudes}}, \href{https://doi.org/10.1007/JHEP03(2011)025}{\emph{JHEP}
  {\bfseries 03} (2011) 025} [\href{https://arxiv.org/abs/1011.1485}{{\ttfamily
  1011.1485}}].

\bibitem{maldacena2015looking}
J.~Maldacena, D.~Simmons-Duffin and A.~Zhiboedov, \emph{{Looking for a bulk
  point}}, \href{https://doi.org/10.1007/JHEP01(2017)013}{\emph{JHEP}
  {\bfseries 01} (2017) 013}
  [\href{https://arxiv.org/abs/1509.03612}{{\ttfamily 1509.03612}}].

\bibitem{Entwinement_2015}
V.~Balasubramanian, B.D.~Chowdhury, B.~Czech and J.~de~Boer, \emph{{Entwinement
  and the emergence of spacetime}},
  \href{https://doi.org/10.1007/JHEP01(2015)048}{\emph{JHEP} {\bfseries 01}
  (2015) 048} [\href{https://arxiv.org/abs/1406.5859}{{\ttfamily 1406.5859}}].

\bibitem{susskind1999holography}
L.~Susskind, \emph{{Holography in the flat space limit}},
  \href{https://doi.org/10.1063/1.1301570}{\emph{AIP Conf. Proc.} {\bfseries
  493} (1999) 98} [\href{https://arxiv.org/abs/hep-th/9901079}{{\ttfamily
  hep-th/9901079}}].

\bibitem{banks1999m}
T.~Banks, W.~Fischler, S.H.~Shenker and L.~Susskind, \emph{{M theory as a
  matrix model: A Conjecture}},
  \href{https://doi.org/10.1103/PhysRevD.55.5112}{\emph{Phys. Rev. D}
  {\bfseries 55} (1997) 5112}
  [\href{https://arxiv.org/abs/hep-th/9610043}{{\ttfamily hep-th/9610043}}].

\bibitem{anous2020areas}
T.~Anous, J.L.~Karczmarek, E.~Mintun, M.~Van~Raamsdonk and B.~Way, \emph{{Areas
  and entropies in BFSS/gravity duality}},
  \href{https://doi.org/10.21468/SciPostPhys.8.4.057}{\emph{SciPost Phys.}
  {\bfseries 8} (2020) 057} [\href{https://arxiv.org/abs/1911.11145}{{\ttfamily
  1911.11145}}].

\bibitem{Balasubramanian_2005}
V.~Balasubramanian, P.~Kraus and M.~Shigemori, \emph{{Massless black holes and
  black rings as effective geometries of the D1-D5 system}},
  \href{https://doi.org/10.1088/0264-9381/22/22/010}{\emph{Class. Quant. Grav.}
  {\bfseries 22} (2005) 4803}
  [\href{https://arxiv.org/abs/hep-th/0508110}{{\ttfamily hep-th/0508110}}].

\bibitem{BTZ_1992}
M.~Banados, C.~Teitelboim and J.~Zanelli, \emph{{The Black hole in
  three-dimensional space-time}},
  \href{https://doi.org/10.1103/PhysRevLett.69.1849}{\emph{Phys. Rev. Lett.}
  {\bfseries 69} (1992) 1849}
  [\href{https://arxiv.org/abs/hep-th/9204099}{{\ttfamily hep-th/9204099}}].

\bibitem{BTZ_1993}
M.~Banados, M.~Henneaux, C.~Teitelboim and J.~Zanelli, \emph{{Geometry of the
  (2+1) black hole}},
  \href{https://doi.org/10.1103/PhysRevD.48.1506}{\emph{Phys. Rev. D}
  {\bfseries 48} (1993) 1506}
  [\href{https://arxiv.org/abs/gr-qc/9302012}{{\ttfamily gr-qc/9302012}}].

\bibitem{Deser1984}
S.~Deser, R.~Jackiw and G.~'t~Hooft, \emph{{Three-Dimensional Einstein Gravity:
  Dynamics of Flat Space}},
  \href{https://doi.org/10.1016/0003-4916(84)90085-X}{\emph{Annals Phys.}
  {\bfseries 152} (1984) 220}.

\bibitem{Deser19842}
S.~Deser and R.~Jackiw, \emph{{Three-Dimensional Cosmological Gravity: Dynamics
  of Constant Curvature}},
  \href{https://doi.org/10.1016/0003-4916(84)90025-3}{\emph{Annals Phys.}
  {\bfseries 153} (1984) 405}.

\bibitem{Tsujimura_2020}
J.~Tsujimura and Y.~Nambu, \emph{{Null Wave Front and
  Ryu\textendash{}Takayanagi Surface}},
  \href{https://doi.org/10.3390/e22111297}{\emph{Entropy} {\bfseries 22} (2020)
  1297} [\href{https://arxiv.org/abs/2003.13374}{{\ttfamily 2003.13374}}].

\bibitem{Umemoto_2019}
K.~Umemoto, \emph{{Quantum and Classical Correlations Inside the Entanglement
  Wedge}}, \href{https://doi.org/10.1103/PhysRevD.100.126021}{\emph{Phys. Rev.
  D} {\bfseries 100} (2019) 126021}
  [\href{https://arxiv.org/abs/1907.12555}{{\ttfamily 1907.12555}}].

\bibitem{Kent_2012}
A.~Kent, \emph{{Quantum Tasks in Minkowski Space}},
  \href{https://doi.org/10.1088/0264-9381/29/22/224013}{\emph{Class. Quant.
  Grav.} {\bfseries 29} (2012) 224013}
  [\href{https://arxiv.org/abs/1204.4022}{{\ttfamily 1204.4022}}].

\bibitem{Wall_2014}
A.C.~Wall, \emph{{Maximin Surfaces, and the Strong Subadditivity of the
  Covariant Holographic Entanglement Entropy}},
  \href{https://doi.org/10.1088/0264-9381/31/22/225007}{\emph{Class. Quant.
  Grav.} {\bfseries 31} (2014) 225007}
  [\href{https://arxiv.org/abs/1211.3494}{{\ttfamily 1211.3494}}].

\bibitem{Witten_2020}
E.~Witten, \emph{{Light Rays, Singularities, and All That}},
  \href{https://doi.org/10.1103/RevModPhys.92.045004}{\emph{Rev. Mod. Phys.}
  {\bfseries 92} (2020) 045004}
  [\href{https://arxiv.org/abs/1901.03928}{{\ttfamily 1901.03928}}].

\bibitem{Hubeny:2012wa}
V.E.~Hubeny and M.~Rangamani, \emph{{Causal Holographic Information}},
  \href{https://doi.org/10.1007/JHEP06(2012)114}{\emph{JHEP} {\bfseries 06}
  (2012) 114} [\href{https://arxiv.org/abs/1204.1698}{{\ttfamily 1204.1698}}].

\bibitem{Maldacena:2001kr}
J.M.~Maldacena, \emph{{Eternal black holes in anti-de Sitter}},
  \href{https://doi.org/10.1088/1126-6708/2003/04/021}{\emph{JHEP} {\bfseries
  04} (2003) 021} [\href{https://arxiv.org/abs/hep-th/0106112}{{\ttfamily
  hep-th/0106112}}].

\bibitem{Orus:2018dya}
R.~Or\'us, \emph{{Tensor networks for complex quantum systems}},
  \href{https://doi.org/10.1038/s42254-019-0086-7}{\emph{APS Physics}
  {\bfseries 1} (2019) 538} [\href{https://arxiv.org/abs/1812.04011}{{\ttfamily
  1812.04011}}].

\bibitem{Banuls:2022vxp}
M.C.~Ba\~nuls, \emph{{Tensor Network Algorithms: A Route Map}},
  \href{https://doi.org/10.1146/annurev-conmatphys-040721-022705}{\emph{Ann.
  Rev. Condensed Matter Phys.} {\bfseries 14} (2023) 173}
  [\href{https://arxiv.org/abs/2205.10345}{{\ttfamily 2205.10345}}].

\bibitem{Basile:2023ycy}
I.~Basile, A.~Campoleoni and J.~Raeymaekers, \emph{{A note on the admissibility
  of complex BTZ metrics}},
  \href{https://doi.org/10.1007/JHEP03(2023)187}{\emph{JHEP} {\bfseries 03}
  (2023) 187} [\href{https://arxiv.org/abs/2301.11883}{{\ttfamily
  2301.11883}}].

\bibitem{Horowitz_1998}
G.T.~Horowitz and R.C.~Myers, \emph{{A}d{S}-{CFT} correspondence and a new
  positive energy conjecture for general relativity},
  \href{https://doi.org/10.1103/physrevd.59.026005}{\emph{Physical Review D}
  {\bfseries 59} (1998) }.

\bibitem{kent2011quantum}
A.~Kent, W.J.~Munro and T.P.~Spiller, \emph{Quantum tagging: {A}uthenticating
  location via quantum information and relativistic signaling constraints},
  {\emph{Physical Review A} {\bfseries 84} (2011) 012326}
  [\href{https://arxiv.org/abs/1008.2147}{{\ttfamily 1008.2147}}].

\bibitem{buhrman2014position}
H.~Buhrman, N.~Chandran, S.~Fehr, R.~Gelles, V.~Goyal, R.~Ostrovsky et~al.,
  \emph{{Position-Based Quantum Cryptography: Impossibility and
  Constructions}}, \href{https://doi.org/10.1137/130913687}{\emph{SIAM J.
  Comput.} {\bfseries 43} (2014) 150}.

\bibitem{allerstorfer2023relating}
R.~Allerstorfer, H.~Buhrman, A.~May, F.~Speelman and P.V.~Lunel,
  \emph{{Relating non-local quantum computation to information theoretic
  cryptography}},  \href{https://arxiv.org/abs/2306.16462}{{\ttfamily
  2306.16462}}.

\bibitem{Maycomplexity_2022}
A.~May, \emph{{Complexity and entanglement in non-local computation and
  holography}}, \href{https://doi.org/10.22331/q-2022-11-28-864}{\emph{Quantum}
  {\bfseries 6} (2022) 864} [\href{https://arxiv.org/abs/2204.00908}{{\ttfamily
  2204.00908}}].

\bibitem{dolev2022holography}
K.~Dolev and S.~Cree, \emph{{Holography as a resource for non-local quantum
  computation}},  \href{https://arxiv.org/abs/2210.13500}{{\ttfamily
  2210.13500}}.

\bibitem{may2023nonlocal}
A.~May and M.~Xu, \emph{{Non-local computation and the black hole interior}},
  \href{https://doi.org/10.1007/JHEP02(2024)079}{\emph{JHEP} {\bfseries 02}
  (2024) 079} [\href{https://arxiv.org/abs/2304.11184}{{\ttfamily
  2304.11184}}].

\bibitem{Hubeny_2012}
V.E.~Hubeny and M.~Rangamani, \emph{{Causal Holographic Information}},
  \href{https://doi.org/10.1007/JHEP06(2012)114}{\emph{JHEP} {\bfseries 06}
  (2012) 114} [\href{https://arxiv.org/abs/1204.1698}{{\ttfamily 1204.1698}}].

\bibitem{Engelhardt_2015}
N.~Engelhardt and A.C.~Wall, \emph{{Quantum Extremal Surfaces: Holographic
  Entanglement Entropy beyond the Classical Regime}},
  \href{https://doi.org/10.1007/JHEP01(2015)073}{\emph{JHEP} {\bfseries 01}
  (2015) 073} [\href{https://arxiv.org/abs/1408.3203}{{\ttfamily 1408.3203}}].

\bibitem{Hubeny_2013}
V.E.~Hubeny, M.~Rangamani and E.~Tonni, \emph{{Global properties of causal
  wedges in asymptotically AdS spacetimes}},
  \href{https://doi.org/10.1007/JHEP10(2013)059}{\emph{JHEP} {\bfseries 10}
  (2013) 059} [\href{https://arxiv.org/abs/1306.4324}{{\ttfamily 1306.4324}}].

\bibitem{Ben_Ami_2014}
O.~Ben-Ami, D.~Carmi and J.~Sonnenschein, \emph{Holographic entanglement
  entropy of multiple strips},
  \href{https://doi.org/10.1007/jhep11(2014)144}{\emph{Journal of High Energy
  Physics} {\bfseries 2014} (2014) }.

\bibitem{Fonda_2015}
P.~Fonda, L.~Giomi, A.~Salvio and E.~Tonni, \emph{On shape dependence of
  holographic mutual information in ads4},
  \href{https://doi.org/10.1007/jhep02(2015)005}{\emph{Journal of High Energy
  Physics} {\bfseries 2015} (2015) }.

\end{thebibliography}\endgroup

\end{document}